\documentclass[twocolumn]{aastex631}
\usepackage{graphicx}
\usepackage{amssymb,amsfonts,amsmath,amstext,amsgen,amsopn,amsxtra,indentfirst,times}

\usepackage{savesym}
\savesymbol{tablenum}
\usepackage{siunitx}
\DeclareSIUnit \parsec {pc}
\DeclareSIUnit \mag {mag}
\DeclareSIUnit \Rjup {R_J}
\restoresymbol{SIX}{tablenum}
\usepackage{chemformula}
\usepackage{booktabs}
\usepackage{rotating}
\usepackage{longtable}
\usepackage{xspace}

\newcommand{\heliosr}{\texttt{Helios-r2}\xspace}

\shorttitle{Retrieval Study of Brown Dwarfs}
\shortauthors{Lueber et al.}

\begin{document}

\title{Retrieval Study of Brown Dwarfs Across the L-T Sequence}

\correspondingauthor{Anna Lueber, Daniel Kitzmann, Kevin Heng}
\email{anna.lueber@unibe.ch, daniel.kitzmann@unibe.ch, kevin.heng@unibe.ch}

\author{Anna Lueber}
\affiliation{University of Bern, Center for Space and Habitability, Gesellschaftsstrasse 6, CH-3012, Bern, Switzerland}

\author[0000-0003-4269-3311]{Daniel Kitzmann}
\affiliation{University of Bern, Center for Space and Habitability, Gesellschaftsstrasse 6, CH-3012, Bern, Switzerland}

\author[0000-0003-2649-2288]{Brendan P. Bowler}
\affil{University of Texas at Austin, Department of Astronomy, 2515 Speedway, Stop C1400, Austin, TX 78712, U.S.A.}

\author[0000-0002-6523-9536]{Adam J. Burgasser}
\affil{Center for Astrophysics and Space Science, University of California San Diego, La Jolla, CA 92093, U.S.A.}

\author[0000-0003-1907-5910]{Kevin Heng}
\affiliation{University of Bern, Center for Space and Habitability, Gesellschaftsstrasse 6, CH-3012, Bern, Switzerland}
\affiliation{University of Warwick, Department of Physics, Astronomy \& Astrophysics Group, Coventry CV4 7AL, United Kingdom}
\affiliation{Ludwig Maximilian University, University Observatory Munich, Scheinerstr. 1, Munich D-81679, Germany}

\begin{abstract}
A large suite of 228 atmospheric retrievals is performed on a curated sample of 19 brown dwarfs spanning the L0 to T8 spectral types using the open-source \heliosr retrieval code, which implements the method of short characteristics for radiative transfer and a finite-element description of the temperature-pressure profile. Surprisingly, we find that cloud-free and cloudy (both gray and non-gray) models are equally consistent with the archival SpeX data from the perspective of Bayesian model comparison. Only upper limits for cloud properties are inferred if log-uniform priors are assumed, but the cloud optical depth becomes constrained if a uniform prior is used.

Water is detected in all 19 objects and methane is detected in all of the T dwarfs, but no obvious trend exists across effective temperature. As carbon monoxide is only detected in a handful of objects, the inferred carbon-to-oxygen ratios are unreliable. The retrieved radius generally decreases with effective temperature, but the values inferred for some T dwarfs are implausibly low and may indicate missing physics or chemistry in the models. For the early L dwarfs, the retrieved surface gravity depends on whether the gray or non-gray cloud model is preferred. Future data are necessary for constraining cloud properties and the vertical variation of chemical abundances, the latter of which is needed for distinguishing between the chemical instability versus traditional cloud interpretation of the L-T transition.
\end{abstract}


\section{Introduction}
\label{sect:intro}

Brown dwarfs are sub-stellar objects that are intermediate in mass ($\sim 13$--$80 M_{\rm J}$) between exoplanets and stars \citep{Burrows1993RvMP}. The exact mass threshold depends on the deuterium abundance and may range from $\sim 11$--$16 M_{\rm J}$ \citep{Spiegel2011ApJ}. In terms of spectral type, brown dwarfs are late-M dwarfs at young ages, L, T or Y dwarfs (see \citealt{Kirkpatrick2005ARA&A,Kirkpatrick2011ASPC,Cushing2011ApJ} for a review).

\begin{figure}
\begin{center}
\includegraphics[width=\columnwidth]{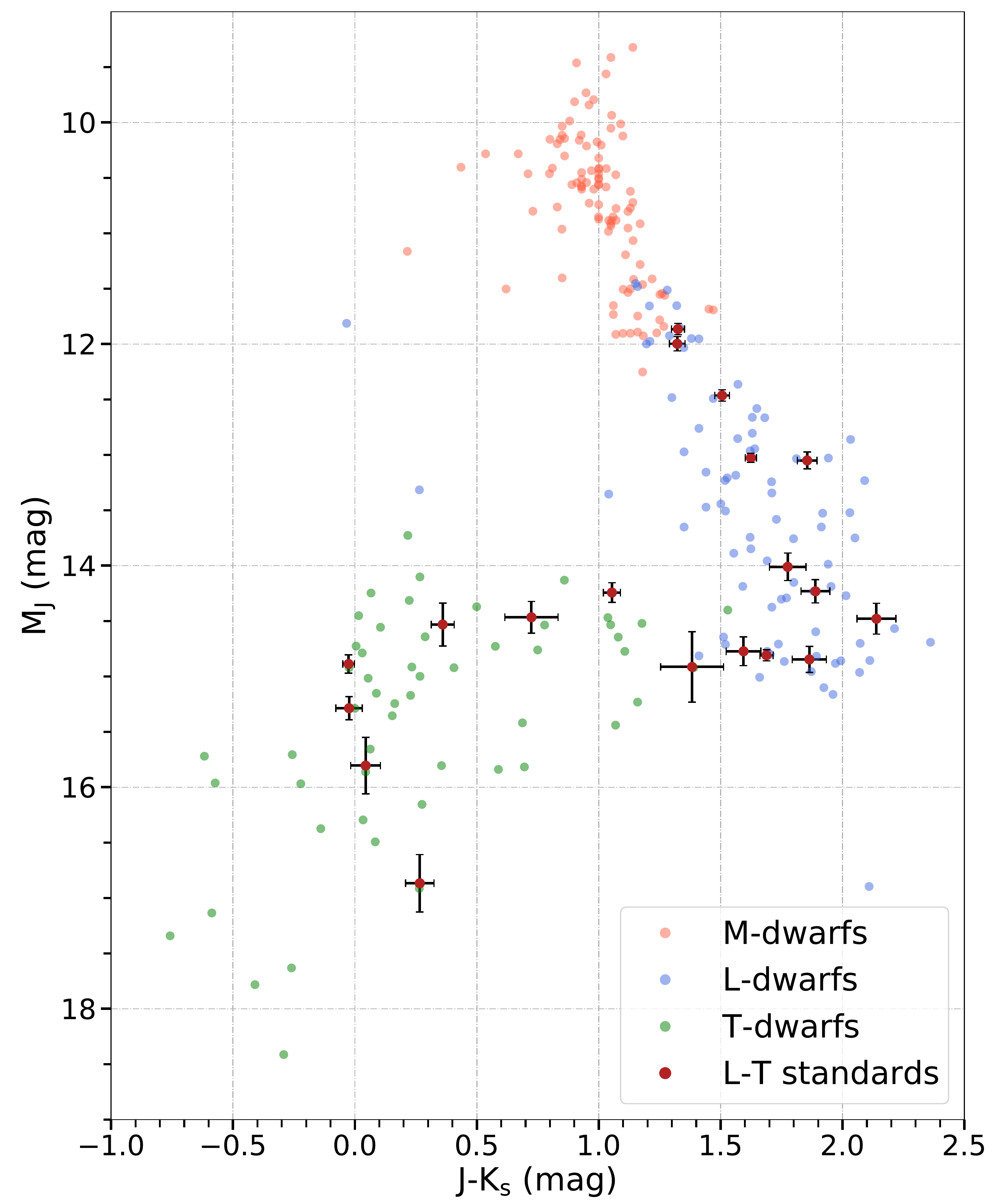}
\end{center}
\vspace{-0.2in}
\caption{Color-magnitude diagram of brown dwarfs. The brown circles with associated uncertainties form the standard L and T dwarfs of our curated sample. In the background are data taken from \citet{Dupuy2012ApJS} with measured parallaxes: M dwarfs (red circles), L dwarfs (blue circles) and T dwarfs (green circles).}
\label{fig:ColorMagnitude}
\end{figure}

Traditionally, brown dwarfs have been studied in the context of color-magnitude diagrams. Figure~\ref{fig:ColorMagnitude} shows our rendition, including a sample of 19 L and T dwarfs we have curated. The L-T transition has traditionally been interpreted as a variation in the apparent cloudiness of a brown dwarf as cloud layers recede below its photosphere with decreasing temperature \citep{Tsuji2003ApJ, Burrows2006ApJ, SaumonMarley2008ApJ}. Furthermore, variable brown dwarfs are typically more variable on the L-T transition \citep{Radigan2014ApJ}. More recently, the L-T transition has been interpreted as being caused by a chemical instability \citep{Tremblin2015ApJ, Tremblin2016ApJ}, although this interpretation has been challenged \citep{Leconte2018ApJ}.

Independent of the controversy surrounding the mechanism behind the L-T transition, the spectra of brown dwarfs as measured by ground-based telescopes are an excellent training ground for atmospheric retrieval, as they are of a comparable quality to future spectra of exoplanets obtained using the James Webb Space Telescope (JWST). Atmospheric retrieval provides a complementary approach to the traditional one of analysing brown dwarf spectra using pre-computed grids of atmospheric models (e.g., \citealt{Marley1996Sci, Burrows1997ApJ, Chabrier2000ApJ, Allard2001ApJ, Ackerman2001ApJ, Baraffe2002A&A, Burrows2003ApJ, Burrows2011ApJ, Morley2014ApJ, Zhang2021ApJ, Zhang2021arXiv210505256Z}). 

A major advantage of a retrieval analysis is its ability to constrain the abundances of chemical species beyond the assumptions that are usually made in atmospheric models, such as the validity of equilibrium chemistry, for example. The atmosphere's overall metallicity or C/O ratio are then an outcome of the retrieval rather than an input parameter like in a self-consistent atmospheric model. This allows one to directly obtain information on the enrichment of directly imaged planets or brown dwarfs, comparable to what we see in the atmospheres of the Solar System's gas and ice giants (see \citealt{Madhusudhan2016SSRv} for a detailed review).

The first comprehensive retrieval study of two benchmark T dwarfs was performed by \citet{Line2015ApJ}.
This pioneering work was continued further in \citet{Line2017ApJ} with an analysis of 11 T dwarfs. An important outcome of this study was the lack of a significant trend associated with the abundances of water, methane or ammonia with the brown dwarfs' equilibrium temperatures. On the other hand, decreasing abundances of the alkali metals sodium and potassium with the effective temperature were found. 
\citet{Burningham2017MNRAS} performed atmospheric retrievals for spectra of two L dwarfs, but were unable to draw decisive conclusions on their cloud properties. The same retrieval framework of \citet{Burningham2017MNRAS} was also used by \citet{Gonzales2020ApJ} to analyze a brown dwarf binary system, consisting of a L7 and a T7.5 dwarf. This study especially noted that disregarding the data blueward of $\SI{1.2}{\micro \metre}$ avoids potential issues with the shapes of the alkali resonance line wings. A similar conclusion was drawn by \citet{Oreshenko2020AJ}, who used pre-computed models as training sets for performing atmospheric retrieval using a supervised machine learning method.

None of the aforementioned studies have performed a suite of atmospheric retrievals on a sample of brown dwarfs spanning the L-T transition, which is the approach of the current study. Several key questions we wish to address include:
\begin{itemize}

    \item Are the retrieved chemical abundances and properties of brown dwarfs robust to assumptions about whether the atmospheres are cloud-free or populated with gray or non-gray clouds?
    
    \item Are cloudy models required to fit the spectra of L dwarfs?
    
    \item Are there trends in the retrieved chemical abundances across the L-T transition?
    
    \item Do the retrieved cloud properties vary across the L-T transition?

\end{itemize}

In Section~\ref{sect:methods}, we describe the ingredients of our Bayesian retrieval framework, as well as the curated set of 19 spectra. In Section~\ref{sect:results}, we report outcomes from benchmarking tests as well as answers to the aforementioned questions. In Section~\ref{sect:discussion}, we discuss limitations to our approach, which motivate opportunities for future work.\\

\section{Sample of L and T dwarf spectra}

For this study, we analyze several different brown dwarfs across the L-T sequence. Similar to previous studies (e.g., \citealt{Line2015ApJ,Kitzmann2020ApJ}), we use data taken by the ground-based SpeX instrument \citep{Rayner2003PASP} on the NASA Infrared Telescope Facility. The spectra for this work are taken directly from the SpeX Prism Libraries \citep{Burgasser2014ASInC..11....7B}\footnote{\url{http://www.browndwarfs.org/spexprism}}. SpeX Prism spectra typically cover a wavelength range from $\SI{0.85}{\micro\metre}$ to about $\SI{2.45}{\micro\metre}$, with a spectral resolution $\lambda/\Delta \lambda$ that varies between 85 and 300.

From the SpeX Prism Libraries we curate a sample of 19 brown dwarfs that represent a clean spectral sequence of L and T dwarfs spanning from class L0 to T8. Several standard (as defined by the SpeX spectral library) brown dwarfs that were previously found to be close binaries have been excluded. One such example is Kelu 1 that was considered to be a L2 standard but has been revealed to be an early and mid L binary by \citet{Liu2005ApJ...634..616L}. Our L0 and T0 templates (2MASS J03454316+2540233 and SDSS J120747.17+024424.8, respectively) are suspected to be unresolved binaries, but so far no conclusive observations on their potential binary nature has been obtained \citep{Dahn2017AJ,Burgasser2010ApJ}. Therefore, we choose to keep them in our sample.

\begin{deluxetable*}{lccccccc}
\tablecaption{Set of brown dwarfs and their observational characteristics used in this study.}
\label{tab:objects}
\tablehead{
\colhead{Object} & \colhead{$d$} & \colhead{$\Delta d$} & \colhead{\textit{J}} & \colhead{\textit{H}} & \colhead{\textit{K}$_S$} & \colhead{NIR SpT} & \colhead{References} \\ \tableline
 & \colhead{($\SI{}{\parsec}$)} & \colhead{($\SI{}{\parsec}$)} &  \colhead{($\SI{}{\mag}$)}  & \colhead{($\SI{}{\mag}$)} &  \colhead{($\SI{}{\mag}$)}  &  \colhead{---}  & \colhead{---} 
}
\startdata
2MASS J03454316+2540233 & 26.6955 & 0.2985 & 13.997 $\pm$ 0.027 & 13.211 $\pm$ 0.030 & 12.672 $\pm$ 0.024 & L0 & [1], [12] \\
2MASS J21304464-0845205 & 26.7916 & 0.3163 & 14.137 $\pm$ 0.032 & 13.334 $\pm$ 0.032 & 12.815 $\pm$ 0.033 &  L1 & [2], [12] \\
SSSPM J0829-1309 & 11.6899 & 0.0235 & 12.803 $\pm$ 0.03 & 11.851 $\pm$ 0.022 & 11.297 $\pm$ 0.021 & L2 & [3], [12] \\
2MASS J15065441+1321060 & 11.6848 & 0.0393 & 13.365 $\pm$ 0.023 & 12.380 $\pm$ 0.021 & 11.741 $\pm$ 0.019 & L3 & [4], [12] \\
2MASS J21580457-1550098 & 25.0 & 5.0 & 15.040 $\pm$ 0.040 & 13.867 $\pm$ 0.033 & 13.185 $\pm$ 0.036 & L4 & [2], [13] \\
SDSSJ083506.16+195304.4 & 26.1 & 5.1 & 16.094 $\pm$ 0.075 & 14.889 $\pm$ 0.057 & 14.319 $\pm$ 0.049 &  L5 & [5], [14] \\
2MASS J10101480-0406499 & 18.0 & 2.0 & 15.508 $\pm$ 0.059 & 14.385 $\pm$ 0.037 & 13.619 $\pm$ 0.046 & L6 & [6], [13] \\
2MASS J01033203+1935361 & 23.0 & 2.0 & 16.288 $\pm$ 0.080 & 14.897 $\pm$ 0.056 & 14.149 $\pm$ 0.059 & L7 & [7], [13]\\
2MASS J16322911+1904407 & 16.0 & 3.3 & 15.867 $\pm$ 0.070 & 14.612 $\pm$ 0.038 & 14.003 $\pm$ 0.047 & L8 & [4], [14] \\
DENIS-P J0255-4700 & 4.868 & 0.004 & 13.246 $\pm$ 0.027 & 12.204 $\pm$ 0.024 & 11.558 $\pm$ 0.024 & L9 & [8], [12] \\
SDSS J120747.17+024424.8  & 14.5 & 2.9 & 15.580 $\pm$ 0.071 & 14.561 $\pm$ 0.065 & 13.986 $\pm$ 0.059 & T0 & [9], [14]\\
SDSS J015141.69+124429.6  & 21.4 & 1.6 & 16.566 $\pm$ 0.129 & 15.603 $\pm$ 0.112 & 15.183 $\pm$ 0.189 & T1 & [10], [13]\\
SDSS J125453.90-012247.4  & 13.48 & 0.419 & 14.891 $\pm$ 0.035 & 14.090 $\pm$ 0.025 & 13.837 $\pm$ 0.054 & T2 & [10], [12]\\
SDSS J120602.51+281328.7  & 26.0 & 2.0 & 16.541 $\pm$ 0.109 & 15.815 $\pm$ 0.126 & 15.817 $\pm$ 0.034 & T3 & [5], [13] \\
2MASS J22541892+3123498   & 14.0 & 2.0 & 15.262 $\pm$ 0.047 & 15.018 $\pm$ 0.081 & 14.902 $\pm$ 0.147 & T4 & [10], [13] \\
2MASS J15031961+2525196  & 6.4549 & 0.0459 & 13.937 $\pm$ 0.024 & 13.856 $\pm$ 0.031 & 13.963 $\pm$ 0.059 & T5 & [10], [12] \\
SDSS J162414.37+002915.6  & 11.0 & 0.1 & 15.494 $\pm$ 0.054 & 15.524 $\pm$ 0.100 & 15.518 $\pm$ 0.050 & T6 & [11], [13]\\
2MASS J07271824+1710012  & 9.1 & 0.2 & 15.600 $\pm$ 0.061 & 15.756 $\pm$ 0.171 & 15.556 $\pm$ 0.194 & T7 & [11], [13]\\
2MASS J04151954-0935066  & 5.83 & 1.26 & 15.695 $\pm$ 0.058 & 15.537 $\pm$ 0.113 & 15.429 $\pm$ 0.201 & T8 & [10], [15]\\
\enddata
\tablerefs{[1]: \citet{Burgasser2006AJ}, [2]: \citet{Kirkpatrick2010ApJS}, [3]: \citet{Marocco2013AJ}, [4]: \citet{Burgasser2007ApJ...659..655B}, [5]: \citet{Chiu2006AJ}, [6]: \citet{Reid2006ApJ}, [7]: \citet{Cruz2004ApJ}, [8]: \citet{Burgasser2006ApJ...637.1067B}, [9]: \citet{Looper2007AJ},  [10]: \citet{Burgasser2004AJ}, [11]: \citet{Burgasser2006ApJ...639.1095B}, [12]: \citet{Gaia2016AA}, [13]: \citet{Faherty2009AJ}, [14]: \citet{Schmidt2010AJ}, [15]: \citet{Lodieu2012AA}.}
\end{deluxetable*}

The spectra are flux-calibrated by using 2MASS photometric data \citep{Skrutskie2006AJ} and the associated multiplicative scale-factor is calculated separately for the \textit{J} (15.32 $\pm$ 0.05 $\SI{}{\mag}$), \textit{H} (15.27 $\pm$ 0.09 $\SI{}{\mag}$), and \textit{K}$_S$ (15.24 $\pm$ 0.16 $\SI{}{\mag}$) bandpasses following the approach described in \citet{Cushing2005ApJ}. The scale factor takes into account spectral measurement errors and photometric uncertainties. We use the weighted average of these three values for our final scale factor for the flux calibration of each object in our sample. 

Where possible, we use distances derived from the Gaia parallax measurements \citep{Gaia2016AA}. In all other cases, less accurate parallaxes from ground-based telescopes or estimates based on spectroscopy are used.

\begin{figure*}
\begin{center}
\includegraphics[width=0.8\textwidth]{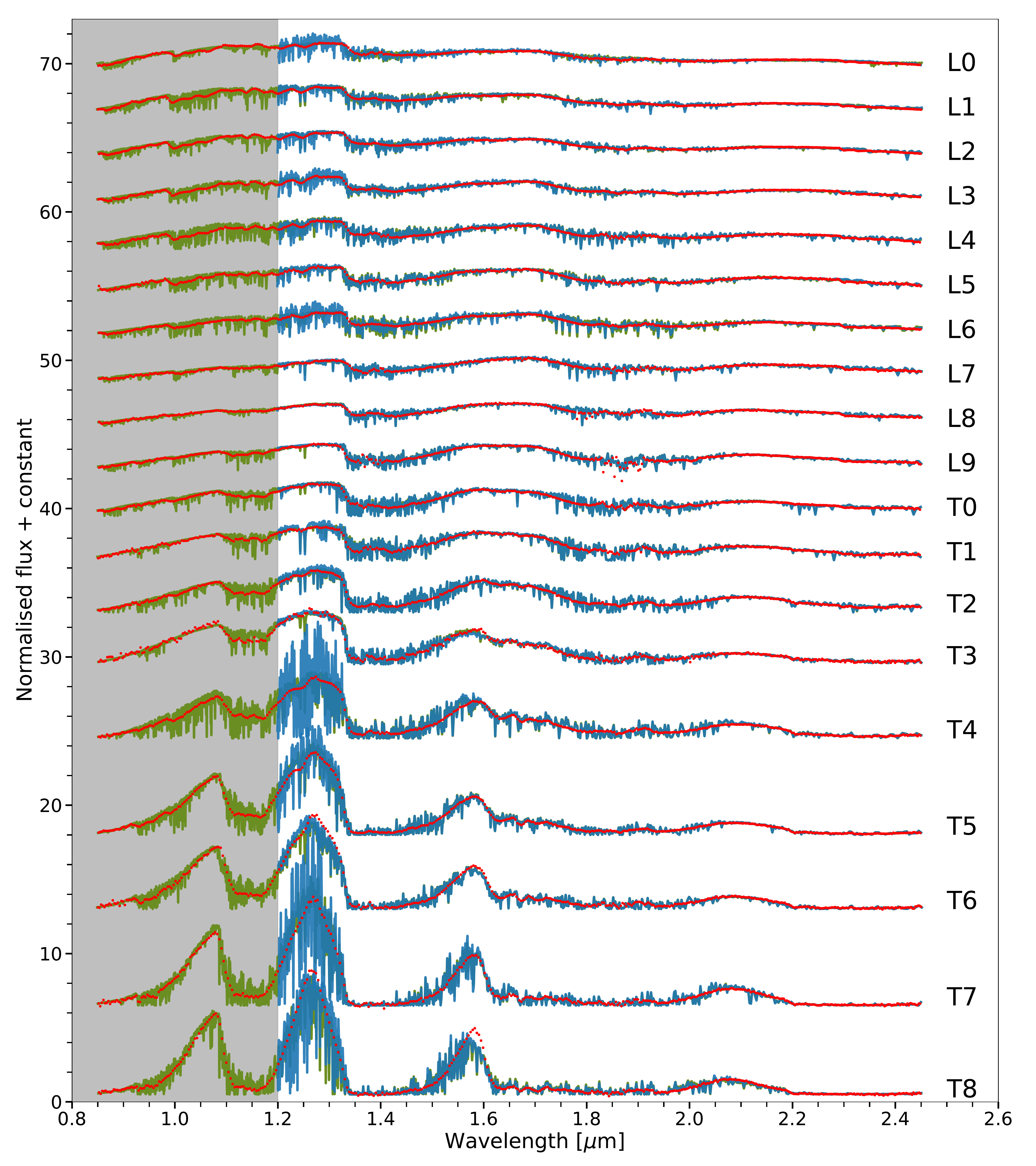}
\end{center}
\vspace{-0.1in}
\caption{Near-infrared spectra of the standard L and T dwarfs in our curated sample shown as red dots. For clarity, we have omitted the associated data uncertainties, but these are shown for individual objects in Figures~\ref{fig:spectra_appendix} \& \ref{fig:spectra_appendix2} in the Appendix. Spectra associated with the retrieved best fit values are overlaid for comparison with the green and blue curves corresponding to the full and restricted range of wavelengths, respectively; the gray area represents the omitted part of the spectra for the restricted wavelength range. All spectra shown are for non-gray cloud models with a reduced set of atoms/molecules (see text for details).}
\label{fig:L-T Spectra non-gray}
\end{figure*}

Our set of 19 brown dwarfs and their known parameters are given in Table~\ref{tab:objects}. Following the approach by \citet{Kitzmann2020ApJ} and \citet{Line2015ApJ}, we use only every third value of the extracted spectrum to prevent oversampling of non-independent flux density values in each resolution element and diminish the effect of correlated uncertainties. An overview of all spectra is shown in Figure~\ref{fig:L-T Spectra non-gray}.

\section{Retrieval Model}
\label{sect:methods}

For the retrieval analysis of the brown dwarf spectra in this study we employ an updated version of the Bayesian retrieval code \heliosr, first introduced in \citet{Kitzmann2020ApJ}. It is part of the open-source Exoclimes Simulation Platform (ESP) (\url{https://github.com/exoclime}). The computationally expensive parts of the model run on a graphics card processor (GPU). Since GPUs have in general thousands of computational cores, the time for performing a forward model calculation is substantially decreased compared to running it on a traditional CPU.

So far, \heliosr has been successfully used to characterize atmospheres of brown dwarfs and exoplanets by analyzing their emission spectra. This includes, for example, the ultra-hot Jupiter WASP-121b \citep{Bourrier2020A&A...637A..36B}, where \heliosr was able to constrain the abundances of the important hydrogen anion, as well as a non-vertical, atmospheric abundance of water in a spectrum taken by the WFC3 instrument on the Hubble Space Telescope (HST). It was also used to analyze spectra of brown dwarfs, such as KELT-1b \citep{Wong2021AJ....162..127W} or HD 19467B \citep{Mesa2020MNRAS.495.4279M}, for example. In the following, we briefly summarize the important parts of \heliosr and describe the updates that have been made to the model.

\subsection{Radiative transfer: emission spectra}

We solve the radiative transfer equation for a one-dimensional, plane-parallel atmosphere using the method of short characteristics \citep{OlsonKunasz1987}, as previously implemented within the \heliosr code \citep{Kitzmann2020ApJ}. The method of short characteristics allows for a stable and efficient solution of the transfer equation in the absence of scattering. As stated in \cite{Kitzmann2020ApJ}, we use the first-order version of short characteristic method presented in \citet{OlsonKunasz1987}. Due to the neglect of scattering, we only need to calculate the spectral intensity in the upward direction. It is calculated for a total of two different, discrete polar angles, which is equivalent to a four-stream radiative transfer method. The result is then numerically integrated over these angles with a Gaussian quadrature to yield the outgoing flux $F_{\nu}^+$.

Due to the low resolution of the SpeX instrument, we calculate the theoretical high-resolution spectra with a constant step size of $\SI{1}{ \per \centi \metre}$ in wavenumber space \citep{Line2015ApJ, Kitzmann2020ApJ}. The resulting spectrum is convolved with an appropriate instrument line profile before it is binned down to the resolution of the measured spectra. 

An additional scaling factor $f$ is used in the the radius-distance relation to scale the outgoing flux $F_{\nu}^{+}$ of the brown dwarf to the one measured by the observer ($F_{\nu}$):
\begin{equation}\label{radius-distance relation}
    F_{\nu}=F_{\nu}^+ f\left(\frac{R_p}{d}\right)^2 \, ,
\end{equation}
where $d$ is the distance between the observer and the brown dwarf and $R_p$ the (prior) radius. For $R_p$ we choose a fixed value of $R_p = 1\,R_J$ throughout this study. 
In practice, $f$ serves as a ``catch all" scaling factor that absorbs uncertainties in the radius and distance, as well as inaccuracies in the atmospheric models and flux calibration of the measured spectra. Since all of these quantities are essentially degenerate, we choose to combine them all in a single factor. 

Not all of our distances listed in Table~\ref{tab:objects} are based on measured parallaxes, but partly on estimated values based on spectroscopic data, such as our T8 (2MASS J04151954-0935066). For our T8 brown dwarf, we have used the distance $d$ based on spectroscopy \citep{Lodieu2012AA}. Other studies have published distances based on parallax measurements (e.g., \citealt{Faherty2012ApJ}; d = 5.736 $\pm$ 0.362 pc). These small differences in $d$ are absorbed into the retrieved value of $f$ and thus do not affect the final outcome.

If $f$ is assumed to only contain the deviations from the assumed prior radius of $1\,R_J$, it can be converted into the actual radius of the brown dwarf in units of Jupiter radii via
\begin{equation}
\label{eq:derived_radius}
  R = \sqrt{f} \ .
\end{equation}
In practice, however, $f$ usually also involves other uncertainties and missing or inaccurate model physics, such that the derived radius $R$ should not always be considered as the true radius of the brown dwarf \citep{Kitzmann2020ApJ}.

\subsection{Temperature-pressure profile}
\label{sec:temperature_profile}

For this study, we divide the atmosphere into a total of 70 levels (i.e. 69 layers). The levels are distributed equidistantly in log pressure-space. 
The specification of a temperature-pressure profile in atmospheric retrievals has a long history. Approaches include the specification of a temperature value in each model atmospheric layer \citep{Irwin2008JQSRT}, the use of a 9-parameter, ad-hoc fitting function \citep{Madhusudhan2009ApJ} and self-consistent but simplified profiles \citep{Guillot2010A&A, Parmentier2014A&A, Heng2012MNRAS, Heng2014ApJS}. In particular, the self-consistent temperature-pressure profiles invoke strong assumptions that can produce an artificially isothermal atmosphere at low pressures \citep{Heng2014ApJS}.

As discussed in \citet{Kitzmann2020ApJ}, \heliosr implements a description of the temperature profile based on a finite element approach. This ensures a continuous temperature-pressure profile, described by a relatively small number of free parameters. Unless stated otherwise, we use six first-order elements for the temperature profile, which results in a total of seven free parameters. Due to inherent, continuous nature of this finite element approach, we can evaluate the temperature for any given pressure in the forward model.

Since this study is focused on brown dwarfs for which temperature inversions are not anticipated, we force the profiles to be monotonically decreasing with pressure \citep{Kitzmann2020ApJ}. As shown in \citet{Bourrier2020A&A...637A..36B}, \heliosr is also able to retrieve temperature inversions if the aforementioned assumption of a monotonically-decreasing profile is not employed.

\subsection{Cloud description}
\label{subsect:clouds}

In its original version \heliosr has has the option of adding a gray cloud layer to the atmosphere. This approximation usually assumes that the cloud particles are large compared to the wavelength range of the measured spectrum. 

For this study we extend \heliosr to also optionally use non-gray cloud layers. Here, we only aim at parameterizing the particles' extinction coefficients. No attempt is made to actually model the formation of these clouds.

We follow the work of \citet{Kitzmann2018MNRAS} to describe the extinction efficiency of the cloud particles as a function of wavelength. Their approach assumes cloud particles with single radii and then approximate the extinction efficiencies $Q_{\mathrm{ext}}$ resulting from Mie theory calculations with a simply, analytic equation:
\begin{equation}
    Q_{\mathrm{ext}}(\lambda) =\frac{Q_{1}}{Q_{0} x_\lambda^{-a_0}+x_\lambda^{0.2}} \, ,
\end{equation}
where $Q_1$ is a normalization constant, $Q_0$ determines the $x$-value at which $Q_{\mathrm{ext}}$ is peaking, $x_\lambda = 2\pi a/ \lambda$ is the dimensionless size parameter, $a$ is the particle radius, and $a_0$ is the power-law index in the small particle limit, where Mie theory converges to the limit of Rayleigh scattering. 

The equation is not supposed to describe the exact behaviour of the extinction efficiencies, but should rather serve as a first-order approximation. While the full Mie absorption and scattering efficiencies of single particles usually exhibit low and high-frequency oscillations, the analytic fit provides a smooth description. It is worth noting that $Q_0$ can be a proxy for the cloud particle composition (see Table 2 of \citealt{Kitzmann2018MNRAS}).

The optical depth $\tau$ of the cloud layer is then given by
\begin{equation}
    \tau(\lambda) = \sigma_\mathrm{ext}(\lambda) n_c \Delta z = Q_{\mathrm{ext}}(\lambda) \pi a^2 n_c \Delta z \, 
\end{equation}
with the extinction cross section $\sigma_\mathrm{ext}$ = $Q_{\rm ext} \pi a^2$, the cloud particle number density $n_c$, and the vertical extent of the cloud layer $\Delta z$.

Since it is difficult to estimate a good prior for $n_c$, we replace it with an optical depth at a reference wavelength of $\lambda_\mathrm{ref} = 1 \, \mu \mathrm{m}$. The optical depth is then given as
\begin{equation}
    \tau(\lambda) = \tau_\mathrm{ref} \frac{Q_\mathrm{ext}(\lambda)}{Q_\mathrm{ext}(\lambda_\mathrm{ref})} = \tau_\mathrm{ref} \frac{Q_{0} x_{\lambda_\mathrm{ref}}^{-a_0}+x_{\lambda_\mathrm{ref}}^{0.2}}{Q_{0} x_\lambda^{-a_0}+x_\lambda^{0.2}}
\end{equation}

The position of the cloud layer in the atmosphere and its vertical extent $\Delta z$ are described by two more free parameters: the cloud-top pressure $p_{\mathrm{t}}$ and its bottom pressure $p_{\mathrm{b}}$. Instead of using $p_{\mathrm{b}}$ directly as a free retrieval parameter, we instead use a factor $b_c \geq 1$, such that $p_{\mathrm{b}} = b_c \, p_{\mathrm{t}}$. The factor $b_c$ is limited to a maximum value of 10, such that a cloud layer can at most span over one order of magnitude in pressure.

For the non-gray cloud description we, thus, have six free parameters in total, whereas the gray cloud requires three. A summary of all cloud retrieval parameters and their prior distributions is given in Table~\ref{tab:priors}.

\subsection{Bayesian framework: nested sampling}

\begin{figure}
\begin{center}
\includegraphics[width=0.95\columnwidth]{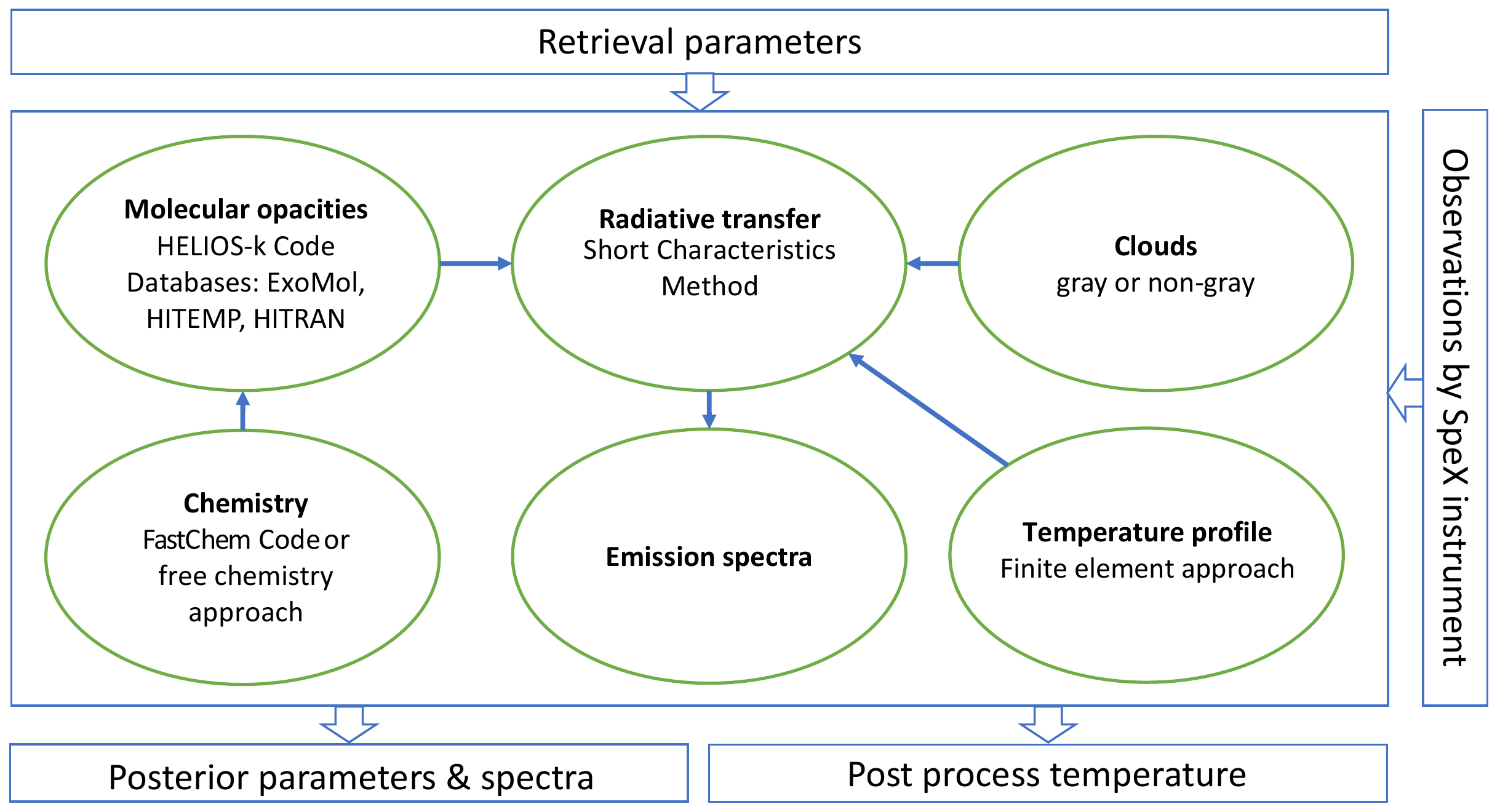}
\end{center}
\vspace{-0.1in}
\caption{Schematic of our brown dwarf retrieval framework.}
\label{fig:Workflow}
\end{figure}

We incorporate all of the components described in the previous subsections into a Bayesian inference framework (see Figure~\ref{fig:Workflow}). Our choice of method to explore the multi-dimensional parameter space is nested sampling \citep{Skilling2006AIPC} in its \texttt{MultiNest} implementation \citep{Feroz2008MNRAS, Feroz2009MNRAS} , which was previously implemented in \heliosr by \citet{Kitzmann2020ApJ}.  Nested sampling was introduced to the exoplanet atmospheric retrieval literature by \citet{Benneke2013ApJ}.  We assume that the prior distributions of our free parameters are either uniform, log-uniform or Gaussian (see Section \ref{sec:priors} and Table~\ref{tab:priors}).

A key ingredient of nested sampling is the specification of the likelihood function ${\mathcal L}$, which is the mathematical relationship between the model $D_{j,\mathrm{m}}$, the observational data $D_j$ and uncertainties $s_j$ associated with the data for all measured data points $j$.  It is common practice to assume a Gaussian likelihood function \citep{Kitzmann2020ApJ}
\begin{equation}
\label{eq:likelihood}
  \ln \mathcal L = -\frac{1}{2} \sum_{j=1}^{J} \frac{\left( D_j - D_{j,\mathrm{m}} \right)^2}{s_j^2} - \frac{1}{2} \ln(2\pi s_j^2) \ .
\end{equation}
Implicitly, this assumes that the data uncertainties $s_j$ are not only Gaussian-distributed, but that they are also uncorrelated for all $j$.


Following \cite{Line2015ApJ} and \cite{Kitzmann2020ApJ}, we account for the possibility that stated uncertainties of the measured fluxes have been under-estimated by implementing the procedure of \cite{Hogg2010arXiv}.  Let the standard deviation of the measured fluxes at each $j$-th data point be $\sigma_j$. The effective standard deviation of the $j$-th data point is then given by
\begin{equation}
  s_{j}^{2}=\sigma_{j}^{2}+e^{2 \,\ln \delta} \ ,
\end{equation}
where the parameter $\ln\delta$ is part of the fit. Specifying $\ln\delta$ as a fitting parameter ensures that $\delta>0$ \citep{ForemanMackey2013PASP}. For example, $\ln\delta=-4$ corresponds to $\delta \approx 2\%$.

\begin{deluxetable}{cccc}
\tablecaption{Correspondence of the Bayes factor $B_{ij}$ to the number of standard deviations $\sigma$. Reproduced from \cite{Trotta2008ConPh}.}
\label{tab:Bayes factors}
\tablehead{
\colhead{\hspace{0.25cm} $B_{ij}$ \hspace{0.25cm} } & \colhead{ \hspace{0.25cm} $\ln B_{ij}$ \hspace{0.25cm} } & \colhead{ \hspace{0.25cm} $\sigma$ \hspace{0.25cm} } & \colhead{ \hspace{0.25cm} category \hspace{0.25cm} }
}
\startdata
2.5 & 0.9 & 2.0 & \\
2.9 & 1.0 & 2.1 & 'weak' at best \\
8.0 & 2.1 & 2.6 & \\
12 & 2.5 & 2.7 & 'moderate' at best \\
21 & 3.0 & 3.0 & \\
53 & 4.0 & 3.3 & \\
150 & 5.0 & 3.6 & 'strong' at best\\
43000 & 11 & 5.0 &\\
\enddata
\end{deluxetable}

The key advantage of nested sampling is that it allows for the calculation of the marginalized likelihood or Bayesian evidence, which may be used to implement a formal form of Occam's Razor known as Bayesian model comparison \citep{Trotta2008ConPh}. 

A pair of models of differing complexity (characterized by different numbers of parameters or prior distributions) is compared by taking the ratio of their Bayesian evidences, which is known as the Bayes factor \citep{Trotta2008ConPh}. There is an established correspondence between the Bayes factor and the number of standard deviations that one of the models is disfavoured by the data, which we reproduce in Table~\ref{tab:Bayes factors}. It is worth pointing out, though, that Bayesian model comparison may fail to exclude unphysical scenarios (e.g. \citealt{Fisher2019ApJ}).\\

\subsection{Retrieval parameters and derived quantities}
\label{sec:priors}

The Bayesian framework requires to define prior distributions for all free parameters of the forward model. All parameters and their priors are shown in Table \ref{tab:priors}. 

For the general description of the brown dwarf atmosphere we require the surface gravity\footnote{Unless stated otherwise, values of $\log g$ are given in cgs units throughout this study. $\log g$, the distance $d$, and the calibration factor $f$.} We use the measured distances and the corresponding errors with a Gaussian prior (see Table~\ref{tab:objects}). This procedure propagates the error in the measured distances through all other retrieval parameters. The temperature profile is described by seven free parameters in total (see Section~\ref{sec:temperature_profile}).

For the abundances of the chemical species we make the usual assumptions that they are isoprofiles throughout the atmosphere. Each considered chemical species, therefore, requires one free parameter for its mixing ratio $x_i$. 

Specifically, we retrieve for mixing ratios of the following species: \ch{H2O}, \ch{CH4}, \ch{NH3}, \ch{CO2}, \ch{CO}, \ch{H2S}, \ch{CrH}, \ch{FeH}, \ch{CaH}, \ch{TiH}, and \ch{K}. The mixing ratio of \ch{Na} is determined from the one of potassium by using the solar element abundance ratio of K and Na (see also \citealt{Kitzmann2020ApJ}). Since the SpeX spectra are only sensitive to the far line wings of the strong alkali resonance lines, the abundances of \ch{K} and \ch{Na} are essentially degenerate. It is, therefore, only possible to directly constrain one of them \citep{Line2015ApJ}. The abundances of \ch{H2} and \ch{He} are derived from the remaining background atmosphere, assuming a solar \ch{H}/\ch{He} element abundance ratio.

Additionally, for the cloudy models we require the cloud parameters as discussed in Section \ref{subsect:clouds}. For the non-gray clouds we use six free parameters, while the gray cloud needs three parameters in total.

Besides the free retrieval parameters that are used directly within the nested sampling, \heliosr also provides posterior distributions for a set of derived quantities. One is the effective temperature of the brown dwarf. This quantity is obtained by integrating the high-resolution spectra of all posterior samples over wavelengths and then converting the resulting total flux to an effective temperature via the Stefan-Boltzmann law.

Two other derived quantities are the C/O ratio and the overall metallicity [M/H]. The C/O ratios are calculated by counting the amount of carbon and oxygen atoms using the retrieved mixing ratios of all carbon and oxygen carriers (e.g. equation (19) of \citealt{Line2013ApJ...775..137L}):
\begin{equation}
  \mbox{C/O} = \frac{\ch{CH4} + \ch{CO} + \ch{CO2}}{\ch{CO} + 2\ch{CO2} + \ch{H2O} } \ .
\end{equation}

The metallicity [M/H], on the other hand, is approximated by summing up the constant mixing ratios for each species weighted by the number of metal atoms and divided by the abundance of hydrogen. The result is then compared to the sum of solar metals relative to hydrogen.

\begin{deluxetable}{lccc}
\tablecaption{Summary of retrieval parameters and prior distributions for the free chemistry approach used in the cloud-free, gray cloud and non-gray cloud models.}
\label{tab:priors}
\tablehead{
\colhead{Parameter \hspace{1cm}} &\colhead{\hspace{2cm}} & \colhead{\hspace{0.5cm} Prior \hspace{3cm}}\\ \cmidrule{2-4}
& \colhead{Type} & & \colhead{Value}
}
\startdata
$\log{g}$    & uniform &  & 3.5 to 6.0 cm/s$^2$ \\
$d$             & Gaussian &  & measured \tablenotemark{i}  \\
$f$                   & uniform &  &  0.1 to 5.0 \\
$T_1$              & uniform &  &  1000 to 5000 K\\
$b_i$               & uniform &  & 0.1 to 0.95  \\
$\ln{\delta}$         & uniform &  &  -10 to 1.0 \\
$x_i$           & log-uniform &  &  $10^{-12}$ to 0.1 \\
\hline
\textit{gray clouds} &  &  &   \\
$p_{\mathrm{t}}$  & log-uniform &  & $10^{-2}$ to 50 bar  \\
$b_{\mathrm{c}}$     & log-uniform &  &  1 to 10 \\
$\tau$              & log-uniform &  &  $10^{-5}$ to 20 \\
\hline
\textit{non-gray clouds} &  &  &   \\
$p_{\mathrm{t}}$  & log-uniform &  & $10^{-2}$ to 50 bar\\
$b_{\mathrm{c}}$  & log-uniform &  &  1 to 10 \\
$\tau_{\mathrm{ref}}$  & log-uniform &  & $10^{-5}$ to 20 \\
$Q_0$                 & log-uniform &  & 1 to 100  \\
$a_0$                 & uniform &  &  3 to 7 \\
$a$ & log-uniform &  & 0.1 to 50 $\SI{}{\micro\metre}$ \\
\enddata
\tablenotetext{i}{Note that all the measured distances $d$ can be found in Table~\ref{tab:objects}.}
\end{deluxetable}

\subsection{Opacity calculations and line lists}
\label{subsect:opacity}

Major absorbers (\ch{H2O}, \ch{CH4}, \ch{NH3}, \ch{CO2}, \ch{CO}, \ch{H2S}, \ch{CrH}, \ch{FeH}, \ch{CaH}, \ch{TiH}, as well as the alkali metals \ch{Na} and \ch{K}) are considered in this study to cover the wavelength range of the SpeX instrument, which is $\SI{0.85}{\micro\metre}$ to about $\SI{2.45}{\micro\metre}$. 

Most opacities are calculated by the open-source \texttt{HELIOS-K} opacity calculator \citep{Grimm2015ApJ, Grimm2021ApJS}. The line list data for these molecules are taken from the ExoMol database \citep{Barber2006MNRAS, Yurchenko2011MNRAS, Yurchenko2014, Azzam2016MNRAS} and the HITEMP database \citep{Rothman2010JQSRT}. The collision-induced absorption coefficients for \ch{H2}–\ch{H2} and \ch{H2}–\ch{He} are based on \citet{Abel2011JPCA} and \citet{Abel2012JChPh}, respectively. We refer the reader to \citet{Tennyson2017MolAs} for a review of the spectroscopic databases. 

For the alkali metals \ch{K} and \ch{Na}, we use the descriptions of their resonance line wings published by \citet{Allard2016A&A} and \citet{Allard2019A&A}, respectively. The computation of the \ch{Na} and \ch{K} opacities are described in \citet{Kitzmann2020ApJ}.

\section{Results}
\label{sect:results}

We perform a suite of atmospheric retrievals on the curated sample of all L and T dwarfs listed in Table~\ref{tab:objects}. In Section~\ref{sec:case_study} we present a more detailed analysis of the results obtained for the L5 and T5 dwarfs. Section~\ref{sec:result_comparison} and Table~\ref{tab:data comparison posteriors T7 and T8 dwarfs} provide a comparison of our results for the T7 and T8 with those of previous publications on the same objects. In Section~\ref{sec:trends} we discuss the trends of the retrieved parameters across the L-T sequence, including the surface gravity, chemical abundances, and clouds. Appendix \ref{sect: Supplementary Figures} provides the posterior distributions of the cloud-free and non-gray cloud models for the spectra cut at below $\SI{1.2}{\micro\metre}$ for all brown dwarfs in our sample. In Appendix~\ref{sect: Supplementary Data}, the Tables~\ref{tab:data posteriors L dwarfs} \& \ref{tab:data posteriors T dwarfs} provide a detailed overview of the outcomes of a large suite of retrievals (six models for each object) for all L and T dwarfs in our sample. Additionally, Appendix~\ref{sect: Impact of prior choice} shortly addresses the impact of the prior choice.

\subsection{Terminology, Overview of Spectra and Model Fits}

Before we discuss the outcome of our retrieval calculations, we first introduce the following terminology and the reasoning behind it:
\begin{itemize}
    \item \citet{Oreshenko2020AJ} previously demonstrated that disregarding data in each spectrum bluewards of $\SI{1.2}{\micro\metre}$ circumvents unresolved issues with the shapes of the alkali metal resonance line wings, but retains enough information in the spectrum to constrain the surface gravity spectroscopically. Following this approach, spectra with and without this cut are referred to as ``restricted" and ``full", respectively.
    
    \item If the full set of chemical species is used in the retrieval (see Section \ref{subsect:opacity}), the label ``all" is employed.  In a follow-up retrieval, species that only have upper limits on their abundances are removed; we refer to this as the ``reduced" set of species. We explicitly check that inferred quantities from the ``all" versus ``reduced" retrievals are consistent with each other.
    
    \item We consider both cloud-free and cloudy atmospheres. For the cloudy cases, we either use gray or non-gray clouds as described in Section \ref{subsect:clouds}.
\end{itemize}

In total, there are twelve permutations of models for each spectrum. However, we find that the Bayesian evidence consistently favors models with a reduced set of chemical species. This is to be expected since the Bayesian evidence effectively penalizes models with more free parameters over those with fewer if both models provide an adequate fit to the measured spectrum. Therefore, we have only listed six models for each object in Tables \ref{tab:data posteriors L dwarfs} \& \ref{tab:data posteriors T dwarfs}.

All 19 spectra, as well as a subset of the best-fit models (full versus restricted models), are shown in Figure~\ref{fig:L-T Spectra non-gray} as an overview. For presentational reasons, only models with non-gray clouds and a reduced set of atoms and molecules are shown. 

The L4 to T1 brown dwarf spectra contain considerable scatter in the measured spectral flux between 1.8--2.1~$\SI{}{\micro\metre}$ due to strong telluric absorption, where the Earth's atmosphere is nearly opaque. This, however, does not impact the overall fit of the spectrum because the elevated error bars in these spectral regions considerably decrease their weight in the computed likelihood (see Equation~\eqref{eq:likelihood}). Other wavelength ranges where telluric absorption might be an issue also include the 1.3--1.5~$\SI{}{\micro\metre}$ and 1.75--2.0~$\SI{}{\micro\metre}$ regions.

\begin{figure}
\includegraphics[width=\columnwidth]{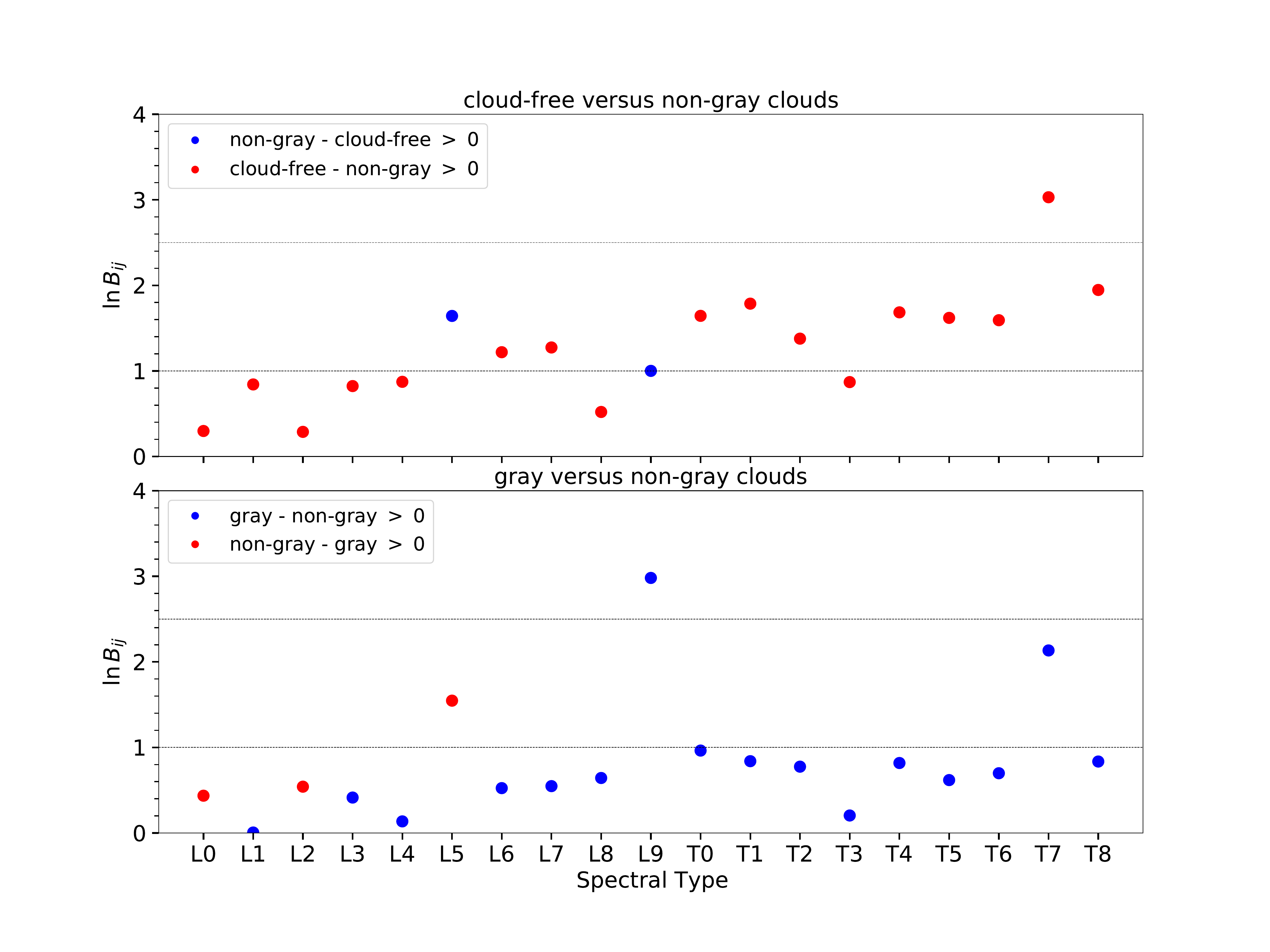}
\caption{Bayes factors derived from our suite of nested sampling retrievals with a restricted wavelength range and a reduced set of atoms/molecules. The top panel shows the Bayesian model comparison between cloud-free and non-gray-cloud models. The bottom panel compares models with gray and non-gray clouds. Denoting the Bayes factor by $B_{ij}$, $\ln{B_{ij}}=1$ and 2.5 correspond to weak and moderate evidence for one model versus the other (see Table~\ref{tab:Bayes factors}). The ratio of Bayesian evidences is taken such that $\ln{B_{ij}}>0$.}
\label{fig:ln_Bayes}
\end{figure}

Figure~\ref{fig:ln_Bayes} summarizes the outcome of a suite of 57 retrievals performed on all 19 observed brown dwarf spectra with restricted wavelength ranges and reduced sets of chemical species. 

Surprisingly, the logarithm of the Bayes factor $(\ln{B_{ij}})$ has values of about unity when comparing cloud-free versus cloudy models (whether gray or non-gray). Thus, the Bayesian evidence does not allow us to favor one class of model over the other. This essentially implies that all considered models are consistent with the data and that the SpeX spectra do not contain enough information to adequately distinguish between the different model scenarios.

Since the study of clouds in brown dwarfs has an established history (e.g., \citealt{Tsuji2003ApJ, Burrows2006ApJ, SaumonMarley2008ApJ}), we will often refer to the results and retrieved quantities from the models with non-gray clouds. 

\subsection{A pair of case studies: L5 and T5 dwarfs}
\label{sec:case_study}

In this subsection we discuss the results for the L5 dwarf SDSSJ083506.16+195304.4 and the T5 dwarf 2MASS J15031961+2525196 dwarfs in greater detail.

\begin{figure*}
\begin{center}
\includegraphics[width=0.98\columnwidth]{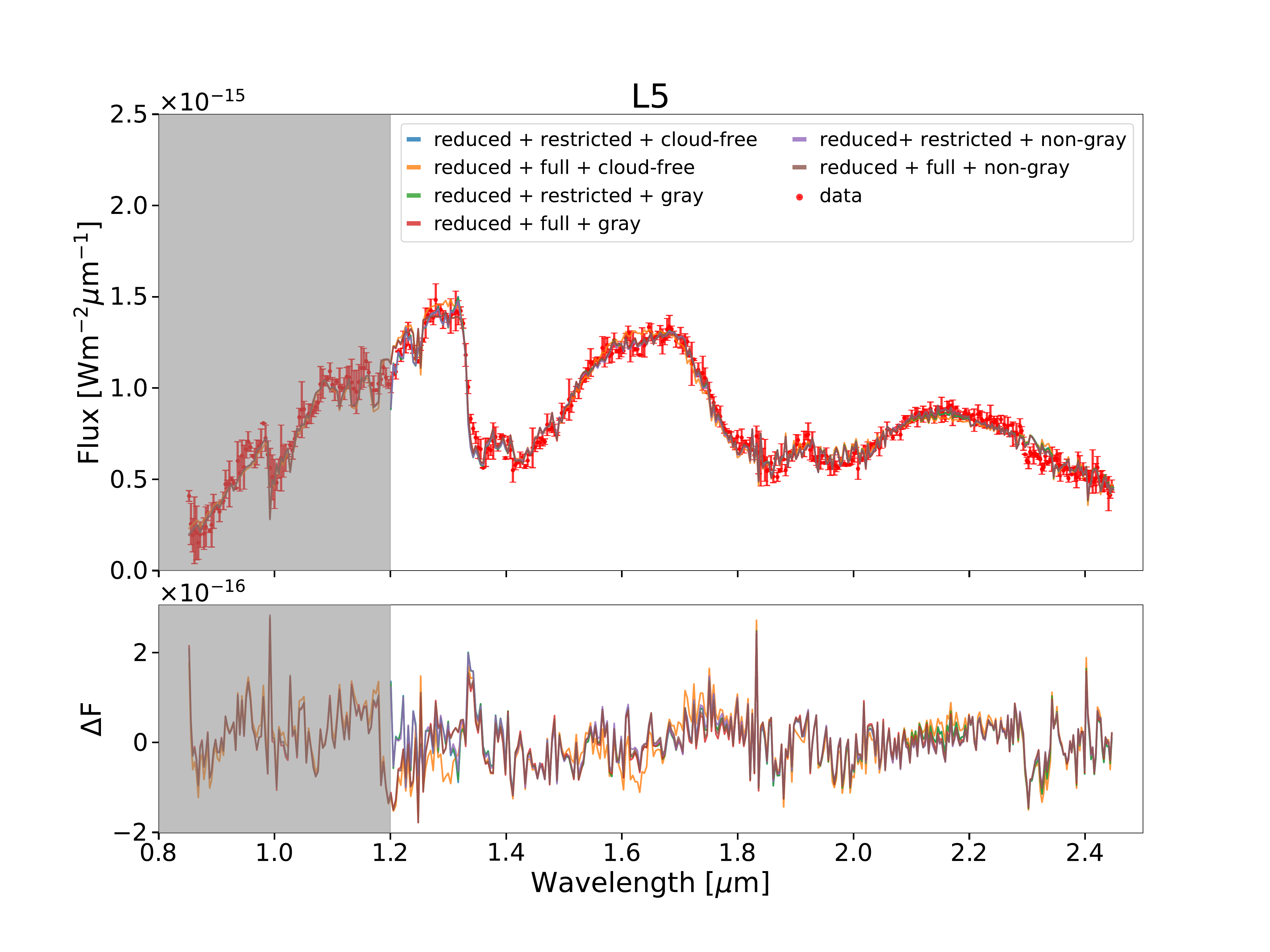}
\includegraphics[width=0.98\columnwidth]{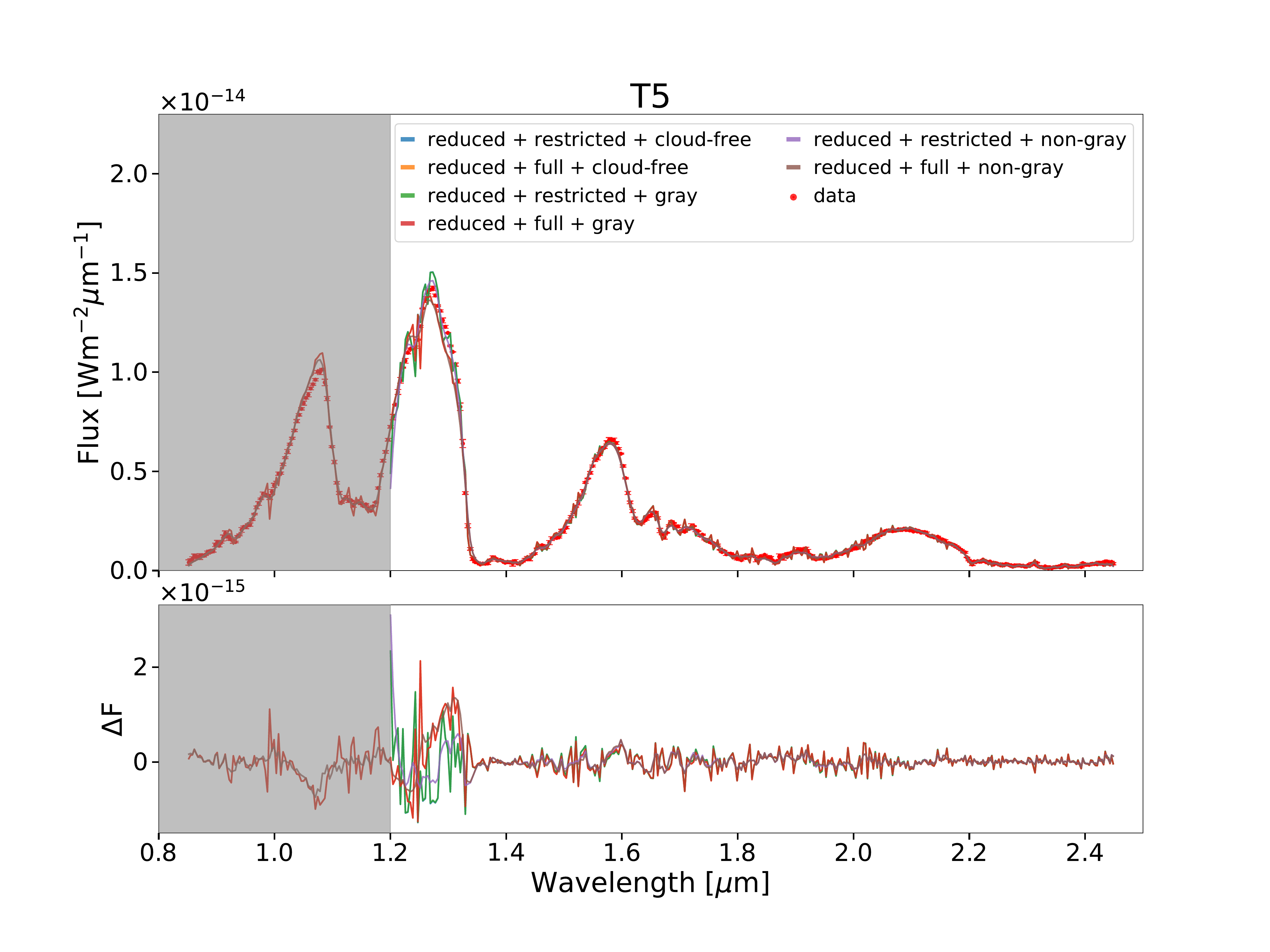}
\end{center}
\vspace{-0.1in}
\caption{Comparing data (red dots with associated uncertainties) and various retrieved spectra (curves) of the L5 (left panel) and T5 (right panel) brown dwarfs of our curated sample. For each model, the median curve of the retrieved set of spectra is displayed. The associated residuals between the data and various models are shown. Gray areas represent the omitted part of the spectra for the restricted wavelength range.}
\label{fig:spectra comparison}
\end{figure*}

The aforementioned inability of the atmospheric retrievals on SpeX spectra to distinguish between cloud-free and cloudy models is further illustrated in Figure~\ref{fig:spectra comparison} for the L5 and T5 dwarfs of our curated sample. The figure shows the posterior spectra and the observed data for all six different models and both objects.  

The different posterior spectra and their residuals clearly indicate that all considered model scenarios provide almost equally good fits to the data. This also explains why the Bayes factors between the models are close to unity for all cases.

The median values of the retrieval parameters for the two objects are summarized in Tables  \ref{tab:data posteriors L dwarfs} \& \ref{tab:data posteriors T dwarfs}. Plots of the entire posterior distributions and the retrieved temperature profiles can be found in the Appendix.\\

For the L5 dwarf, the reduced set of chemical species consists of water, methane, carbon monoxide, hydrogen sulfide, and iron hydride for the restricted range of wavelengths. If the full wavelength range is considered, potassium (and, thus, sodium) is additionally detected. The models with restricted wavelength range will in general not allow a constraint of the abundances of K and Na because their important resonance line wings are not contained in these spectra. Since the inferred quantities are generally in good agreement between the ``all" versus ``reduced" retrievals, we will compare outcomes based on the reduced set of species in the following. 

For the L5 dwarf, if we compare the non-gray-cloud retrievals performed on the restricted versus full spectra, the retrieved surface gravities are $\log{g}=5.83_{-0.22}^{+0.30}$ and $\log{g}=5.61_{-0.23}^{+0.21}$ , respectively. The derived carbon-to-oxygen ratios are $0.59 \pm 0.10$ and $0.70 \pm 0.12$. This result is, thus, roughly consistent with solar element abundances for carbon and oxygen. The outcome also demonstrates that the C/O ratio can be robustly inferred if the main carbon and oxygen carriers can be constrained.

The derived radii and effective temperatures are also consistent for both the restricted and full cases: $0.77_{-0.09}^{+0.08} \,\si{\Rjup}$ versus $0.79_{-0.09}^{+0.08}\,\si{\Rjup}$; $1432.86_{-32.94}^{+33.65} \,\si{\kelvin}$ versus $1493.11_{-23.13}^{+22.76} \,\si{\kelvin}$. These radii are consistent with the lower range of measured values for this class of object (e.g., \citealt{Burrows2011ApJ}; Table 6 of \citealt{Bayliss2017AJ}).\\

For the T5 dwarf, the reduced set of molecules consists of \ch{H2O}, \ch{CH4}, \ch{NH3}, and \ch{FeH} for the restricted range of wavelengths, as well as K and TiH for the full spectrum. The derived C/O ratios are $0.49 \pm 0.03$ and $0.51 \pm 0.03$, respectively, which again is roughly consistent with solar element abundances of C and O.

For the T5 dwarf, if we compare the non-gray-cloud retrievals performed on the restricted versus full spectra, the retrieved surface gravities are $\log{g}=4.73_{-0.12}^{+0.12}$ and $4.82_{-0.09}^{+0.11}$, respectively. The retrieved radii and effective temperatures are also consistent with each other: $0.69_{-0.05}^{+0.05}\,\si{\Rjup}$ versus $0.71_{-0.04}^{+0.04}\,\si{\Rjup}$; $964.77_{-30.74}^{+33.75} \,\si{\kelvin}$ versus $1048.70_{-24.88}^{+27.85} \,\si{\kelvin}$. With the exception of the effective temperatures, the other quantities are comfortably consistent for the pair of restricted versus full retrievals.

\subsection{Comparison with previous studies: T7 and T8 dwarfs}
\label{sec:result_comparison}

The T7 (2MASS J07271824+1710012) and T8 (2MASS J04151954-0935066) dwarfs of our curated sample provide an opportunity to compare our results with previous publications. Table~\ref{tab:data comparison posteriors T7 and T8 dwarfs} summarises outcomes from the current and previous studies.

\begin{deluxetable}{ccccccccc}
\tabletypesize{\scriptsize}
\tablecolumns{7}
\tablewidth{0.8\columnwidth}
\tablecaption{Comparison of retrieval outcomes with values from the published literature for the T7 and T8 dwarfs.  Only models with the reduced set of molecules are tabulated.  Variations of the models shown are: full spectra cloud-free (FC), full spectra gray (FG), full spectra non-gray (FN), restricted spectra cloud-free (RC), restricted spectra gray (RG) and restricted spectra non-gray (RN).}
\label{tab:data comparison posteriors T7 and T8 dwarfs}
\tablehead{
 \colhead{Obj.} & \colhead{Model} & \colhead{$\lambda$} & \colhead{$T_\mathrm{eff}$} & \colhead{log g}  & \colhead{Radius} & \colhead{Ref.}\\ \hline
 &  & \colhead{($\si{\micro\metre}$)} & \colhead{(K)} & \colhead{($\si{\centi\metre \per \square \second)}$}  & \colhead{(R$_\mathrm{J}$)} &
}
\startdata 
T7 & --- & 1.00-2.50 & $807_{-19}^{+17}$ & $5.13_{-0.10}^{+0.10}$ & $1.12_{-0.06}^{+0.07}$ & [1]\\
T7 & --- & 1.00-2.10 & $900-940$ & $4.8-5.0$ & --- &  [2]\\ 
T7 & --- & 0.30–14.50 & $845^{+71}_{-71}$ & $4.95^{+0.49}_{-0.49}$ & $0.94^{+0.16}_{-0.16}$ &  [3]\\ \hline
T7 & F C & 0.85-2.45 & $847.62_{-20.41}^{+21.15}$ & $4.92_{-0.13}^{+0.16}$ & $0.69_{-0.04}^{+0.04}$ & [*]	\\
T7 & F G & 0.85-2.45 & $845.48_{-19.99}^{+20.98}$ & $4.91_{-0.12}^{+0.15}$ & $0.69_{-0.04}^{+0.04}$ & [*] \\
T7 & F N & 0.85-2.45 & $852.59_{-18.69}^{+20.49}$ & $4.59_{-0.15}^{+0.17}$  & $0.68_{-0.03}^{+0.03}$ & [*]\\
T7 & R C & 1.20-2.45 & $712.02_{-22.57}^{+26.18}$ & $3.91_{-0.17}^{+0.20}$ & $0.81_{-0.06}^{+0.06}$ & [*]\\
T7 & R G & 1.20-2.45 & $691.80_{-20.36}^{+26.09}$ & $3.82_{-0.14}^{+0.16}$ & $0.86_{-0.06}^{+0.06}$ & [*]\\
T7 & R N & 1.20-2.45 & $710.72_{-21.75}^{+23.84}$ & $3.91_{-0.16}^{+0.18}$ & $0.81_{-0.05}^{+0.06}$ & [*]\\ \hline 
T8 & --- & 1.00-2.50 & $680_{-18}^{+13}$ & $5.04_{-0.20}^{+0.20}$ & $1.06_{-0.06}^{+0.05}$ & [1] \\
T8 & --- & 1.15-2.25 & $600-800$ & $4.0-5.5$ & $0.89-1.33$ &  [4]\\
T8 & --- & 1.00-2.10 & $740-760$ & $4.9-5.0$ & --- &  [2] \\
T8 & --- & 0.30–14.50 & $677^{+56}_{-56}$ & $4.83^{+0.51}_{-0.51}$ & $0.95^{+0.16}_{-0.16}$ &  [3]\\\hline
T8 & F C & 0.85-2.45 & $717.17_{-30.67}^{+32.00}$ & $3.67_{-0.10}^{+0.12}$ & $0.63_{-0.12}^{+0.13}$ & [*]\\
T8 & F G & 0.85-2.45 & $715.34_{-30.18}^{+31.38}$ & $3.67_{-0.10}^{+0.12}$ & $0.63_{-0.12}^{+0.12}$ & [*]\\
T8 & F N & 0.85-2.45 & $716.90_{-29.83}^{+30.65}$ & $3.67_{-0.10}^{+0.11}$ & $0.63_{-0.12}^{+0.12}$ & [*]\\
T8 & R C & 1.20-2.45 & $722.01_{-40.77}^{+42.74}$ & $3.66_{-0.10}^{+0.16}$ & $0.52_{-0.10}^{+0.11}$ & [*]\\
T8 & R G & 1.20-2.45 & $719.76_{-39.73}^{+41.50}$ & $3.67_{-0.10}^{+0.15}$ & $0.53_{-0.10}^{+0.11}$ & [*]\\
T8 & R N & 1.20-2.45 & $844.14_{-53.45}^{+53.89}$ & $4.52_{-0.32}^{+0.30}$ & $0.40_{-0.07}^{+0.05}$ & [*]\\ \hline 
\enddata
\tablerefs{[*]: This work, [1]: \citet{Line2017ApJ}, [2]: \citet{Burgasser2006ApJ...639.1095B}, [3]: \citet{Filippazzo2015ApJ}, [4]: \citet{Liu2011ApJ}}
\end{deluxetable}

Based on the analysis using the full T7 brown dwarf spectrum, we obtain a surface gravity of $\log{g} \approx 4.6$ -- 4.9, consistent with the values reported by \citet{Burgasser2006ApJ...639.1095B}, \citet{Line2017ApJ} and \citet{Filippazzo2015ApJ} for the same object.  
Here, \citet{Burgasser2006ApJ...639.1095B} used the best-fit spectrum from a model grid, while \citet{Line2017ApJ} performed an actual retrieval. The results from \citet{Filippazzo2015ApJ}, on the other hand, are based on the evolutionary tracks of \citet{Baraffe2003A&A} (COND03) as well as the cloud-free models published by \citet{SaumonMarley2008ApJ} (SMNC08). 

Our retrievals based on the restricted spectrum, however, yield a $\log{g}$ value of about 3.8 -- 3.9, considerably lower than those based on the full spectra. 
This difference is likely caused by the impact of the resonance line wings of the alkali metals sodium and potassium. Restricting the wavelength range does not only exclude the line wing of the \ch{K} resonance line at $\SI{0.77}{\micro \metre}$, though, but it also removes the $\SI{0.9}{\micro \metre}$ \ch{H2O} and the $\SI{1.0}{\micro \metre}$ \ch{FeH} molecular absorption bands.

All retrieved radii ($R\approx0.7$ -- 0.9 $R_{\rm J}$) are smaller than the one reported by \citet{Line2017ApJ} or \citet{Filippazzo2015ApJ}. The restricted-spectra values lie within the confidence interval of the evolutionary model-derived value by \citet{Filippazzo2015ApJ}. As discussed by \citet{Filippazzo2015ApJ}, obtaining consistent radii from the radius-distance relationship is challenging due to e.g. incomplete molecular line lists and spectrally poorly reproduced regions, especially below $\SI{0.9}{\micro\metre}$ or the \textit{H}-band peak.

Our derived effective temperatures for the full-spectra retrievals fall within the predicted range published in earlier studies. For the restricted-spectra range retrievals, we obtain values that are approximately $\SI{100}{\kelvin}$ below the lower bound estimated by \cite{Line2017ApJ}. This discrepancy might be caused by the different obtained radii that enter the radius-distance relationship, which is used to scale the total emitted flux of the atmosphere. Thus, smaller retrieved radii result in higher total flux values, which yields larger derived effective temperatures. Specifically, in equation (\ref{radius-distance relation}), smaller values of $R$ are compensated by larger values of $F_{\nu}^{+}$ in order to produce the same $F_{\nu}$.

We constrain the abundances of the molecules \ch{H2O}, \ch{CH4}, \ch{K}, and \ch{FeH}. As expected, potassium is not detected in the case of the restricted spectra for most of our models. The individual abundances of molecules are of the same orders of magnitude compared to \citet{Line2017ApJ} with the exception of \ch{FeH} that was not considered in their study. On the other hand, \citet{Line2017ApJ} constrained the abundance of the theoretically expected \ch{NH3}, which was not found by our retrieval. Our upper limit for \ch{NH3} is orders of magnitudes smaller than the lower bound retrieved by \citet{Line2017ApJ}.\\

For the T8 dwarf, we obtain a surface gravity of $\log{g}~\approx~3.7$. This value is below the lower bounds estimated by \citet{Burgasser2006ApJ...639.1095B}, \citet{Line2017ApJ}, \citet{Liu2011ApJ}, or \citet{Filippazzo2015ApJ}. Only the wavelength-restricted, non-gray cloud retrieval (RN) resulted in a higher $\log{g}$ value of $\approx~4.5$. All derived radii ($R\approx0.4$ -- 0.6~$R_{\rm J}$) are again smaller than the ones reported by \citet{Line2017ApJ}, \citet{Filippazzo2015ApJ}, and \citet{Liu2011ApJ}. These small radii might indicate that some model physics is missing in the current version of \heliosr. The effective temperatures fall within the predicted confidence interval published in earlier studies. The only exception is, again, the wavelength-restricted, non-gray cloud case, where $T_\mathrm{eff}$ is somewhat higher. Peaks in the measured fluxes are underestimated by the posterior spectra, especially with the \textit{H}-band being spectrally poorly reproduced. For reasons unclear to us, the fit in the \textit{H} band is noticeably worse for the spectrum of the T8 dwarf compared to other objects, possibly due to a missing opacity source.

\begin{figure*}
\centering
\includegraphics[width=0.75\columnwidth]{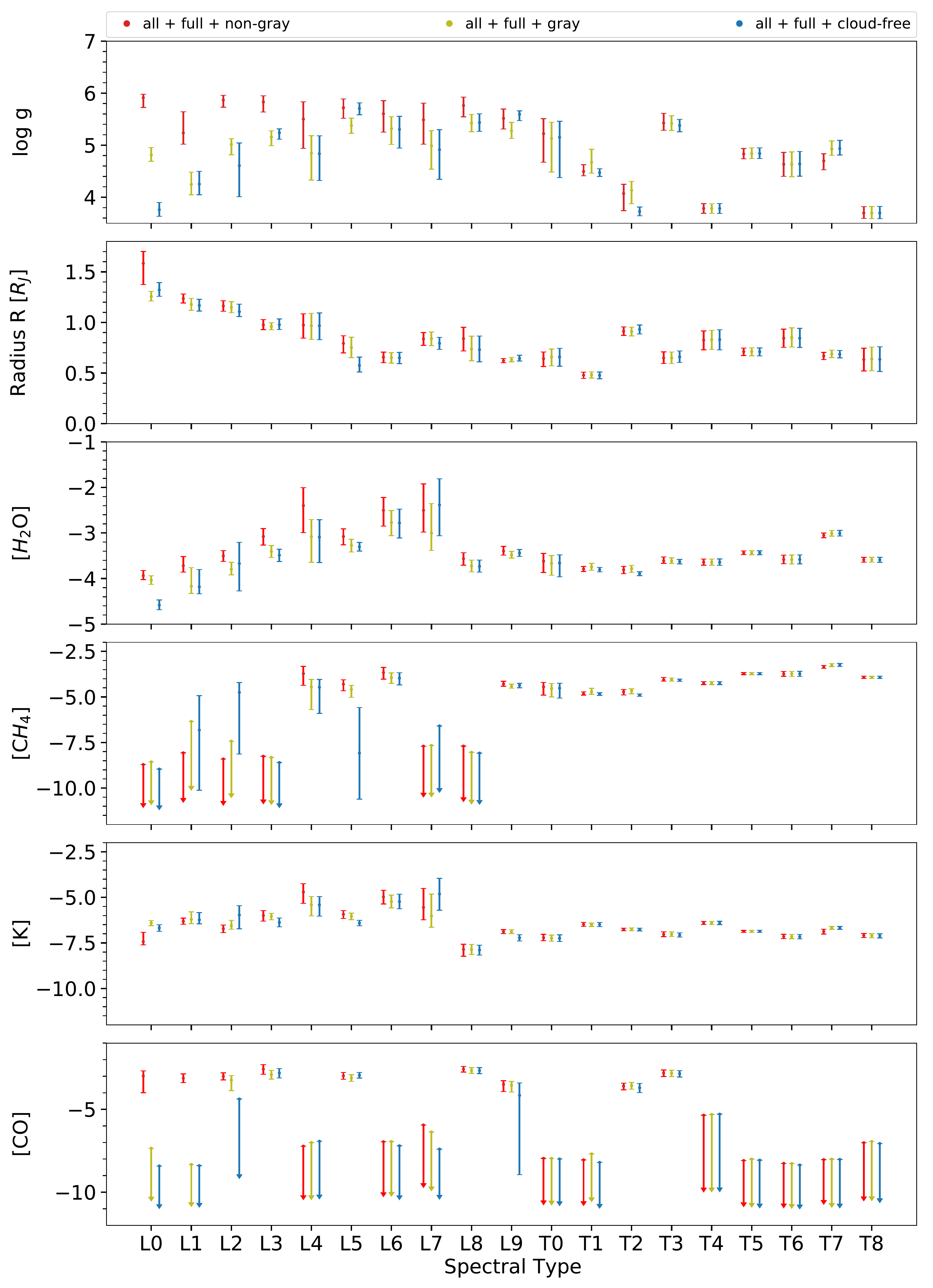}
\includegraphics[width=0.75\columnwidth]{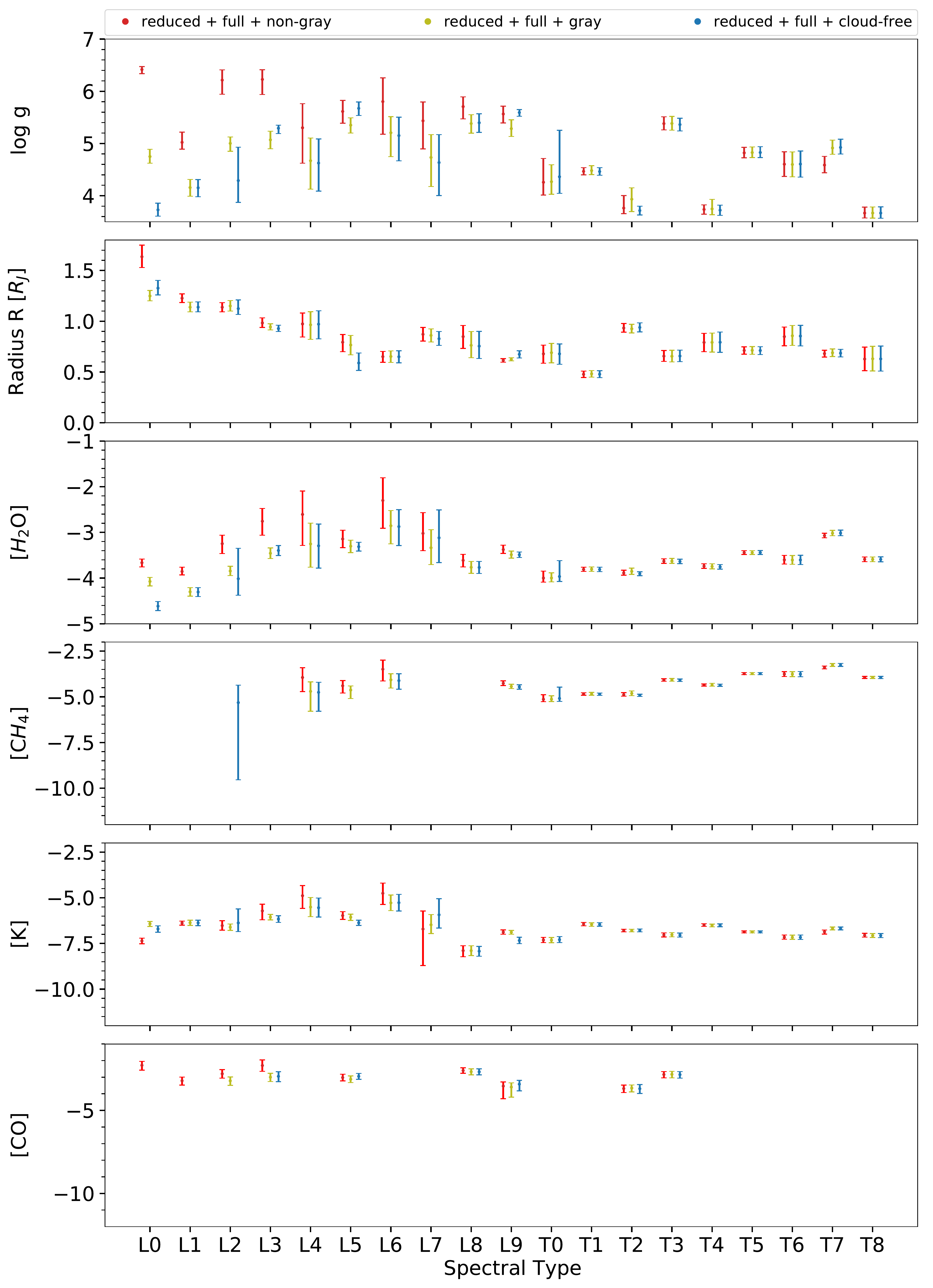} \\
\includegraphics[width=0.75\columnwidth]{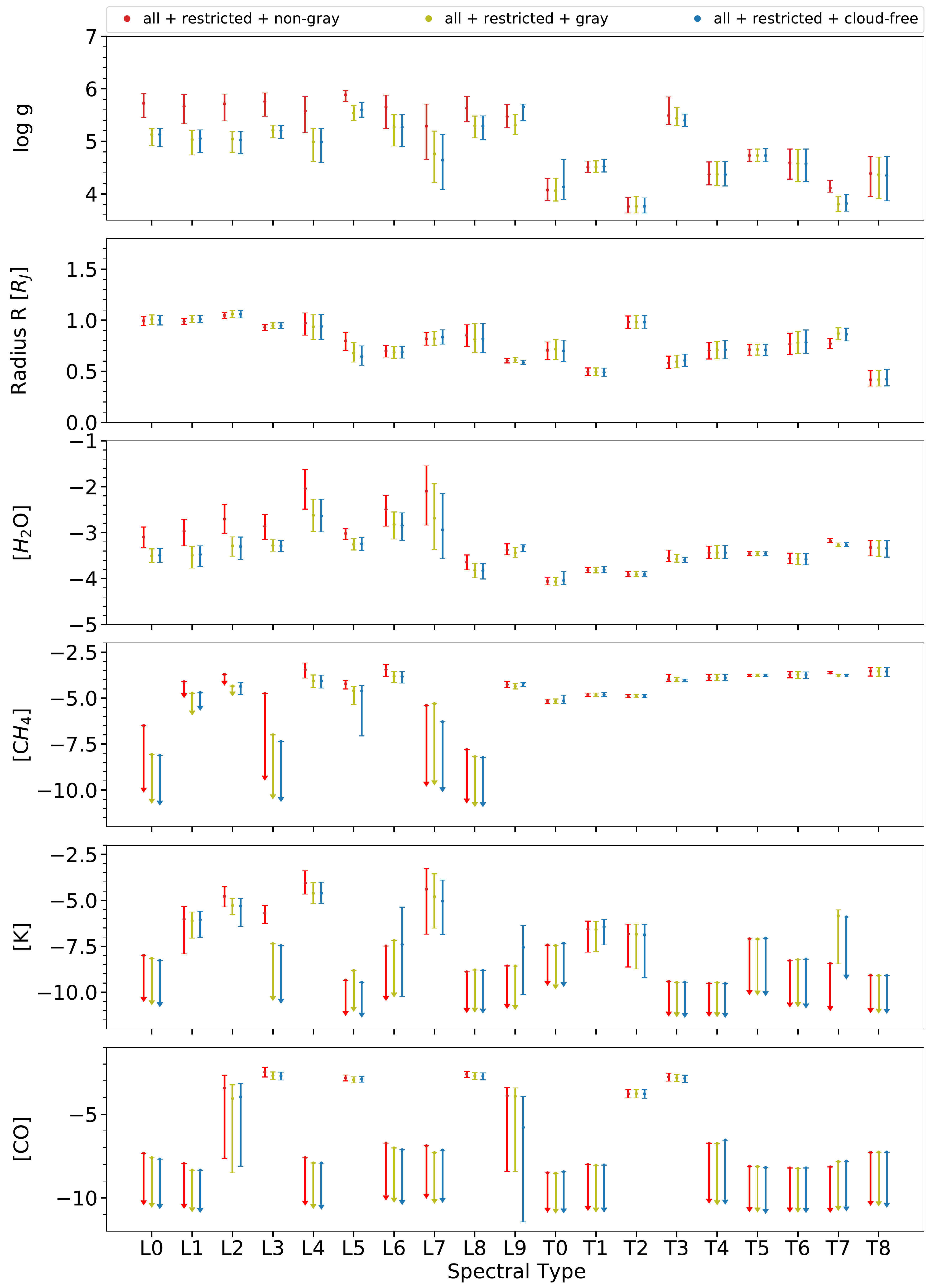}
\includegraphics[width=0.75\columnwidth]{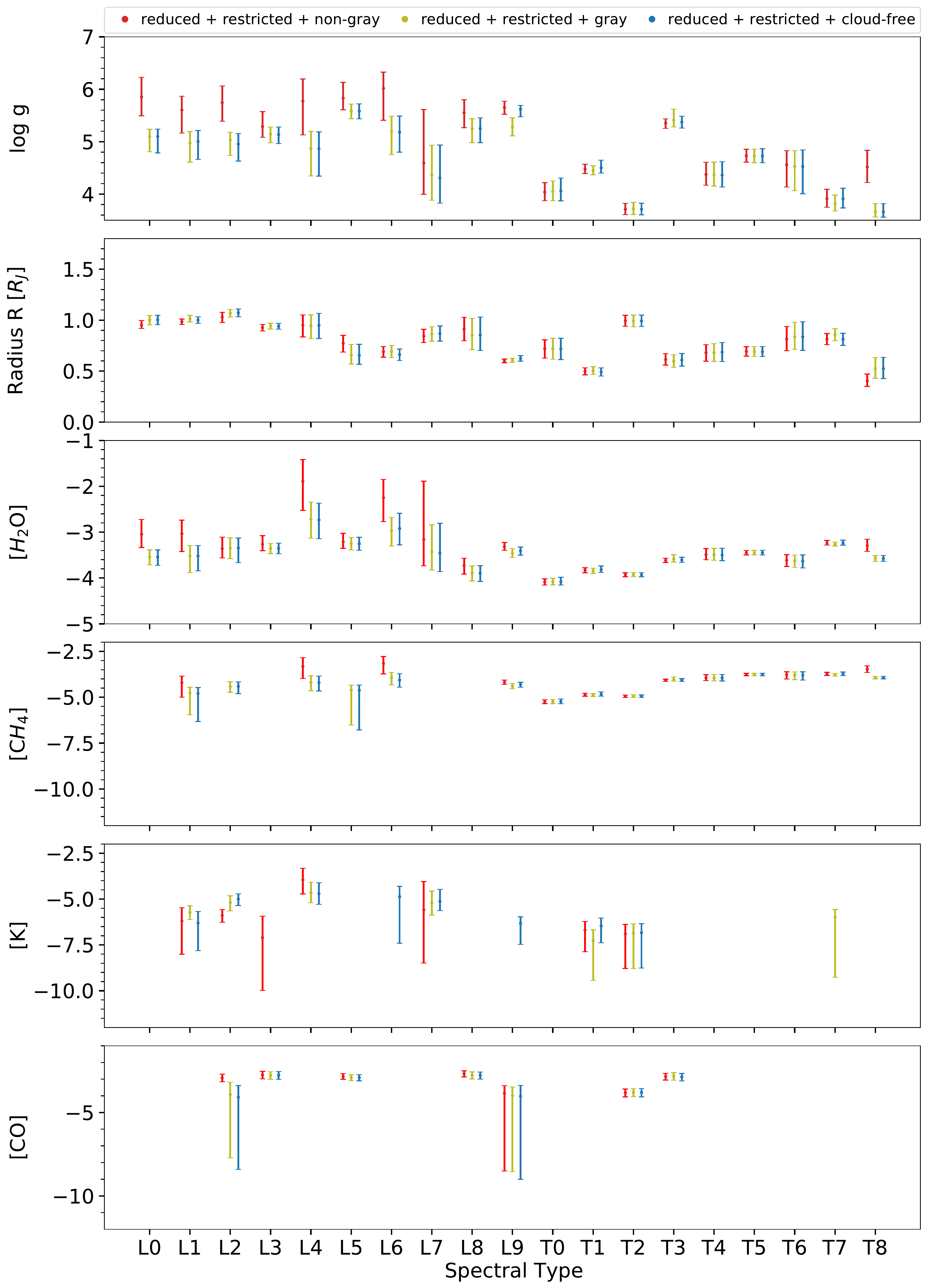}
\caption{Comparing the retrieved parameters from our suite of brown dwarf retrievals across the L-T sequence. Shown are various permutations of all versus a reduced set of atoms/molecules and the full versus restricted range of wavelengths associated with the measured spectra.  In each panel, the cloud-free (blue line), gray-cloud (olive-green line) and non-gray-cloud (red line) models are compared. $1\sigma$ uncertainties are shown.}
\label{fig: robustness test sequence cloudmodels}
\end{figure*}

Our retrieved molecular abundances of \ch{H2O} and \ch{CH4} are of the same order of magnitude, while abundances of \ch{K} are an order of magnitude lower than reported by \citet{Line2017ApJ}. This difference is expected due to the different considerations of sodium and potassium. \citet{Line2017ApJ} used the mixing ratio of \ch{Na} as a free parameter and calculates \ch{K} subsequently, while we derive \ch{Na} from the retrieved \ch{K} mixing ratio by using their solar elemental abundance ratio \citep{Kitzmann2020ApJ}. Similar to the previous case, \citet{Line2017ApJ} constrained the expected \ch{NH3}, which was not found in our retrieval. Overall, we conclude that differences in the input physics and/or chemistry probably account for deviations in the retrieved surface gravities and radii. As already noted by \cite{Kitzmann2020ApJ}, it remains unclear how to set a physically motivated prior on the radius and therefore to judge if a retrieved value of the radius is unphysical.

\subsection{Trends across the L-T sequence}
\label{sec:trends}

So far, our results have focused on four individual brown dwarfs. In the following, we discuss trends of retrieved quantities (or lack thereof) across our curated sample of L and T dwarfs. Figures \ref{fig: robustness test sequence cloudmodels} \& \ref{fig: robustness test sequence cloudparams} summarize the retrieved parameter values as a function of the spectral type. The posterior distributions and best-fit spectra are all depicted in the Appendix. For the retrieved molecular abundances, we have chosen to focus on \ch{H2O}, \ch{CH4}, \ch{K}, and \ch{CO}. Figures~\ref{fig:sanity check Teff} \& \ref{fig:T-P sequence} display the effective temperatures and retrieved temperature-pressure profiles, respectively.

The following general conclusions may be drawn:
\begin{itemize}
    \item Retrievals that include all chemical species or just a reduced set of detected atoms and molecules yield parameter values that are in excellent agreement.
    
    \item For the T dwarfs, the retrieved parameter values are robust to the choice of either cloud-free or cloudy (both gray and non-gray) models. This indicates that the outcome is unaffected when clouds are absent.
\end{itemize}

In the following subsections we discuss the results for some of the important retrieval and derived parameters.

\subsubsection{Surface gravities}

For the early L dwarfs, there is considerable scatter in the retrieved surface gravities, regardless of whether the ``all", ``reduced", ``full" or ``restricted" retrievals are employed. The cloud-free models versus the ones with gray clouds are consistent with each other. The strongest difference is between the models with non-gray clouds and the cloud-free/gray-cloud models. 

Overall, there is no obvious trend of the surface gravity across the L-T sequence with $\log{g}$ values varying from about $4$ to $5$. Non-gray-cloud models consistently yield higher values of $\log{g} \approx 6$ for the early L dwarfs.

\subsubsection{Derived brown dwarf radii}

The derived brown dwarf radii overall decrease with decreasing effective temperature, consistent with an evolutionary cooling sequence. Comparing these retrieved radii with evolutionary models, however, requires not only knowledge of the ages, but also the cloud configuration \citep{Burrows2011ApJ}.

It is worth noticing that the retrieved radii are consistent between the trio of cloud-free, gray-cloud and non-gray-cloud models, regardless of whether ``all", ``reduced", ``full" or ``restricted" retrievals are employed.  One important outcome of our retrievals is that the derived radii for the L dwarfs decrease monotonically with effective temperature only for the retrievals performed on the full spectra. This suggests that the resonance line wings of the alkali metals have a decisive impact on the analysis of brown dwarf spectra.  

Some of the retrievals, especially for the T dwarfs, result in implausibly small values of $R \approx 0.5 \, R_{\rm J}$. As already mentioned in Section~\ref{sect:methods}, the radius is derived from the calibration factor $f$, assuming that $f$ only contains contributions with respect to the assumed prior radius of 1 R$_\mathrm{J}$. Thus, these small values of $f$, and thus $R$, might indicate, for example, missing physics and chemistry or problems in the spectra calibration as already noted by \cite{Kitzmann2020ApJ}.

\subsubsection{Derived effective temperatures}

\begin{figure}
\centering
\includegraphics[width=0.95\columnwidth]{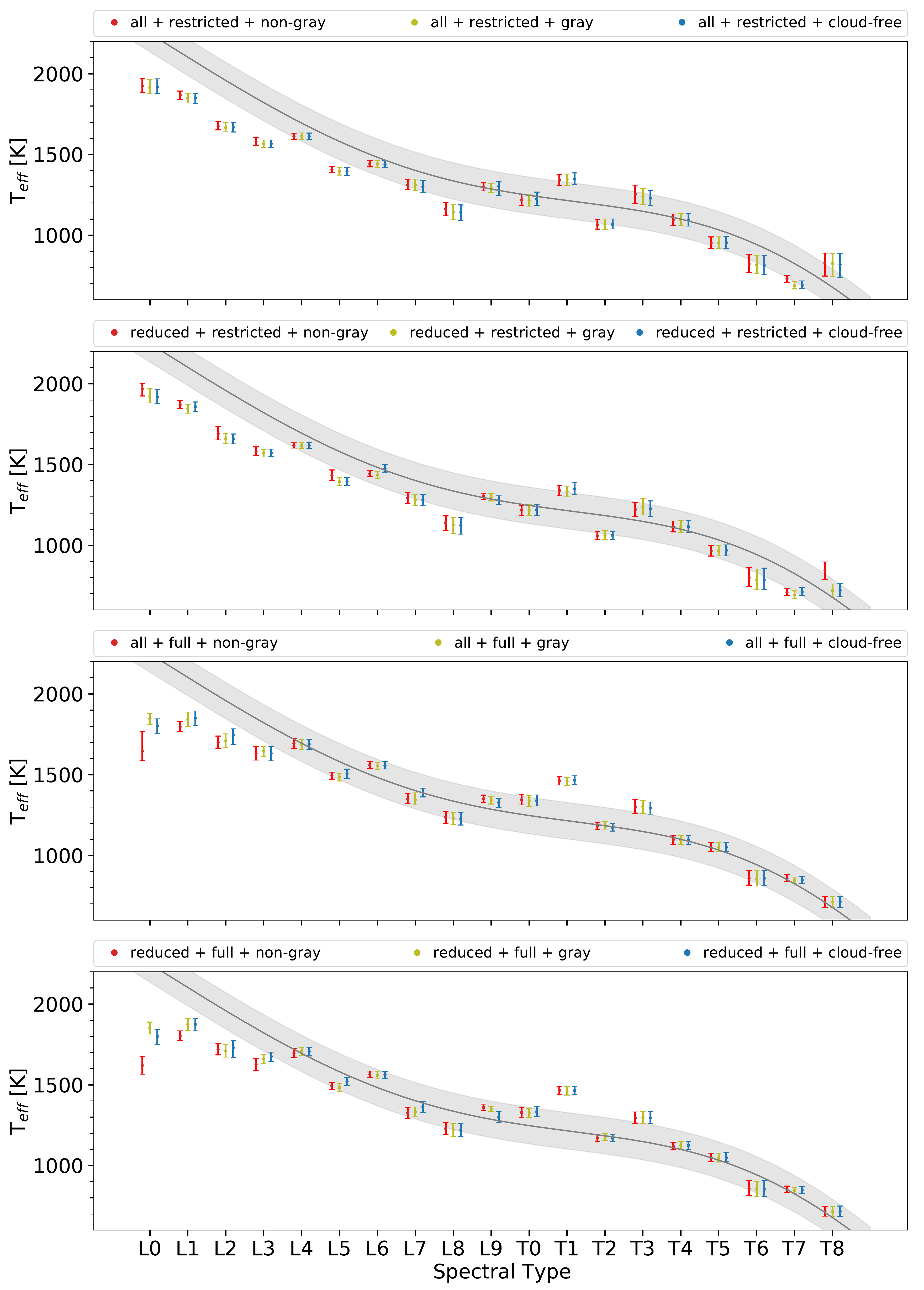}
\caption{Comparison of all inferred effective temperatures over the entire L-T sequence. $1\sigma$ uncertainties are shown. The sixth order field age polynomial fit of \citet{Filippazzo2015ApJ} including its rms is shown in gray.}
\label{fig:sanity check Teff}
\end{figure}

As mentioned in Section~\ref{sect:methods}, the effective temperature $T_{\rm eff}$ is obtained in a post-processing step by calculating the total outgoing flux from all posterior spectra and convert the result into an effective temperature by using the Stefan-Boltzmann law. The derived $T_{\rm eff}$ for all spectral classes are shown in Figure~\ref{fig:sanity check Teff}.

The resulting effective temperatures show a general decrease from about 1800~K for the L0 dwarf to 800~K (T8) across the entire L-T sequence. This outcome is found for all different model scenarios, regardless of whether all or a reduced set of chemical species is used or whether the full or restricted spectra are being analyzed. The general trend of decreasing effective temperatures with spectral class and, thus, age, is consistent with theoretical expectations and observations (e.g., \citealt{Kirkpatrick2005ARA&A, Filippazzo2015ApJ, Kirkpatrick2021ApJS}). Our results are in particular also robust with respect to the choice of cloud model

\subsubsection{Abundances of chemical species}

There is no clear trend in the water abundances across the L-T sequence with the volume mixing ratio being roughly constant at $x_{\rm H_2O} \sim 10^{-4}$--$10^{-3}$, regardless of whether ``all", ``reduced", ``full" or ``restricted" retrievals are employed.  We obtain the same behaviour for the potassium abundance. Here, $x_{\rm K}$ is roughly constant with values of about $10^{-7}$--$10^{-6}$.  
Methane is constrained for all of the T dwarfs and additionally also the L9 dwarf with a roughly constant mixing ratio of $x_{\rm CH_4} \sim 10^{-5}$--$10^{-4}$. With the exception of the L9 case, \ch{CH4} is rarely found in the other L dwarfs, as expected from theoretical predictions. Carbon monoxide, on the other hand, is only detected in a few objects, most notably in the L-dwarf spectra.\\

No obvious trend for the C/O ratios can be found. The considerable scatter in C/O is likely a consequence of the inability to constrain all import carbon and oxygen-bearing molecules in every spectrum. While water is consistently detected in all 19 objects, other major carbon and oxygen carriers, such as \ch{CH4}, \ch{CO}, and \ch{CO2}, are not always retrieved with sufficient constraints to provide good estimates on their mixing ratios. It is possible that this is caused by the $\SI{3.3}{\micro \metre}$ \ch{CH4} and $\SI{4.3}{\micro \metre}$ CO absorption bands not being covered by the SpeX instrument. Consequently, deriving the C/O from the retrieved molecular abundances for these cases becomes unreliable.

\subsubsection{Inability to retrieve cloud properties}
\label{sec:lt_clouds}

\begin{figure}
\centering
\includegraphics[width=0.98\columnwidth]{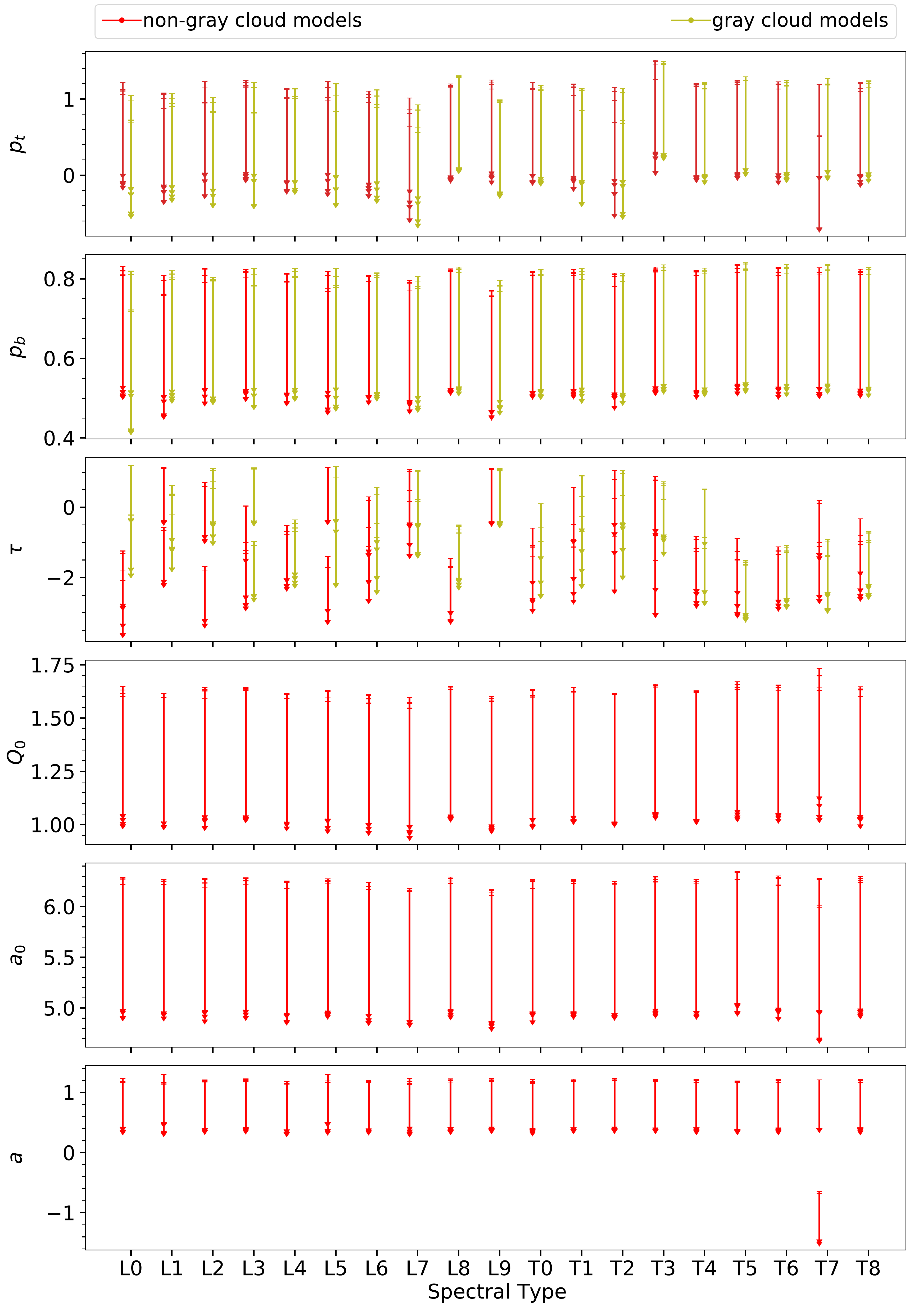}
\caption{Comparing the retrieved cloud parameters from our suite of brown dwarf retrievals across the L-T sequence. Parameters $p_{\mathrm{t}}$ and $p_{\mathrm{b}}$ represent the cloud top and bottom pressures, $\tau_{\rm ref}$ is optical depth at a reference wavelength, $Q_0$ is the proxy for the cloud particle composition, $a$ is the monodisperse particle radius and $a_0$ is the power-law index that describes wavelength variation. In each panel, gray-cloud (olive-green line) and non-gray-cloud (red line) models are compared. Overall, only upper limits are obtained.}
\label{fig: robustness test sequence cloudparams}
\end{figure}

As already noted, the Bayesian evidence does not strongly favor cloud-free or cloudy models. The Bayes factor of these two model scenarios is usually around unity, suggesting that neither model is preferred from a data-driven point of view. An overview of our retrieved cloud properties is presented in Figure~\ref{fig: robustness test sequence cloudparams} for all brown dwarfs in our sample.

Perhaps one of the most surprising outcomes of this study is that in most instances only upper limits for the cloud properties are obtained.

The cloud optical depth is unconstrained when a log-uniform prior is used (see Figure~\ref{fig: robustness test sequence cloudparams}). However, when a uniform prior is used for $\tau$ it becomes constrained (see Figure~\ref{fig:cloud sequence newtau} in Appendix~\ref{sect: Impact of prior choice}). Other cloud properties remain unconstrained. Consequently we are unable to obtain any clear trends in cloud properties across the L-T sequence as suggested by the retrieved cloud parameters depicted in Figure~\ref{fig: robustness test sequence cloudparams}. We obtain these results for both the gray and the non-gray cloud scenarios.

\subsubsection{Temperature-pressure profiles}

\begin{figure}
\centering
\includegraphics[width=\columnwidth]{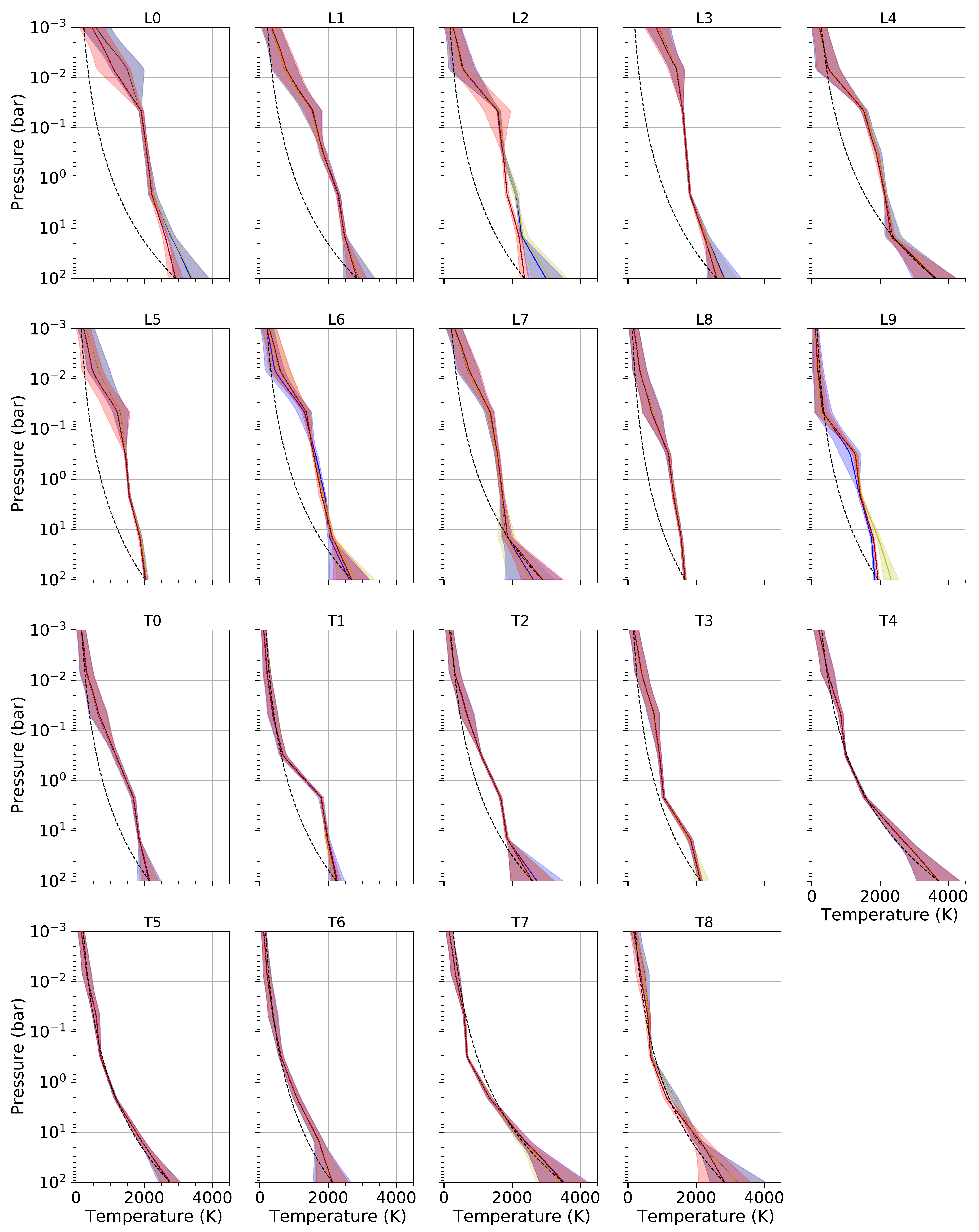}
\caption{Retrieved median temperature-pressure profiles with their associated $1\sigma$ uncertainties. In each panel, we compare the cloud-free (blue line), gray-cloud (olive-green line), and non-gray-cloud (red line) models. Only retrieved profiles from models using a reduced set of chemical species and analyzing spectra with a restricted wavelength range are shown. Adiabatic profiles (black dashed line) are indicated for comparison.}
\label{fig:T-P sequence}
\end{figure}

The temperature pressure profiles for all brown dwarfs in our sample are shown in Figure~\ref{fig:T-P sequence}. The figure also additionally depicts the dry adiabatic lapse rates for comparison. These adiabates are given by 
\begin{equation}
    \Gamma_\mathrm{ad} = - \frac{\mathrm d T_\mathrm{ad}}{\mathrm d z} = \frac{g}{c_p} \ ,
\end{equation}
where $c_p$ is the heat capacity at constant pressure.

Most of the T dwarfs show temperature profiles that follow those of theoretical models, with a potentially convective, lower atmosphere and an upper atmosphere that does not follow the adiabatic lapse rate. In theoretical brown dwarf models, the latter part of the atmosphere would be governed by radiative equilibrium \citep{Marley2015ARA&A}. 

The L dwarfs and some of the T dwarfs, however, result in temperature profiles that are unexpectedly shallow in the lower atmospheres. The temperature is in many cases almost isothermal, with lapse rates of the median profile of about $\Gamma \approx 5\times 10^{-3}$ K km$^{-1}$. With adiabatic lapse rates of about $\Gamma_\mathrm{ad} \approx$ 40 K km$^{-1}$ in these regions, the atmosphere is convectively stable. 

The same behavior was already noted by \cite{Kitzmann2020ApJ} in their analysis of the spectrum of the brown dwarf $\epsilon$ Indi Ba. Such an isothermal behaviour could be explained by the absence of clouds. A thick cloud layer would effectively block the emission of the lower atmosphere. If the spectrum is dominated by clouds, a retrieval without clouds would replicate this cloud impact by making the lower atmosphere more or less isothermal, with a temperature corresponding to the location of the cloud layer.

However, as already explained in the last subsection, even when clouds are added to the model, the retrieval is unable to constrain any cloud properties, except for their optical depths if a uniform prior is used. The retrieval seems to prefer isothermal temperature profiles over the potential existence of cloud layers. One reason for this behaviour might be caused by the description of the temperature profile being too flexible and allowing it to deviate strongly from the expected, adiabatic lapse rates in the lower atmosphere.

On the other hand, this outcome might also be explained by a chemical instability, as discussed in \cite{Tremblin2015ApJ} and \citet{ Tremblin2016ApJ}. This instability would effectively change the adiabatic index of the atmosphere in a way to allow for very small lapse rates even in the lower parts of the atmosphere.

\section{Discussion}
\label{sect:discussion}

\subsection{Summary}

In the current study, we subjected a curated sample of 19 SpeX spectra (0.85 to 2.45~$\SI{}{\micro\metre}$) of brown dwarfs, from spectral type L0 to T8, to a large suite of atmospheric retrievals.  Our findings include
\begin{itemize}

    \item From the perspective of Bayesian model comparison, cloud-free and cloudy models (with both gray and non-gray clouds) fit the data equally well. In other words, the SpeX data is consistent with both the absence and presence of clouds from atmospheric retrieval analysis. Consequently, only upper limits on cloud properties are retrieved. However, when a uniform (rather than log-uniform) prior is used for the cloud optical depth it becomes constrained.
    
    \item Water and potassium are detected in all 19 objects and their abundances are roughly constant across the L-T sequence.  Methane is detected in all of the T dwarfs, while carbon monoxide is only detected less than half of the sample. Consequently, the retrieved C/O ratios are unreliable and heterogeneous across the L-T sequence.
    
    \item For early L dwarfs, the retrieved surface gravity depends on whether the gray or non-gray cloud model is used with the resulting uncertainty sometimes spanning an order of magnitude or more.  
    
    \item The retrieved radius is robust to whether the cloud-free, gray-cloud or non-gray-cloud model is used, but the values associated with T dwarfs are often implausibly low, possibly indicating missing physics or chemistry \citep{Kitzmann2020ApJ}.
    
    \item All models are generally consistent in their T-P profiles and their atmospheres are stable to convection. We obtain shallow temperature gradients with the lower atmosphere being almost isothermal, especially when looking at L dwarfs. T dwarfs mostly follow the adiabatic lines where radiation pressure becomes inefficient.
    
\end{itemize}

\subsection{How may we test the chemical instability hypothesis?}

Instead of clouds, \cite{Tremblin2015ApJ, Tremblin2016ApJ} previously proposed that the variation in observed color across the L-T sequence may alternatively be explained by a chemical instability. One of the signatures of this instability is the \textit{vertical/radial} variation of the adiabatic index. Such a variation requires the relative abundances of atoms and molecules to vary across height/pressure. In the current suite of retrievals, we have assumed chemical abundances that are constant across height/pressure, which imply that the adiabatic index is constant throughout. Given the inability of these retrievals to distinguish between cloud-free and cloudy models, it is unlikely that retrievals with vertically/radially varying chemical abundances will be adequately constrained by the SpeX spectra. Spectra measured by the Hubble Space Telescope \citep{Apai2013ApJ} and the upcoming James Webb Space Telescope will be decisive for addressing this question.

\subsection{Looking towards the future: better data or better models?}

The formation and evolution of clouds in brown dwarfs, as well as their observational manifestation, remains incompletely understood. Forward models of brown dwarfs continue to be developed (e.g., \citealt{Marley2021arXiv}). Retrieval models have a useful role to play as they may offer hints on future directions for forward models, while incorporating the latest ideas on first-principles cloud formation models.

\begin{figure}[ht]
\begin{center}
  \includegraphics[width=\columnwidth]{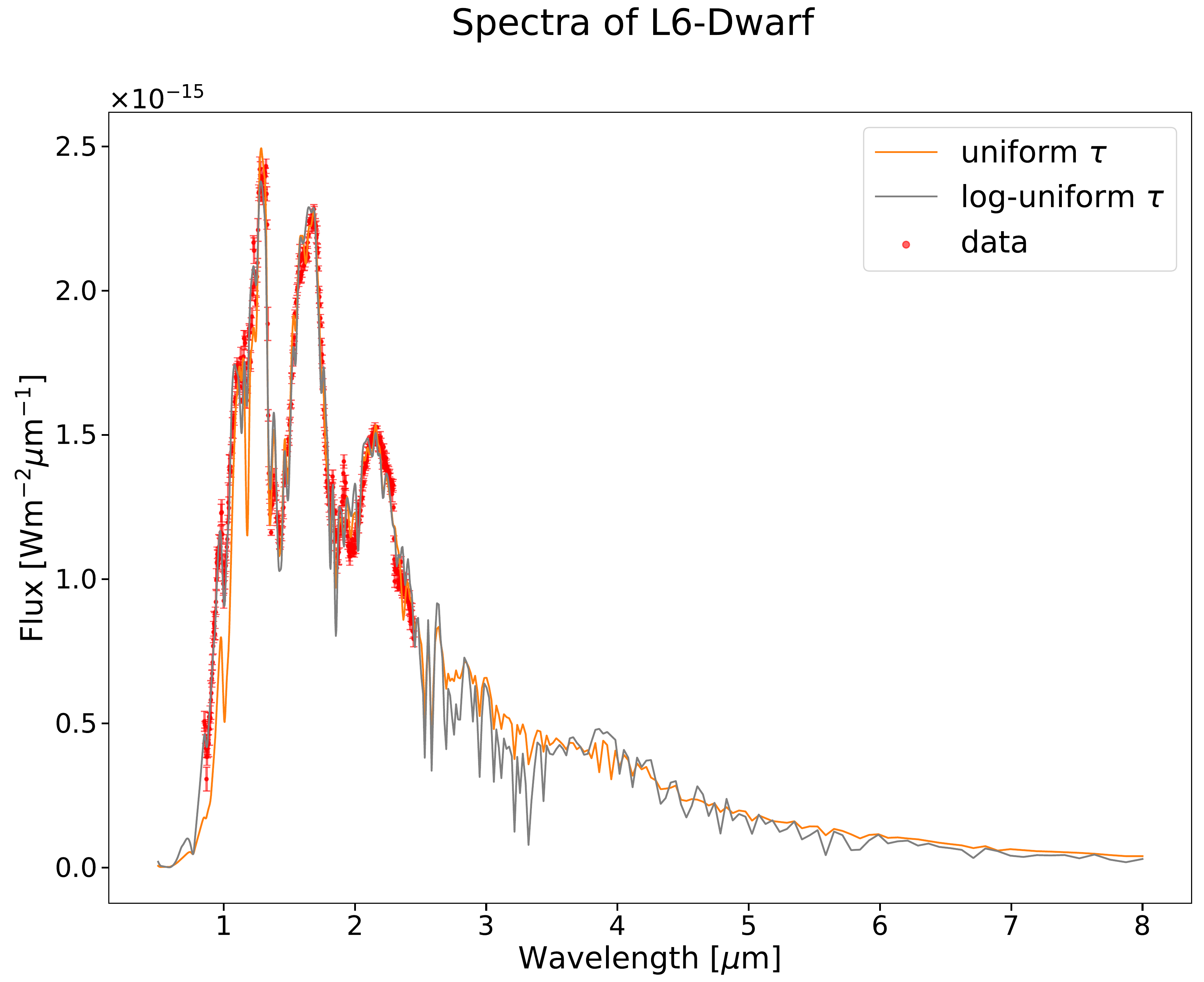}
\end{center}
\caption{Median restricted spectra ($F$) associated with the L6 dwarf of our curated sample, comparing the gray-cloud models with the original prior of optical depth (gray line) and the adjusted one (orange line) models. Data are shown as dots with associated uncertainties.}
\vspace{0.1in}
\label{fig:wavelength range increased}
\end{figure}

Our retrieval study based on SpeX spectra alone is unable to constrain the cloud properties of brown dwarfs. Nonetheless, letting our model to predict the spectra within a larger wavelength region towards the IR, we spot a slight difference in Figure~\ref{fig:wavelength range increased} when considering a change in the cloud prior distributions and thus, retrieving optical depths that indicate a cloud existence (see Appendix~\ref{sect: Impact of prior choice}). Clouds seem to diminish spectral features within the IR.

A plausible next step is to perform retrieval analyses of Hubble Space Telescope (HST) observations of brown dwarfs \citep{Apai2013ApJ, Madhusudhan2016arXiv}, where the enhanced signal-to-noise of the data may allow both cloud properties and vertical variation of chemical abundances to be constrained. It is possible that viewing geometry and variability may play a role in data procurement and interpretation \citep{Vos2017ApJ...842...78V,Bowler2020ApJ}. Analysis of the HST data will provide a glimpse of what to expect with spectra from the James Webb Space Telescope, which will potentially offer 0.6--28 $\SI{}{\micro\metre}$ coverage in addition to exquisite signal-to-noise.\\

\begin{acknowledgments}
We acknowledge partial financial support from the Swiss National Science Foundation, the European Research Council (via a Consolidator Grant to KH; grant number 771620) and the Center for Space and Habitability (CSH). This research has benefitted from the SpeX Prism Library, maintained by Adam Burgasser at \url{http://www.browndwarfs.org/spexprism}.
\end{acknowledgments}

\bibliographystyle{aasjournal}
\bibliography{references}

\appendix

\section{Supplementary Figures}
\label{sect: Supplementary Figures}

\begin{figure}[ht]
\begin{center}
\vspace{-0.1in}
  \includegraphics[width=0.75\textwidth]{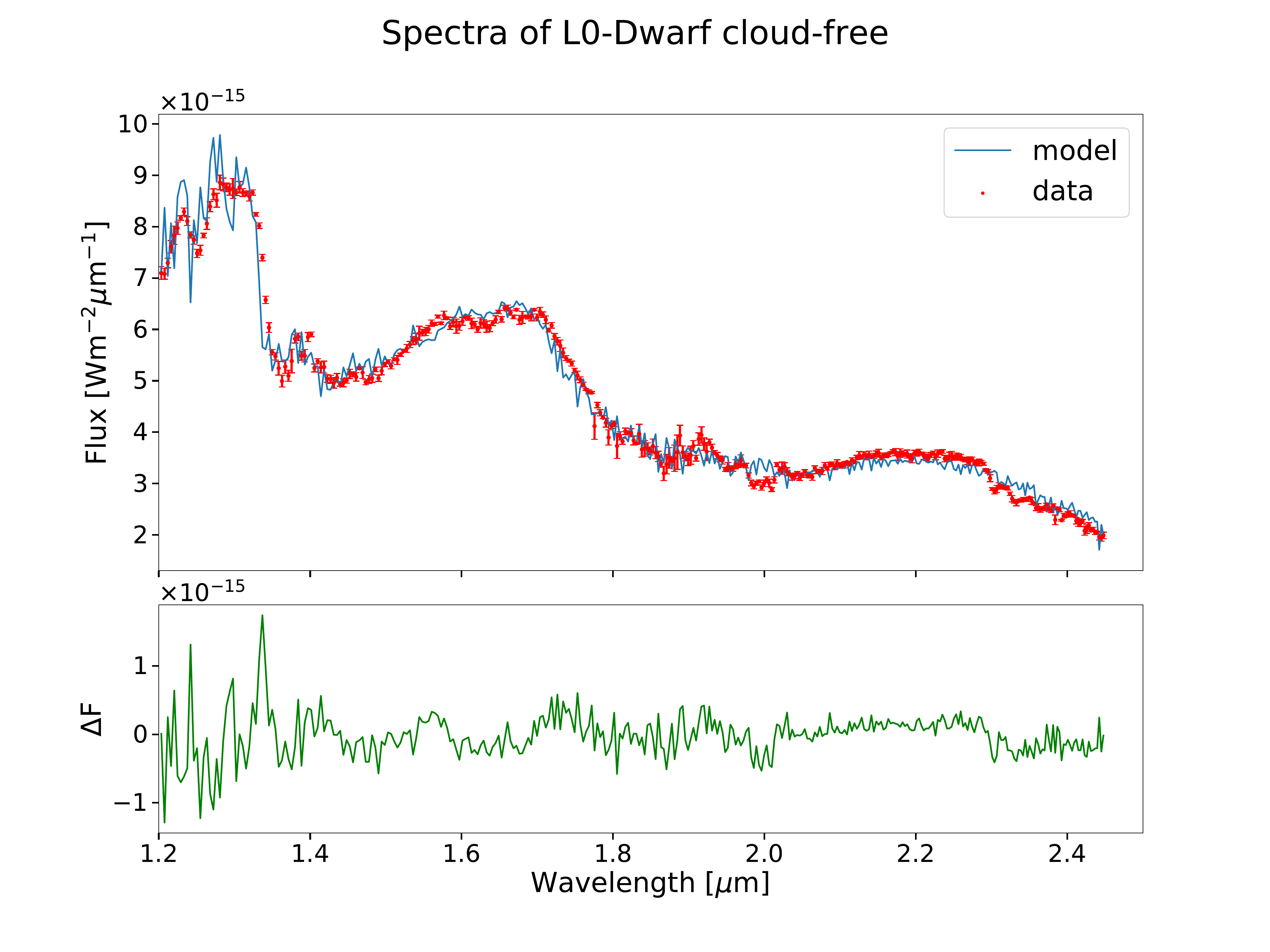}
\end{center}
\vspace{-0.2in}
\caption{Median restricted spectra ($F$) and residuals ($\Delta F$) associated with the L0 dwarf of our curated sample for a cloud-free model with a reduced set of molecules. Data are shown as dots with associated uncertainties. The complete figure set for the entire L0 to T8 sequence (19 images) is available in the online journal.}
\label{fig:spectra_appendix}
\end{figure}

\begin{figure}[ht]
\begin{center}
\vspace{-0.1in}
  \includegraphics[width=0.75\textwidth]{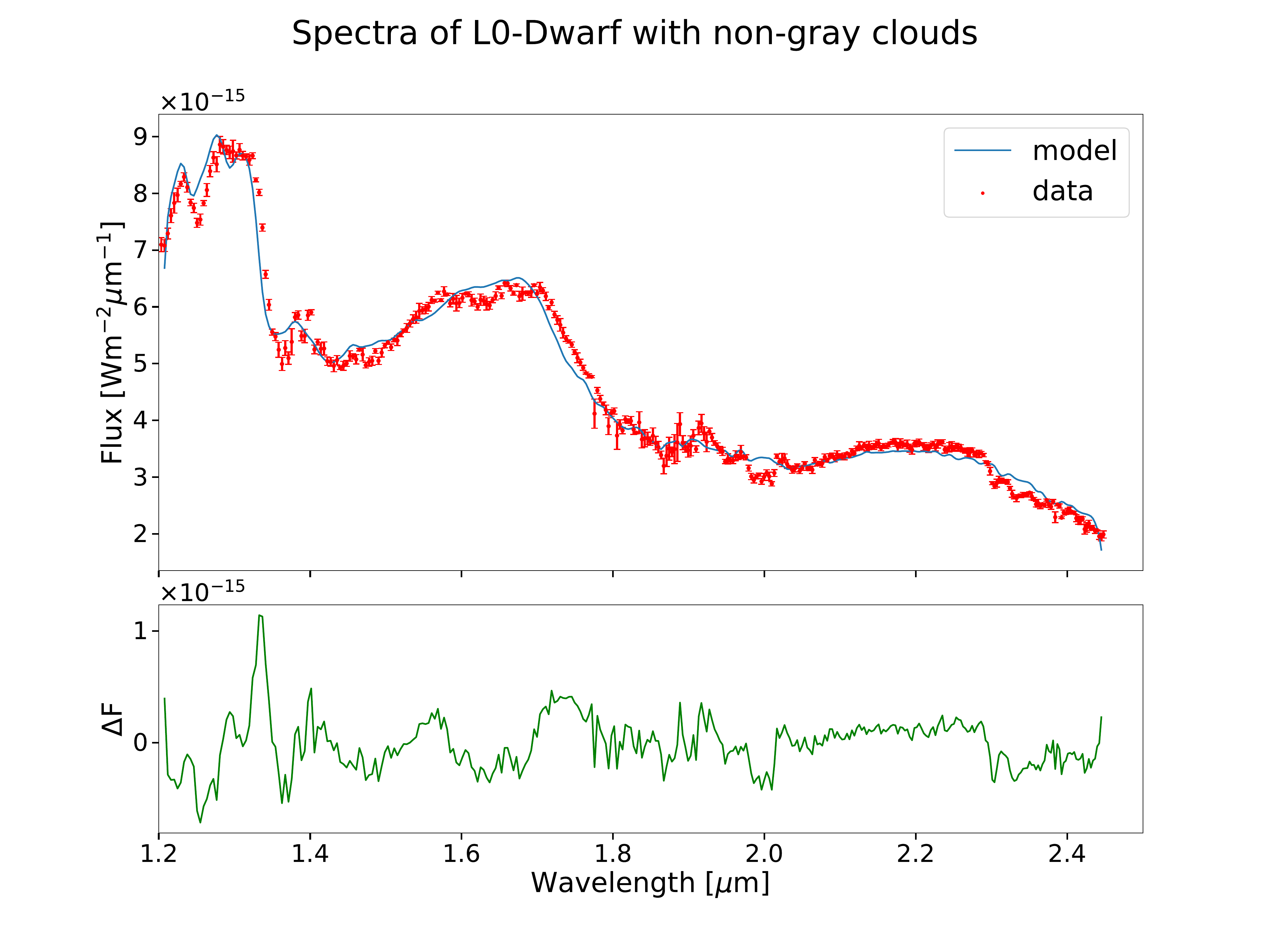}
\end{center}
\vspace{-0.2in}
\caption{Median restricted spectra ($F$) and residuals ($\Delta F$) associated with the L0 dwarf of our curated sample for a non-gray-cloud model with a reduced set of molecules. Data are shown as dots with associated uncertainties. The complete figure set for the entire L0 to T8 sequence (19 images) is available in the online journal.}
\label{fig:spectra_appendix2}
\end{figure}

\begin{figure}[!h]
\centering
\includegraphics[width=0.9\textwidth]{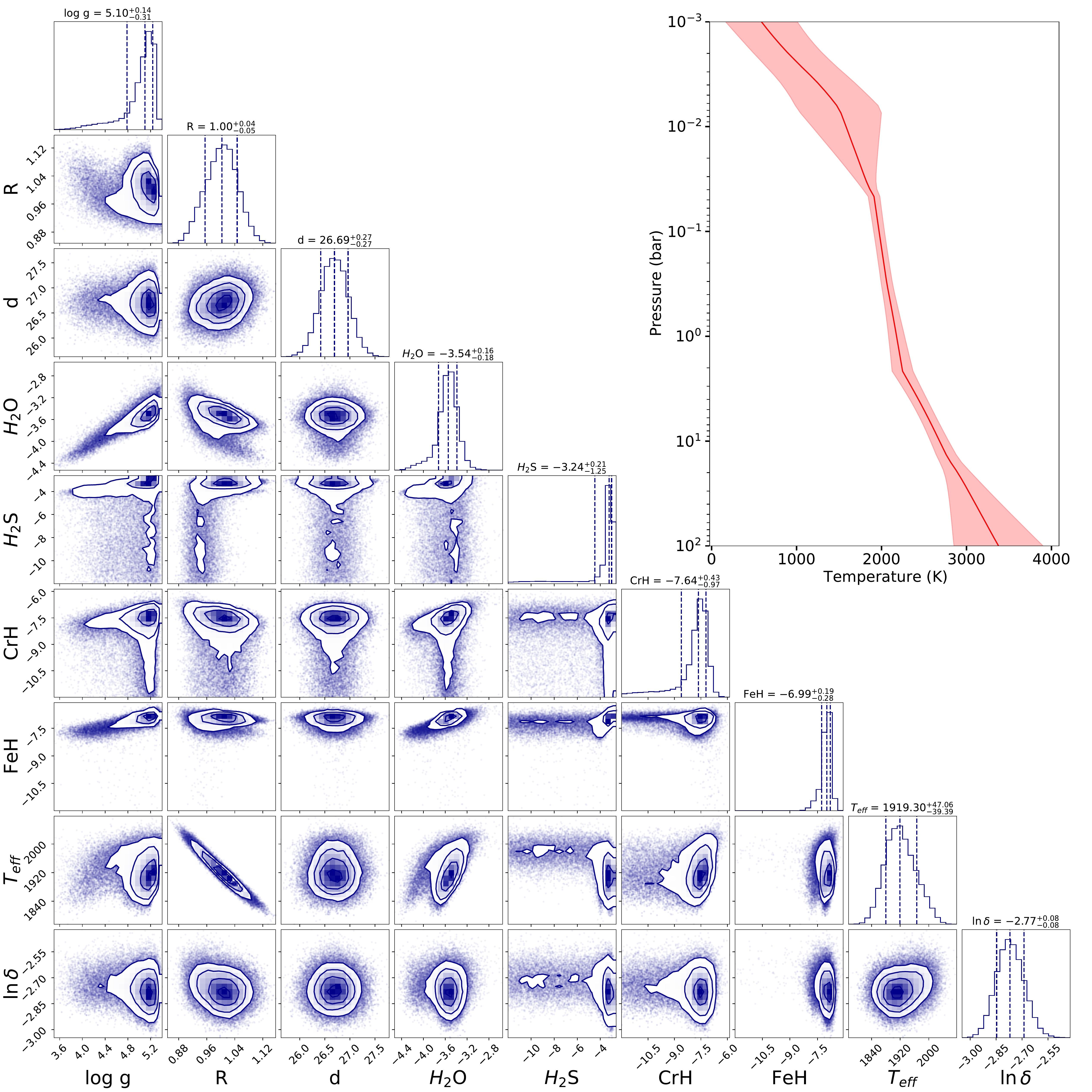}
\vspace{-0.1in}
\caption{Joint posterior distributions from free-chemistry retrieval analyses of the spectrum with a restricted wavelength range and cloud-free model for the L0 standard brown dwarf of our curated sample. The vertical dashed lines correspond to the median parameter values and their $1\sigma$ uncertainties. Accompanying each montage of joint posterior distributions is the retrieved median temperature-pressure profile and its associated $1\sigma$ uncertainties. The effective temperature $T_{\mathrm{eff}}$ is the only parameter that is determined via post-processing (see text for details). The complete figure set for the entire L0 to T8 sequence (19 images) is available in the online journal.}
\label{fig:posteriors_appendix}
\end{figure}

\begin{figure}[!h]
\centering
\includegraphics[width=0.9\textwidth]{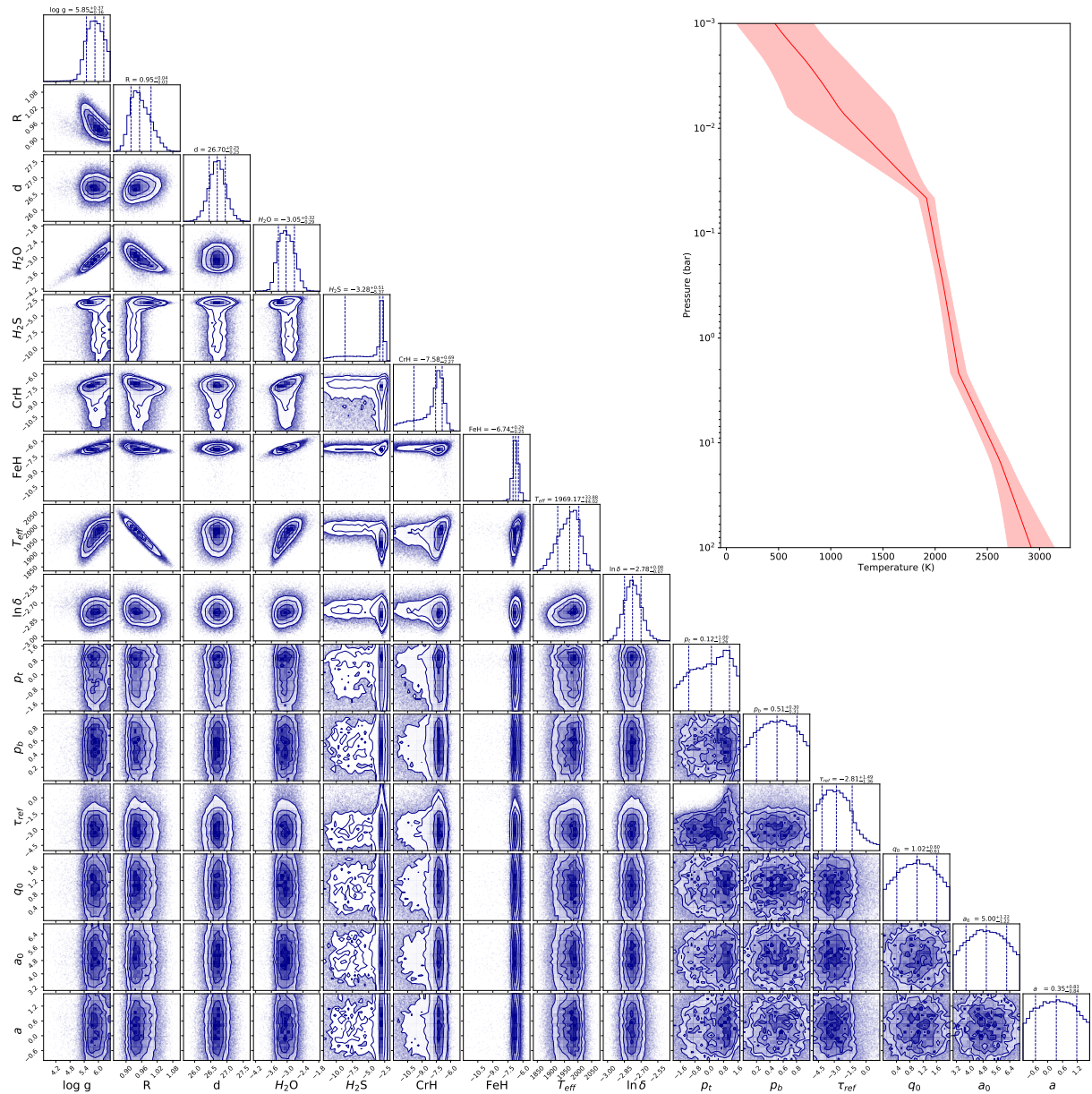}
\caption{Joint posterior distributions from free-chemistry retrieval analyses of the spectrum with a restricted wavelength range and non-gray-cloud model for the L0 standard brown dwarf of our curated sample. The vertical dashed lines correspond to the median parameter values and their $1\sigma$ uncertainties. Accompanying each montage of joint posterior distributions is the retrieved median temperature-pressure profile and its associated $1\sigma$ uncertainties. The effective temperature $T_{\mathrm{eff}}$ is the only parameter that is determined via post-processing (see text for details). The complete figure set for the entire L0 to T8 sequence (19 images) is available in the online journal.}
\label{fig:posteriors_appendix2}
\end{figure}

For completeness, Figures~\ref{fig:spectra_appendix} \& \ref{fig:spectra_appendix2} compare the cloud-free and non-gray-cloud models for restricted spectra with reduced sets of chemical species. The corresponding full set of posterior distributions of the parameters are shown in Figures~\ref{fig:posteriors_appendix} \& \ref{fig:posteriors_appendix2}.\\

\figsetstart
\figsetnum{1}
\figsettitle{Cloud-free model spectra associated with the L0 to T8 dwarfs of our curated sample}
\figsetgrpstart
\figsetgrpnum{1.0}
\figsetgrptitle{L0 cloud-free}
\figsetplot{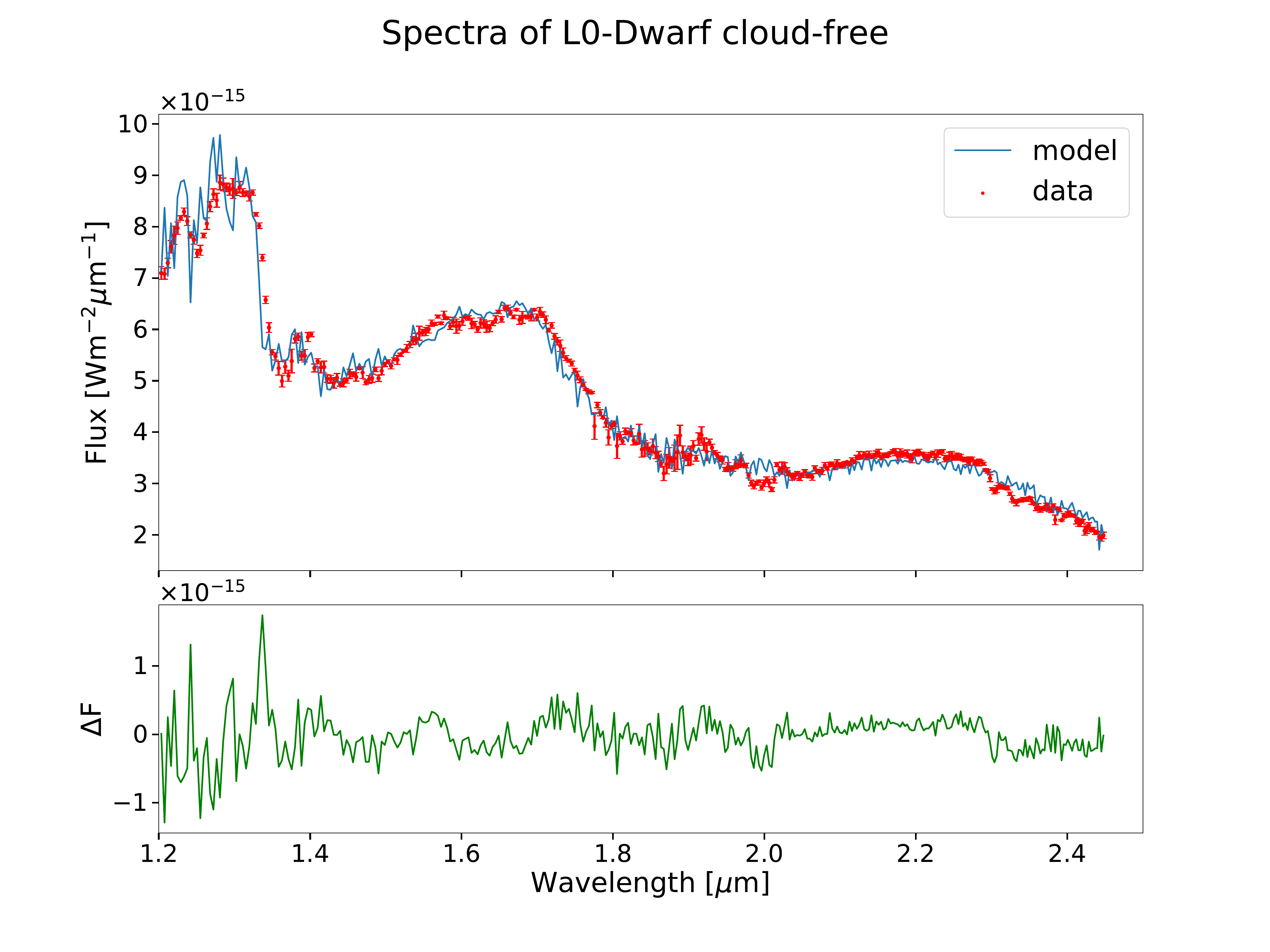}
\figsetgrpnote{Median restricted spectra ($F$) and residuals ($\Delta F$) associated with the L0 dwarf of our curated sample for a cloud-free model with a reduced set of molecules. Data are shown as dots with associated uncertainties.}
\figsetgrpend
\figsetend

\clearpage
\figsetstart
\figsetnum{2}
\figsettitle{Non-gray-cloud model spectra associated with the L0 to T8 dwarfs of our curated sample}
\figsetgrpstart
\figsetgrpnum{2.0}
\figsetgrptitle{L0 non-gray-cloud}
\figsetplot{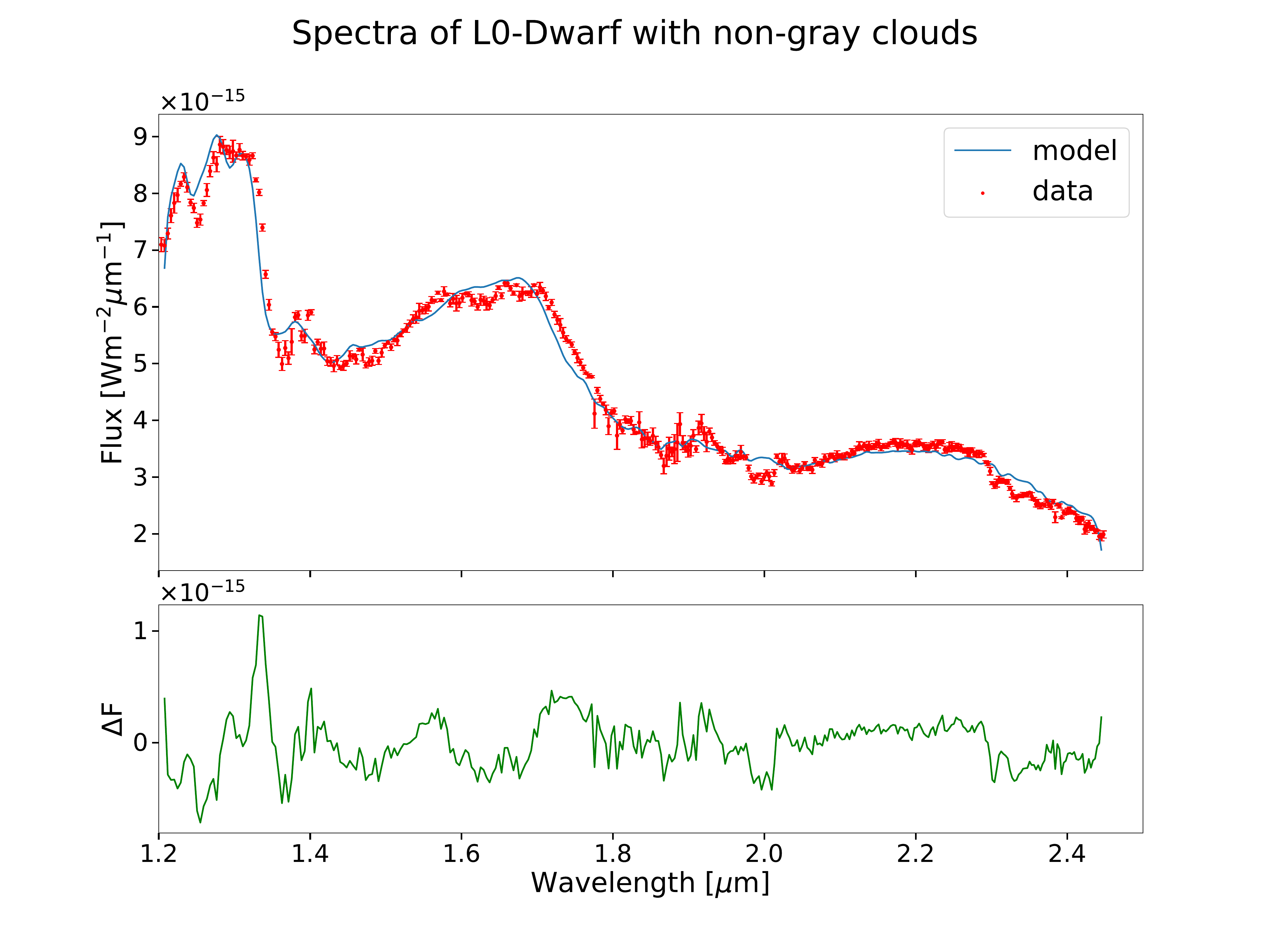}
\figsetgrpnote{Median restricted spectra ($F$) and residuals ($\Delta F$) associated with the L0 dwarf of our curated sample for a non-gray-cloud model with a reduced set of molecules. Data are shown as dots with associated uncertainties.}
\figsetgrpend
\figsetend

\clearpage
\figsetstart
\figsetnum{3}
\figsettitle{Cloud-free model posterior distributions associated with the L0 to T8 dwarfs of our curated sample}
\figsetgrpstart
\figsetgrpnum{3.0}
\figsetgrptitle{L0 posterior cloud-free}
\figsetplot{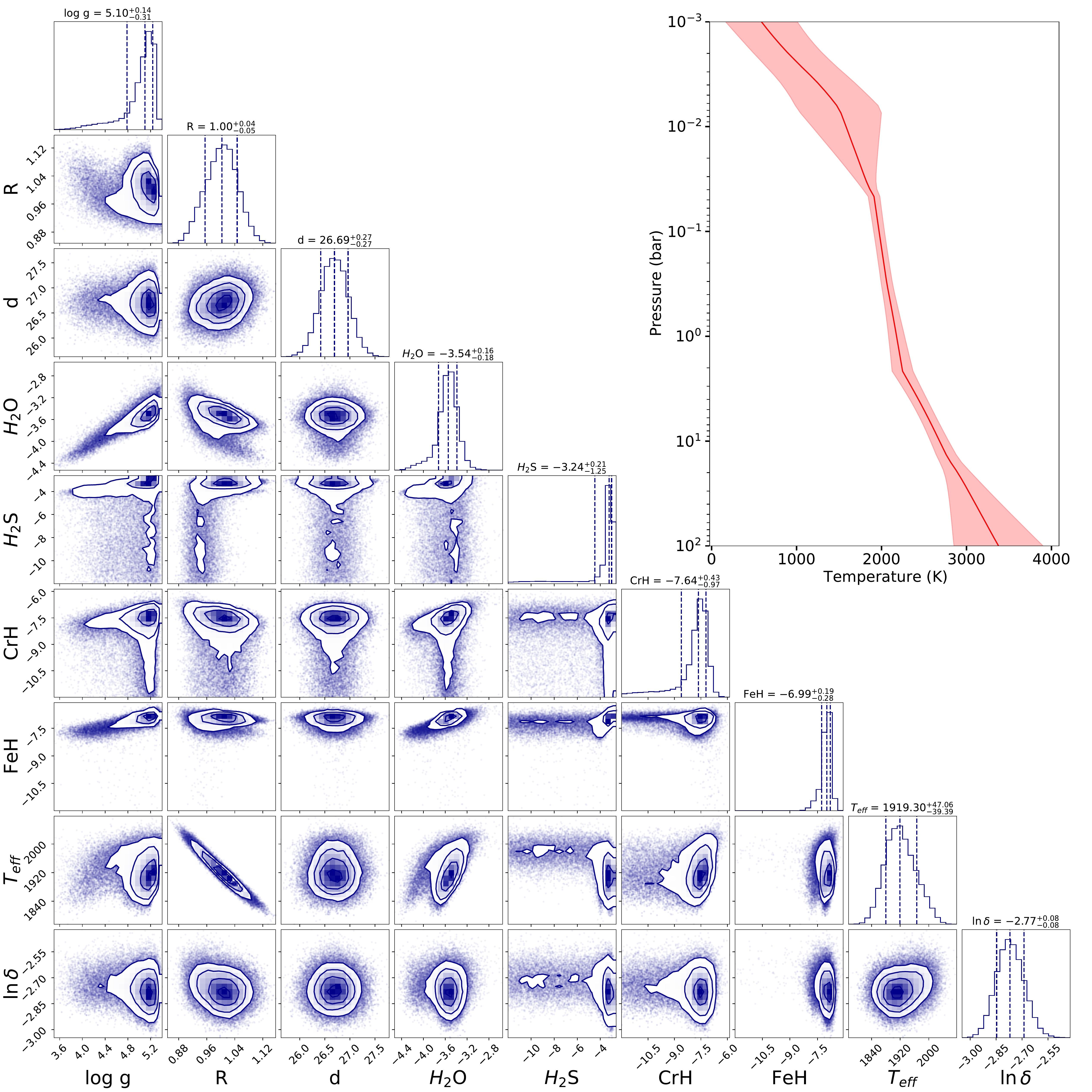}
\figsetgrpnote{Joint posterior distributions from free-chemistry retrieval analyses of the spectrum with a restricted wavelength range and cloud-free model for the L0 standard brown dwarf of our curated sample. The vertical dashed lines correspond to the median parameter values and their $1\sigma$ uncertainties. Accompanying each montage of joint posterior distributions is the retrieved median temperature-pressure profile and its associated $1\sigma$ uncertainties. The effective temperature $T_{\mathrm{eff}}$ is the only parameter that is determined via post-processing (see text for details).}
\figsetgrpend
\figsetgrpstart
\figsetgrpnum{3.9}
\figsetgrptitle{L9 posterior cloud-free}
\figsetplot{Posteriors and Spectra/L0_restricted_spectra_posterior.pdf}
\figsetgrpnote{Joint posterior distributions from free-chemistry retrieval analyses of the spectrum with a restricted wavelength range and cloud-free model for the L9 standard brown dwarf of our curated sample. The vertical dashed lines correspond to the median parameter values and their $1\sigma$ uncertainties. Accompanying each montage of joint posterior distributions is the retrieved median temperature-pressure profile and its associated $1\sigma$ uncertainties. The effective temperature $T_{\mathrm{eff}}$ is the only parameter that is determined via post-processing (see text for details).}
\figsetgrpend
\figsetend

\clearpage
\figsetstart
\figsetnum{4}
\figsettitle{Non-gray-cloud model posterior distributions associated with the L0 to T8 dwarfs of our curated sample}
\figsetgrpstart
\figsetgrpnum{4.0}
\figsetgrptitle{L0 posterior non-gray-cloud}
\figsetplot{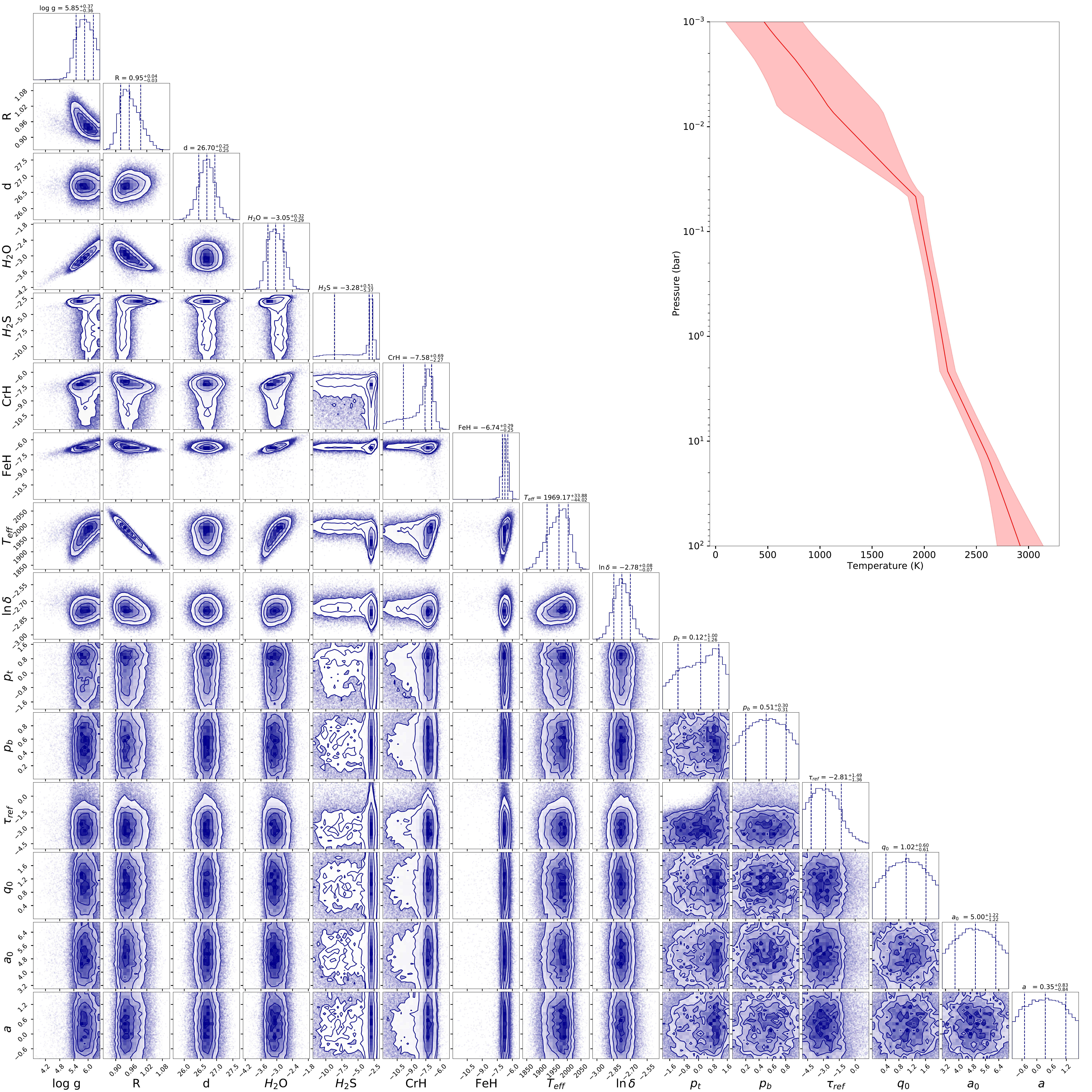}
\figsetgrpnote{Joint posterior distributions from free-chemistry retrieval analyses of the spectrum with a restricted wavelength range and non-gray-cloud model for the L0 standard brown dwarf of our curated sample. The vertical dashed lines correspond to the median parameter values and their $1\sigma$ uncertainties. Accompanying each montage of joint posterior distributions is the retrieved median temperature-pressure profile and its associated $1\sigma$ uncertainties. The effective temperature $T_{\mathrm{eff}}$ is the only parameter that is determined via post-processing (see text for details).}
\figsetgrpend
\figsetend

\section{Supplementary Data}
\label{sect: Supplementary Data}

For completeness, Tables \ref{tab:data posteriors L dwarfs} \& \ref{tab:data posteriors T dwarfs} record the outcomes of a large suite of retrievals (6 models for each object).

\begin{longrotatetable}
\begin{deluxetable*}{cccccccccccc}
\tabletypesize{\scriptsize}
\tablecolumns{12}
\tablewidth{\columnwidth}
\tablecaption{Summary of retrieval outcomes for the standard L dwarfs. Only models with the reduced set of molecules are tabulated.  Variations of the models shown are: full spectra cloud-free (FC), full spectra gray (FG), full spectra non-gray (FN), restricted spectra cloud-free (RC), restricted spectra gray (RG) and restricted spectra non-gray (RN).}
\label{tab:data posteriors L dwarfs}
\tablehead{
 \colhead{Model} & \colhead{Parameter} & \colhead{L0} & \colhead{L1} & \colhead{L2} & \colhead{L3} & \colhead{L4} & \colhead{L5} & \colhead{L6} & \colhead{L7} & \colhead{L8} & \colhead{L9}	\\
	}
\startdata 
F C & log g & $3.73_{-0.12}^{+0.13}$ & $4.15_{-0.17}^{+0.16}$ & $4.29_{-0.42}^{+0.64}$ & $5.29_{-0.10}^{+0.06}$ & $4.63_{-0.54}^{+0.46}$ & $5.67_{-0.14}^{+0.12}$ & $5.15_{-0.48}^{+0.35}$ & $4.63_{-0.63}^{+0.53}$ & $5.40_{-0.18}^{+0.17}$ & $5.59_{-0.07}^{+0.06}$	\\
F G & log g & $4.75_{-0.13}^{+0.14}$ & $4.16_{-0.17}^{+0.15}$ & $5.00_{-0.15}^{+0.12}$ & $5.07_{-0.17}^{+0.17}$ & $4.67_{-0.54}^{+0.44}$ & $5.35_{-0.15}^{+0.14}$ & $5.21_{-0.46}^{+0.31}$ & $4.73_{-0.56}^{+0.44}$ & $5.38_{-0.18}^{+0.17}$ & $5.28_{-0.15}^{+0.17}$	\\
F N & log g & $6.42_{-0.08}^{+0.06}$ & $5.02_{-0.13}^{+0.20}$ & $6.21_{-0.27}^{+0.20}$ & $6.23_{-0.29}^{+0.18}$ & $5.30_{-0.68}^{+0.46}$ & $5.61_{-0.23}^{+0.21}$ & $5.80_{-0.62}^{+0.45}$ & $5.43_{-0.54}^{+0.36}$ & $5.71_{-0.24}^{+0.19}$ & $5.56_{-0.17}^{+0.15}$	\\
R C & log g & $5.10_{-0.31}^{+0.14}$ & $5.00_{-0.34}^{+0.21}$ & $4.95_{-0.33}^{+0.20}$ & $5.13_{-0.17}^{+0.14}$ & $4.86_{-0.52}^{+0.32}$ & $5.58_{-0.15}^{+0.14}$ & $5.18_{-0.38}^{+0.31}$ & $4.31_{-0.48}^{+0.63}$ & $5.25_{-0.27}^{+0.20}$ & $5.62_{-0.14}^{+0.08}$	\\
R G & log g & $5.10_{-0.29}^{+0.14}$ & $4.98_{-0.37}^{+0.22}$ & $5.03_{-0.29}^{+0.15}$ & $5.14_{-0.16}^{+0.14}$ & $4.87_{-0.52}^{+0.33}$ & $5.58_{-0.14}^{+0.14}$ & $5.19_{-0.44}^{+0.29}$ & $4.36_{-0.48}^{+0.57}$ & $5.25_{-0.27}^{+0.19}$ & $5.28_{-0.17}^{+0.17}$	\\
R N & log g & $5.85_{-0.36}^{+0.37}$ & $5.60_{-0.44}^{+0.26}$ & $5.75_{-0.36}^{+0.32}$ & $5.29_{-0.20}^{+0.28}$ & $5.77_{-0.64}^{+0.42}$ & $5.83_{-0.22}^{+0.30}$ & $6.02_{-0.61}^{+0.31}$ & $4.59_{-0.60}^{+1.02}$ & $5.55_{-0.29}^{+0.25}$ & $5.65_{-0.13}^{+0.12}$	\\  \hline
F C & R & $1.33_{-0.07}^{+0.08}$ & $1.14_{-0.05}^{+0.05}$ & $1.12_{-0.06}^{+0.09}$ & $0.93_{-0.03}^{+0.03}$ & $0.97_{-0.14}^{+0.13}$ & $0.59_{-0.07}^{+0.10}$ & $0.65_{-0.06}^{+0.06}$ & $0.83_{-0.07}^{+0.07}$ & $0.76_{-0.12}^{+0.14}$ & $0.67_{-0.04}^{+0.03}$	\\
F G & R & $1.25_{-0.05}^{+0.05}$ & $1.14_{-0.05}^{+0.05}$ & $1.15_{-0.05}^{+0.05}$ & $0.94_{-0.03}^{+0.03}$ & $0.96_{-0.14}^{+0.13}$ & $0.77_{-0.10}^{+0.09}$ & $0.65_{-0.06}^{+0.05}$ & $0.86_{-0.07}^{+0.06}$ & $0.76_{-0.12}^{+0.14}$ & $0.63_{-0.02}^{+0.02}$	\\
F N & R & $1.64_{-0.11}^{+0.11}$ & $1.23_{-0.04}^{+0.04}$ & $1.14_{-0.05}^{+0.05}$ & $0.98_{-0.04}^{+0.05}$ & $0.97_{-0.13}^{+0.11}$ & $0.79_{-0.09}^{+0.08}$ & $0.65_{-0.06}^{+0.05}$ & $0.87_{-0.07}^{+0.07}$ & $0.85_{-0.12}^{+0.11}$ & $0.62_{-0.02}^{+0.02}$	\\
R C & R & $1.00_{-0.05}^{+0.04}$ & $1.00_{-0.03}^{+0.03}$ & $1.07_{-0.04}^{+0.04}$ & $0.94_{-0.03}^{+0.03}$ & $0.95_{-0.13}^{+0.12}$ & $0.65_{-0.09}^{+0.11}$ & $0.66_{-0.06}^{+0.06}$ & $0.87_{-0.07}^{+0.08}$ & $0.85_{-0.15}^{+0.18}$ & $0.62_{-0.03}^{+0.03}$	\\
R G & R & $1.00_{-0.05}^{+0.04}$ & $1.01_{-0.03}^{+0.03}$ & $1.07_{-0.04}^{+0.04}$ & $0.94_{-0.03}^{+0.03}$ & $0.94_{-0.12}^{+0.11}$ & $0.66_{-0.09}^{+0.11}$ & $0.69_{-0.06}^{+0.06}$ & $0.86_{-0.07}^{+0.07}$ & $0.85_{-0.14}^{+0.17}$ & $0.61_{-0.02}^{+0.02}$	\\
R N & R & $0.95_{-0.03}^{+0.04}$ & $0.99_{-0.03}^{+0.03}$ & $1.03_{-0.05}^{+0.05}$ & $0.93_{-0.03}^{+0.03}$ & $0.95_{-0.12}^{+0.10}$ & $0.77_{-0.09}^{+0.08}$ & $0.69_{-0.05}^{+0.05}$ & $0.84_{-0.06}^{+0.07}$ & $0.91_{-0.11}^{+0.12}$ & $0.60_{-0.02}^{+0.02}$	\\  \hline
F C & $\mathrm{T_{eff}}$ & $1799.16_{-48.86}^{+44.79}$ & $1873.90_{-39.95}^{+37.93}$ & $1730.29_{-62.21}^{+46.20}$ & $1674.93_{-28.44}^{+27.16}$ & $1706.08_{-27.08}^{+24.99}$ & $1521.41_{-25.19}^{+24.26}$ & $1561.84_{-22.67}^{+21.98}$ & $1364.02_{-34.61}^{+32.34}$ & $1220.01_{-39.70}^{+38.87}$ & $1298.89_{-31.48}^{+35.09}$	\\
F G & $\mathrm{T_{eff}}$ & $1852.63_{-38.11}^{+35.47}$ & $1874.60_{-38.86}^{+36.30}$ & $1709.85_{-37.96}^{+38.56}$ & $1660.35_{-26.88}^{+26.70}$ & $1706.56_{-26.30}^{+24.30}$ & $1482.93_{-23.12}^{+23.80}$ & $1558.38_{-21.73}^{+21.98}$ & $1332.57_{-26.97}^{+30.51}$ & $1221.46_{-40.54}^{+39.04}$ & $1349.29_{-16.20}^{+16.82}$	\\
F N & $\mathrm{T_{eff}}$ & $1619.61_{-53.75}^{+54.11}$ & $1804.20_{-29.53}^{+30.18}$ & $1719.26_{-33.55}^{+35.04}$ & $1626.62_{-39.43}^{+37.81}$ & $1696.72_{-29.27}^{+28.59}$ & $1493.11_{-23.13}^{+22.76}$ & $1564.36_{-20.10}^{+21.21}$ & $1326.71_{-33.25}^{+33.91}$ & $1229.04_{-37.71}^{+35.36}$ & $1359.59_{-17.89}^{+21.31}$	\\
R C & $\mathrm{T_{eff}}$ & $1919.30_{-39.39}^{+47.06}$ & $1858.86_{-28.26}^{+28.95}$ & $1658.74_{-29.90}^{+30.73}$ & $1571.92_{-24.01}^{+23.91}$ & $1618.26_{-17.46}^{+17.85}$ & $1394.87_{-23.82}^{+24.06}$ & $1476.70_{-22.48}^{+22.51}$ & $1281.97_{-37.05}^{+33.35}$ & $1123.47_{-53.85}^{+48.80}$ & $1279.36_{-27.83}^{+27.54}$	\\
R G & $\mathrm{T_{eff}}$ & $1920.73_{-38.82}^{+48.32}$ & $1846.83_{-27.99}^{+27.14}$ & $1661.00_{-28.09}^{+30.66}$ & $1570.77_{-23.83}^{+23.23}$ & $1617.90_{-17.32}^{+17.61}$ & $1394.45_{-22.80}^{+23.40}$ & $1435.36_{-21.34}^{+21.65}$ & $1282.35_{-34.47}^{+32.08}$ & $1125.49_{-51.62}^{+46.42}$ & $1297.24_{-21.72}^{+21.88}$	\\
R N & $\mathrm{T_{eff}}$ & $1969.17_{-44.02}^{+33.88}$ & $1871.61_{-22.75}^{+24.09}$ & $1690.84_{-38.11}^{+45.64}$ & $1581.54_{-25.79}^{+28.17}$ & $1618.33_{-16.13}^{+16.60}$ & $1432.86_{-32.94}^{+33.65}$ & $1444.14_{-17.30}^{+18.91}$ & $1294.20_{-34.33}^{+31.21}$ & $1140.83_{-47.93}^{+42.05}$ & $1303.96_{-19.26}^{+19.28}$	\\  \hline
F C & log \ch{H2O} & $-4.62_{-0.10}^{+0.10}$ & $-4.31_{-0.10}^{+0.10}$ & $-4.01_{-0.36}^{+0.67}$ & $-3.40_{-0.11}^{+0.11}$ & $-3.29_{-0.49}^{+0.48}$ & $-3.31_{-0.10}^{+0.10}$ & $-2.87_{-0.41}^{+0.37}$ & $-3.12_{-0.54}^{+0.61}$ & $-3.77_{-0.13}^{+0.13}$ & $-3.49_{-0.06}^{+0.06}$	\\
F G & log \ch{H2O} & $-4.07_{-0.10}^{+0.09}$ & $-4.30_{-0.09}^{+0.09}$ & $-3.85_{-0.10}^{+0.10}$ & $-3.46_{-0.12}^{+0.12}$ & $-3.25_{-0.51}^{+0.46}$ & $-3.30_{-0.14}^{+0.13}$ & $-2.85_{-0.39}^{+0.33}$ & $-3.34_{-0.37}^{+0.40}$ & $-3.76_{-0.13}^{+0.13}$ & $-3.49_{-0.07}^{+0.08}$	\\
F N & log \ch{H2O} & $-3.66_{-0.09}^{+0.08}$ & $-3.86_{-0.07}^{+0.09}$ & $-3.25_{-0.22}^{+0.18}$ & $-2.76_{-0.31}^{+0.28}$ & $-2.61_{-0.68}^{+0.51}$ & $-3.14_{-0.19}^{+0.19}$ & $-2.30_{-0.61}^{+0.50}$ & $-3.02_{-0.38}^{+0.45}$ & $-3.61_{-0.14}^{+0.13}$ & $-3.38_{-0.08}^{+0.10}$	\\
R C & log \ch{H2O} & $-3.54_{-0.18}^{+0.16}$ & $-3.52_{-0.33}^{+0.23}$ & $-3.35_{-0.32}^{+0.22}$ & $-3.35_{-0.12}^{+0.11}$ & $-2.73_{-0.41}^{+0.37}$ & $-3.25_{-0.14}^{+0.14}$ & $-2.92_{-0.36}^{+0.33}$ & $-3.46_{-0.40}^{+0.65}$ & $-3.90_{-0.18}^{+0.16}$ & $-3.41_{-0.09}^{+0.08}$	\\
R G & log \ch{H2O} & $-3.54_{-0.17}^{+0.15}$ & $-3.52_{-0.36}^{+0.23}$ & $-3.35_{-0.22}^{+0.23}$ & $-3.36_{-0.11}^{+0.11}$ & $-2.72_{-0.41}^{+0.37}$ & $-3.25_{-0.14}^{+0.14}$ & $-2.97_{-0.34}^{+0.29}$ & $-3.43_{-0.40}^{+0.59}$ & $-3.89_{-0.18}^{+0.15}$ & $-3.46_{-0.09}^{+0.10}$	\\
R N & log \ch{H2O} & $-3.05_{-0.29}^{+0.32}$ & $-3.04_{-0.38}^{+0.30}$ & $-3.36_{-0.20}^{+0.25}$ & $-3.26_{-0.14}^{+0.19}$ & $-1.89_{-0.63}^{+0.47}$ & $-3.21_{-0.14}^{+0.19}$ & $-2.25_{-0.52}^{+0.40}$ & $-3.16_{-0.57}^{+1.27}$ & $-3.73_{-0.18}^{+0.16}$ & $-3.31_{-0.08}^{+0.09}$	\\  \hline
F C & log \ch{CH4} & - & - & $-5.32_{-4.23}^{+0.95}$ & - & $-4.75_{-1.03}^{+0.55}$ & - & $-4.11_{-0.47}^{+0.38}$ & - & - & $-4.46_{-0.12}^{+0.11}$	\\
F G & log \ch{CH4} & - & - & - & - & $-4.70_{-1.09}^{+0.52}$ & $-4.63_{-0.46}^{+0.23}$ & $-4.07_{-0.44}^{+0.33}$ & - & - & $-4.43_{-0.12}^{+0.12}$	\\
F N & log \ch{CH4} & - & - & - & - & $-3.94_{-0.77}^{+0.53}$ & $-4.39_{-0.40}^{+0.28}$ & $-3.49_{-0.64}^{+0.50}$ & - & - & $-4.26_{-0.13}^{+0.13}$	\\
R C & log \ch{CH4} & - & $-4.80_{-1.52}^{+0.33}$ & $-4.41_{-0.40}^{+0.25}$ & - & $-4.20_{-0.46}^{+0.36}$ & $-4.62_{-2.17}^{+0.28}$ & $-4.06_{-0.39}^{+0.34}$ & - & - & $-4.30_{-0.15}^{+0.12}$	\\
R G & log \ch{CH4} & - & $-4.77_{-1.18}^{+0.31}$ & $-4.41_{-0.31}^{+0.26}$ & - & $-4.19_{-0.45}^{+0.36}$ & $-4.60_{-1.91}^{+0.27}$ & $-3.94_{-0.39}^{+0.29}$ & - & - & $-4.40_{-0.14}^{+0.14}$	\\
R N & log \ch{CH4} & - & $-4.21_{-0.78}^{+0.36}$ & - & - & $-3.32_{-0.66}^{+0.47}$ & - & $-3.16_{-0.56}^{+0.38}$ & - & - & $-4.18_{-0.11}^{+0.12}$	\\  \hline
F C & log \ch{NH3} & - & - & - & - & - & - & $-4.57_{-0.65}^{+0.43}$ & $-4.35_{-0.58}^{+0.58}$ & $-4.45_{-0.15}^{+0.15}$ & $-5.05_{-3.06}^{+0.21}$	\\
F G & log \ch{NH3} & - & - & - & - & - & - & $-4.54_{-0.62}^{+0.40}$ & $-4.45_{-0.46}^{+0.41}$ & $-4.45_{-0.15}^{+0.14}$ & -	\\
F N & log \ch{NH3} & $-3.95_{-0.13}^{+0.11}$ & - & $-4.21_{-0.55}^{+0.28}$ & $-3.86_{-0.44}^{+0.30}$ & - & - & $-4.04_{-0.92}^{+0.61}$ & $-3.99_{-0.43}^{+0.40}$ & $-4.24_{-0.18}^{+0.15}$ & -	\\
R C & log \ch{NH3} & - & - & - & - & - & - & $-4.96_{-3.84}^{+0.60}$ & $-4.68_{-0.46}^{+0.65}$ & $-4.57_{-0.20}^{+0.17}$ & -	\\
R G & log \ch{NH3} & - & - & - & - & - & - & - & $-4.64_{-0.45}^{+0.59}$ & $-4.57_{-0.19}^{+0.16}$ & -	\\
R N & log \ch{NH3} & - & - & - & - & - & - & - & $-4.39_{-0.61}^{+1.23}$ & $-4.37_{-0.22}^{+0.19}$ & -	\\  \hline
F C & log \ch{CO2} & - & - & - & - & - & - & - & - & - & -	\\
F G & log \ch{CO2} & - & - & - & - & - & - & - & - & - & -	\\
F N & log \ch{CO2} & $-2.90_{-0.14}^{+0.13}$ & - & - & $-3.29_{-5.02}^{+0.46}$ & - & - & - & - & - & -	\\
R C & log \ch{CO2} & - & - & - & - & - & - & - & - & - & -	\\
R G & log \ch{CO2} & - & - & - & - & - & - & - & - & - & -	\\
R N & log \ch{CO2} & - & - & - & - & - & - & - & - & - & -	\\  \hline 
F C & log \ch{CO} & - & - & - & $-2.95_{-0.32}^{+0.28}$ & - & $-2.95_{-0.18}^{+0.18}$ & - & - & $-2.68_{-0.18}^{+0.18}$ & $-3.42_{-0.41}^{+0.23}$	\\
F G & log \ch{CO} & - & - & $-3.22_{-0.27}^{+0.23}$ & $-3.00_{-0.26}^{+0.24}$ & - & $-3.12_{-0.20}^{+0.20}$ & - & - & $-2.67_{-0.19}^{+0.18}$ & $-3.60_{-0.60}^{+0.26}$	\\
F N & log \ch{CO} & $-2.29_{-0.29}^{+0.24}$ & $-3.22_{-0.26}^{+0.22}$ & $-2.80_{-0.26}^{+0.25}$ & $-2.30_{-0.35}^{+0.33}$ & - & $-3.02_{-0.20}^{+0.20}$ & - & - & $-2.60_{-0.17}^{+0.16}$ & $-3.54_{-0.76}^{+0.25}$	\\
R C & log \ch{CO} & - & - & $-4.08_{-4.33}^{+0.69}$ & $-2.77_{-0.25}^{+0.22}$ & - & $-2.91_{-0.19}^{+0.18}$ & - & - & $-2.77_{-0.23}^{+0.19}$ & $-4.02_{-4.98}^{+0.65}$	\\
R G & log \ch{CO} & - & - & $-3.92_{-3.79}^{+0.73}$ & $-2.77_{-0.24}^{+0.22}$ & - & $-2.91_{-0.18}^{+0.18}$ & - & - & $-2.76_{-0.23}^{+0.19}$ & $-3.99_{-4.55}^{+0.52}$	\\
R N & log \ch{CO} & - & - & $-2.93_{-0.22}^{+0.23}$ & $-2.75_{-0.23}^{+0.21}$ & - & $-2.84_{-0.16}^{+0.17}$ & - & - & $-2.67_{-0.20}^{+0.17}$ & $-3.85_{-4.66}^{+0.46}$	\\   \hline
F C & log \ch{H2S} & $-4.14_{-0.15}^{+0.13}$ & - & $-3.63_{-0.40}^{+0.69}$ & $-3.67_{-5.02}^{+0.38}$ & - & - & - & - & - & -	\\
F G & log \ch{H2S} & $-3.47_{-0.13}^{+0.11}$ & - & $-3.34_{-0.12}^{+0.11}$ & $-3.35_{-0.19}^{+0.15}$ & - & $-3.20_{-0.21}^{+0.16}$ & - & - & - & -	\\
F N & log \ch{H2S} & - & $-3.37_{-0.11}^{+0.12}$ & $-2.69_{-0.25}^{+0.22}$ & $-2.59_{-1.48}^{+0.35}$ & $-3.70_{-4.98}^{+1.19}$ & $-3.04_{-0.25}^{+0.22}$ & - & - & - & -	\\
R C & log \ch{H2S} & $-3.24_{-1.25}^{+0.21}$ & - & $-3.02_{-0.34}^{+0.23}$ & $-3.09_{-0.14}^{+0.12}$ & - & $-3.06_{-0.24}^{+0.19}$ & - & - & - & -	\\
R G & log \ch{H2S} & $-3.25_{-1.84}^{+0.22}$ & - & $-2.99_{-0.24}^{+0.20}$ & $-3.09_{-0.14}^{+0.12}$ & - & $-3.06_{-0.24}^{+0.18}$ & $-2.87_{-0.41}^{+0.30}$ & - & - & -	\\
R N & log \ch{H2S} & $-3.28_{-5.37}^{+0.51}$ & - & $-3.01_{-0.23}^{+0.21}$ & $-3.06_{-0.18}^{+0.15}$ & - & $-3.37_{-4.24}^{+0.23}$ & $-2.07_{-0.61}^{+0.39}$ & - & - & -	\\  \hline
F C & log \ch{K} & $-6.71_{-0.18}^{+0.17}$ & $-6.37_{-0.15}^{+0.15}$ & $-6.37_{-0.48}^{+0.77}$ & $-6.16_{-0.18}^{+0.18}$ & $-5.54_{-0.51}^{+0.53}$ & $-6.37_{-0.15}^{+0.15}$ & $-5.27_{-0.45}^{+0.46}$ & $-5.93_{-0.72}^{+0.88}$ & $-7.92_{-0.27}^{+0.26}$ & $-7.34_{-0.17}^{+0.18}$	\\
F G & log \ch{K} & $-6.42_{-0.15}^{+0.13}$ & $-6.36_{-0.15}^{+0.14}$ & $-6.61_{-0.18}^{+0.17}$ & $-6.06_{-0.15}^{+0.15}$ & $-5.51_{-0.52}^{+0.52}$ & $-6.06_{-0.18}^{+0.16}$ & $-5.28_{-0.42}^{+0.43}$ & $-6.48_{-0.48}^{+0.56}$ & $-7.89_{-0.27}^{+0.27}$ & $-6.88_{-0.11}^{+0.10}$	\\
F N & log \ch{K} & $-7.36_{-0.16}^{+0.15}$ & $-6.38_{-0.12}^{+0.10}$ & $-6.52_{-0.25}^{+0.27}$ & $-5.72_{-0.49}^{+0.37}$ & $-4.89_{-0.69}^{+0.56}$ & $-5.97_{-0.21}^{+0.22}$ & $-4.75_{-0.61}^{+0.56}$ & $-6.71_{-2.00}^{+0.99}$ & $-7.89_{-0.34}^{+0.27}$ & $-6.88_{-0.11}^{+0.12}$	\\
R C & log \ch{K} & - & $-6.30_{-1.52}^{+0.62}$ & $-5.00_{-0.36}^{+0.29}$ & - & $-4.70_{-0.58}^{+0.60}$ & - & $-4.87_{-2.54}^{+0.56}$ & $-5.13_{-0.51}^{+0.65}$ & - & $-6.32_{-1.15}^{+0.35}$	\\
R G & log \ch{K} & - & $-5.73_{-0.39}^{+0.36}$ & $-5.19_{-0.46}^{+0.36}$ & - & $-4.66_{-0.54}^{+0.58}$ & - & - & $-5.21_{-0.66}^{+0.64}$ & - & -	\\
R N & log \ch{K} & - & $-6.20_{-1.80}^{+0.73}$ & $-5.89_{-0.37}^{+0.32}$ & $-7.10_{-2.88}^{+1.16}$ & $-3.95_{-0.76}^{+0.63}$ & - & - & $-5.59_{-2.89}^{+1.55}$ & - & -	\\  \hline
F C & log \ch{CrH} & $-8.94_{-0.24}^{+0.21}$ & $-9.00_{-0.35}^{+0.27}$ & $-8.25_{-0.43}^{+0.65}$ & $-8.10_{-0.28}^{+0.24}$ & $-7.44_{-0.51}^{+0.52}$ & - & $-7.29_{-0.47}^{+0.44}$ & $-7.68_{-0.68}^{+0.73}$ & $-8.83_{-0.24}^{+0.21}$ & $-8.64_{-0.14}^{+0.13}$	\\
F G & log \ch{CrH} & $-8.46_{-0.21}^{+0.19}$ & $-8.99_{-0.32}^{+0.26}$ & $-8.14_{-0.16}^{+0.16}$ & $-8.36_{-0.31}^{+0.27}$ & $-7.40_{-0.53}^{+0.50}$ & - & $-7.31_{-0.43}^{+0.42}$ & $-7.74_{-0.38}^{+0.43}$ & $-8.83_{-0.23}^{+0.21}$ & $-8.59_{-0.13}^{+0.13}$	\\
F N & log \ch{CrH} & $-8.08_{-0.13}^{+0.13}$ & $-8.63_{-0.23}^{+0.24}$ & $-7.36_{-0.25}^{+0.22}$ & $-7.37_{-0.32}^{+0.28}$ & $-6.78_{-0.70}^{+0.54}$ & - & $-6.73_{-0.63}^{+0.54}$ & $-7.46_{-0.41}^{+0.52}$ & $-8.65_{-0.23}^{+0.20}$ & $-8.46_{-0.13}^{+0.14}$	\\
R C & log \ch{CrH} & $-7.64_{-0.97}^{+0.43}$ & $-8.52_{-2.05}^{+0.99}$ & - & - & $-6.70_{-1.91}^{+0.68}$ & - & - & - & - & -	\\
R G & log \ch{CrH} & $-7.61_{-0.75}^{+0.40}$ & - & $-7.41_{-2.18}^{+0.56}$ & - & $-6.69_{-1.66}^{+0.67}$ & - & - & - & - & -	\\
R N & log \ch{CrH} & $-7.58_{-2.27}^{+0.69}$ & - & $-7.39_{-1.90}^{+0.40}$ & - & $-6.07_{-2.29}^{+0.85}$ & - & - & - & - & -	\\  \hline
F C & log \ch{FeH} & $-9.18_{-0.52}^{+0.32}$ & $-8.68_{-0.27}^{+0.23}$ & $-8.44_{-0.52}^{+0.78}$ & $-7.84_{-0.21}^{+0.20}$ & $-7.62_{-0.49}^{+0.52}$ & $-8.07_{-0.16}^{+0.15}$ & $-7.51_{-0.48}^{+0.46}$ & $-9.37_{-1.55}^{+1.26}$ & - & -	\\
F G & log \ch{FeH} & $-8.74_{-0.40}^{+0.26}$ & $-8.68_{-0.25}^{+0.21}$ & $-8.58_{-0.28}^{+0.25}$ & $-7.76_{-0.17}^{+0.16}$ & $-7.58_{-0.51}^{+0.50}$ & $-7.99_{-0.19}^{+0.18}$ & $-7.50_{-0.45}^{+0.43}$ & - & - & -	\\
F N & log \ch{FeH} & - & $-8.39_{-0.19}^{+0.16}$ & $-8.78_{-1.32}^{+0.46}$ & $-7.42_{-0.47}^{+0.39}$ & $-6.92_{-0.69}^{+0.56}$ & $-7.83_{-0.23}^{+0.22}$ & $-7.00_{-0.62}^{+0.58}$ & - & - & -	\\
R C & log \ch{FeH} & $-6.99_{-0.28}^{+0.19}$ & $-7.81_{-2.20}^{+0.53}$ & - & $-6.83_{-0.13}^{+0.12}$ & - & $-7.06_{-0.20}^{+0.18}$ & $-7.68_{-2.73}^{+0.93}$ & - & - & -	\\
R G & log \ch{FeH} & $-7.02_{-0.32}^{+0.20}$ & - & - & $-6.84_{-0.12}^{+0.12}$ & - & $-7.07_{-0.20}^{+0.18}$ & $-6.43_{-0.35}^{+0.30}$ & - & - & -	\\
R N & log \ch{FeH} & $-6.74_{-0.25}^{+0.29}$ & $-7.06_{-0.48}^{+0.36}$ & - & $-6.89_{-0.18}^{+0.15}$ & - & $-7.21_{-0.16}^{+0.18}$ & $-5.71_{-0.53}^{+0.42}$ & $-8.05_{-1.51}^{+0.96}$ & - & -	\\   \hline
F C & log \ch{CaH} & - & - & - & - & - & - & - & - & - & -	\\
F G & log \ch{CaH} & - & - & - & - & - & - & - & - & - & -	\\
F N & log \ch{CaH} & - & - & - & - & - & - & - & - & - & -	\\
R C & log \ch{CaH} & - & - & - & - & - & - & - & - & - & -	\\
R G & log \ch{CaH} & - & - & - & - & - & - & - & - & - & -	\\
R N & log \ch{CaH} & - & - & - & - & - & - & - & - & - & -	\\   \hline
F C & log \ch{TiH} & - & - & - & - & - & - & - & - & - & -	\\
F G & log \ch{TiH} & - & - & - & - & - & - & - & - & - & -	\\
F N & log \ch{TiH} & - & - & - & - & - & - & - & $-10.54_{-0.88}^{+0.91}$ & - & $-10.61_{-0.79}^{+0.61}$	\\
R C & log \ch{TiH} & - & - & - & - & - & - & - & - & - & -	\\
R G & log \ch{TiH} & - & - & - & - & - & - & - & - & - & -	\\
R N & log \ch{TiH} & - & - & - & - & - & - & - & - & - & -	\\ \hline
F C & $p_{\mathrm{t}}$ & - & - & - & - & - & - & - & - & - & -	\\
F G & $p_{\mathrm{t}}$ & $0.58_{-0.08}^{+0.11}$ & $0.31_{-1.30}^{+0.69}$ & $0.71_{-0.12}^{+0.12}$ & $0.68_{-0.13}^{+0.14}$ & $0.21_{-1.33}^{+0.92}$ & $0.72_{-0.09}^{+0.11}$ & $0.32_{-1.13}^{+0.61}$ & $0.40_{-0.30}^{+0.22}$ & $0.32_{-1.52}^{+0.99}$ & $0.87_{-0.11}^{+0.12}$	\\
F N & $p_{\mathrm{t}}$ & $-0.09_{-1.21}^{+1.20}$ & $0.73_{-0.10}^{+0.14}$ & $0.12_{-1.36}^{+1.10}$ & $0.26_{-1.43}^{+0.99}$ & $0.21_{-1.29}^{+0.92}$ & $0.85_{-0.11}^{+0.12}$ & $0.37_{-1.24}^{+0.65}$ & $0.74_{-0.41}^{+0.27}$ & $0.03_{-1.27}^{+1.19}$ & $1.08_{-0.13}^{+0.13}$	\\
R C & $p_{\mathrm{t}}$ & - & - & - & - & - & - & - & - & - & -	\\
R G & $p_{\mathrm{t}}$ & $0.26_{-1.39}^{+0.79}$ & $0.35_{-1.21}^{+0.59}$ & $0.45_{-1.35}^{+0.50}$ & $0.13_{-1.41}^{+1.09}$ & $0.12_{-1.11}^{+0.91}$ & $0.34_{-1.50}^{+0.86}$ & $0.22_{-1.36}^{+0.90}$ & $0.11_{-1.12}^{+0.81}$ & $0.35_{-1.51}^{+0.92}$ & $0.83_{-0.11}^{+0.12}$	\\
R N & $p_{\mathrm{t}}$ & $0.12_{-1.26}^{+1.00}$ & $0.24_{-1.30}^{+0.83}$ & $0.60_{-1.44}^{+0.48}$ & $0.11_{-1.33}^{+1.06}$ & $0.13_{-1.13}^{+0.89}$ & $0.16_{-1.36}^{+1.08}$ & $0.20_{-1.26}^{+0.90}$ & $0.15_{-1.11}^{+0.72}$ & $0.04_{-1.26}^{+1.13}$ & $1.13_{-0.10}^{+0.12}$	\\ \hline
F C & $p_{\mathrm{b}}$ & - & - & - & - & - & - & - & - & - & -	\\
F G & $p_{\mathrm{b}}$ & $0.37_{-0.24}^{+0.35}$ & $0.49_{-0.32}^{+0.33}$ & $0.46_{-0.30}^{+0.33}$ & $0.45_{-0.29}^{+0.34}$ & $0.50_{-0.32}^{+0.32}$ & $0.47_{-0.28}^{+0.31}$ & $0.50_{-0.32}^{+0.31}$ & $0.49_{-0.30}^{+0.30}$ & $0.50_{-0.33}^{+0.33}$ & $0.45_{-0.28}^{+0.33}$	\\
F N & $p_{\mathrm{b}}$ & $0.51_{-0.33}^{+0.32}$ & $0.44_{-0.27}^{+0.31}$ & $0.51_{-0.32}^{+0.32}$ & $0.50_{-0.32}^{+0.32}$ & $0.50_{-0.31}^{+0.31}$ & $0.47_{-0.28}^{+0.30}$ & $0.50_{-0.30}^{+0.31}$ & $0.48_{-0.29}^{+0.31}$ & $0.51_{-0.32}^{+0.31}$ & $0.44_{-0.28}^{+0.33}$	\\
R C & $p_{\mathrm{b}}$ & - & - & - & - & - & - & - & - & - & -	\\
R G & $p_{\mathrm{b}}$ & $0.50_{-0.32}^{+0.31}$ & $0.50_{-0.31}^{+0.32}$ & $0.49_{-0.30}^{+0.32}$ & $0.50_{-0.32}^{+0.33}$ & $0.50_{-0.30}^{+0.30}$ & $0.51_{-0.33}^{+0.32}$ & $0.50_{-0.31}^{+0.32}$ & $0.50_{-0.31}^{+0.30}$ & $0.50_{-0.32}^{+0.33}$ & $0.48_{-0.29}^{+0.32}$	\\
R N & $p_{\mathrm{b}}$ & $0.51_{-0.31}^{+0.30}$ & $0.51_{-0.31}^{+0.30}$ & $0.50_{-0.30}^{+0.31}$ & $0.50_{-0.31}^{+0.32}$ & $0.49_{-0.29}^{+0.30}$ & $0.50_{-0.32}^{+0.31}$ & $0.50_{-0.29}^{+0.30}$ & $0.51_{-0.29}^{+0.28}$ & $0.50_{-0.32}^{+0.31}$ & $0.46_{-0.27}^{+0.30}$	\\  \hline
F C & $\tau$ & - & - & - & - & - & - & - & - & - & -	\\
F G & $\tau$ & $0.96_{-0.3}^{+0.22}$ & $-2.00_{-1.95}^{+2.34}$ & $0.74_{-0.38}^{+0.36}$ & $-0.80_{-0.37}^{+0.33}$ & $-2.51_{-1.59}^{+1.92}$ & $0.91_{-0.28}^{+0.25}$ & $-1.57_{-2.22}^{+1.92}$ & $0.59_{-0.61}^{+0.43}$ & $-2.52_{-1.64}^{+1.98}$ & $0.65_{-0.33}^{+0.39}$	\\
F N & $\tau$ & $-3.23_{-1.14}^{+1.14}$ & $0.87_{-0.35}^{+0.27}$ & $-3.08_{-1.23}^{+1.27}$ & $-2.71_{-1.49}^{+1.70}$ & $-2.55_{-1.54}^{+1.79}$ & $0.87_{-0.30}^{+0.27}$ & $-2.02_{-1.93}^{+2.21}$ & $0.54_{-0.97}^{+0.47}$ & $-2.97_{-1.29}^{+1.29}$ & $0.74_{-0.42}^{+0.36}$	\\
R C & $\tau$ & - & - & - & - & - & - & - & - & - & -	\\
R G & $\tau$ & $-2.52_{-1.58}^{+2.14}$ & $-1.86_{-2.02}^{+1.24}$ & $-1.76_{-2.10}^{+2.49}$ & $-2.57_{-1.57}^{+1.49}$ & $-2.29_{-1.68}^{+1.93}$ & $-2.66_{-1.53}^{+2.00}$ & $-2.57_{-1.50}^{+1.71}$ & $-2.13_{-1.79}^{+2.35}$ & $-2.54_{-1.61}^{+2.04}$ & $0.82_{-0.30}^{+0.30}$	\\
R N & $\tau$ & $-2.81_{-1.36}^{+1.49}$ & $-2.58_{-1.53}^{+1.92}$ & $-1.91_{-1.96}^{+2.52}$ & $-2.71_{-1.46}^{+1.47}$ & $-2.34_{-1.60}^{+1.82}$ & $-2.98_{-1.28}^{+1.36}$ & $-2.70_{-1.41}^{+1.59}$ & $-2.05_{-1.78}^{+2.21}$ & $-2.91_{-1.31}^{+1.35}$ & $0.76_{-0.37}^{+0.32}$	\\  \hline
F C & $Q_0$ & - & - & - & - & - & - & - & - & - & -	\\
F G & $Q_0$ & - & - & - & - & - & - & - & - & - & -	\\
F N & $Q_0$ & $10.36_{-8.07}^{+34.24}$ & $9.16_{-6.88}^{+30.37}$ & $9.96_{-7.70}^{+34.11}$ & $9.95_{-7.68}^{+34.09}$ & $9.92_{-7.52}^{+31.24}$ & $8.92_{-6.66}^{+28.92}$ & $9.73_{-7.39}^{+30.72}$ & $9.67_{-7.26}^{+29.93}$ & $10.08_{-7.78}^{+33.73}$ & $9.37_{-7.06}^{+30.59}$	\\
R C & $Q_0$ & - & - & - & - & - & - & - & - & - & -	\\
R G & $Q_0$ & - & - & - & - & - & - & - & - & - & -	\\
R N & $Q_0$ & $10.43_{-7.86}^{+30.63}$ & $10.08_{-7.59}^{+31.16}$ & $10.01_{-7.59}^{+30.92}$ & $10.17_{-7.83}^{+33.22}$ & $10.54_{-7.91}^{+30.16}$ & $9.81_{-7.48}^{+31.71}$ & $9.50_{-6.99}^{+27.67}$ & $10.09_{-7.45}^{+27.43}$ & $10.05_{-7.71}^{+32.53}$ & $9.85_{-7.28}^{+28.01}$	\\ \hline
F C & $a_0$ & - & - & - & - & - & - & - & - & - & -	\\
F G & $a_0$ & - & - & - & - & - & - & - & - & - & -	\\
F N & $a_0$ & $0.70_{-0.13}^{+0.10}$ & $0.70_{-0.12}^{+0.09}$ & $0.70_{-0.13}^{+0.10}$ & $0.70_{-0.13}^{+0.10}$ & $0.70_{-0.12}^{+0.10}$ & $0.70_{-0.12}^{+0.09}$ & $0.70_{-0.12}^{+0.09}$ & $0.69_{-0.12}^{+0.10}$ & $0.70_{-0.13}^{+0.10}$ & $0.69_{-0.12}^{+0.10}$	\\
R C & $a_0$ & - & - & - & - & - & - & - & - & - & -	\\
R G & $a_0$ & - & - & - & - & - & - & - & - & - & -	\\
R N & $a_0$ & $0.70_{-0.12}^{+0.10}$ & $0.70_{-0.12}^{+0.09}$ & $0.70_{-0.12}^{+0.10}$ & $0.70_{-0.13}^{+0.10}$ & $0.70_{-0.12}^{+0.09}$ & $0.70_{-0.13}^{+0.10}$ & $0.70_{-0.12}^{+0.09}$ & $0.70_{-0.12}^{+0.09}$ & $0.70_{-0.12}^{+0.10}$ & $0.69_{-0.11}^{+0.09}$	\\ \hline
F C & $a$ & - & - & - & - & - & - & - & - & - & -	\\
F G & $a$ & - & - & - & - & - & - & - & - & - & -	\\
F N & $a$ & $0.34_{-0.87}^{+0.88}$ & $0.66_{-0.66}^{+0.64}$ & $0.31_{-0.88}^{+0.89}$ & $0.36_{-0.86}^{+0.86}$ & $0.36_{-0.85}^{+0.83}$ & $0.65_{-0.67}^{+0.65}$ & $0.37_{-0.82}^{+0.83}$ & $0.39_{-0.81}^{+0.79}$ & $0.40_{-0.87}^{+0.82}$ & $0.36_{-0.79}^{+0.82}$	\\
R C & $a$ & - & - & - & - & - & - & - & - & - & -	\\
R G & $a$ & - & - & - & - & - & - & - & - & - & -	\\
R N &$a$ & $0.35_{-0.84}^{+0.83}$ & $0.34_{-0.84}^{+0.82}$ & $0.37_{-0.85}^{+0.83}$ & $0.34_{-0.87}^{+0.85}$ & $0.33_{-0.79}^{+0.81}$ & $0.31_{-0.85}^{+0.86}$ & $0.37_{-0.81}^{+0.81}$ & $0.37_{-0.77}^{+0.76}$ & $0.30_{-0.83}^{+0.87}$ & $0.47_{-0.83}^{+0.74}$	\\
\enddata
\end{deluxetable*}
\end{longrotatetable}

\begin{longrotatetable}
\begin{deluxetable*}{ccccccccccc}
\tablecolumns{11}
\tablewidth{0.9\textwidth}
\tablecaption{Summary of retrieval outcomes for the standard T dwarfs. Only models with the reduced set of molecules are tabulated.  Variations of the models shown are: full spectra cloud-free (FC), full spectra gray (FG), full spectra non-gray (FN), restricted spectra cloud-free (RC), restricted spectra gray (RG) and restricted spectra non-gray (RN).}
\label{tab:data posteriors T dwarfs}
\tablehead{
 \colhead{Model} & \colhead{Parameter} & \colhead{T0} & \colhead{T1} & \colhead{T2} & \colhead{T3} & \colhead{T4} & \colhead{T5} & \colhead{T6} & \colhead{T7} & \colhead{T8} \\
}
\startdata 
F C & log g & $4.36_{-0.32}^{+0.89}$ & $4.46_{-0.07}^{+0.07}$ & $3.72_{-0.08}^{+0.08}$ & $5.36_{-0.12}^{+0.12}$ & $3.72_{-0.10}^{+0.10}$ & $4.83_{-0.10}^{+0.11}$ & $4.61_{-0.25}^{+0.25}$ & $4.92_{-0.13}^{+0.16}$ & $3.67_{-0.10}^{+0.12}$	\\
F G & log g & $4.27_{-0.24}^{+0.32}$ & $4.48_{-0.08}^{+0.09}$ & $3.93_{-0.24}^{+0.22}$ & $5.38_{-0.13}^{+0.14}$ & $3.75_{-0.11}^{+0.18}$ & $4.83_{-0.10}^{+0.11}$ & $4.60_{-0.24}^{+0.24}$ & $4.91_{-0.12}^{+0.15}$ & $3.67_{-0.10}^{+0.12}$	\\
F N & log g & $4.26_{-0.25}^{+0.45}$ & $4.47_{-0.07}^{+0.07}$ & $3.76_{-0.11}^{+0.24}$ & $5.38_{-0.12}^{+0.13}$ & $3.74_{-0.09}^{+0.09}$ & $4.82_{-0.09}^{+0.11}$ & $4.60_{-0.23}^{+0.24}$ & $4.59_{-0.15}^{+0.17}$ & $3.67_{-0.10}^{+0.11}$	\\
R C & log g & $4.06_{-0.19}^{+0.25}$ & $4.50_{-0.10}^{+0.15}$ & $3.71_{-0.11}^{+0.12}$ & $5.37_{-0.11}^{+0.11}$ & $4.36_{-0.23}^{+0.26}$ & $4.73_{-0.13}^{+0.14}$ & $4.53_{-0.52}^{+0.32}$ & $3.91_{-0.17}^{+0.20}$ & $3.66_{-0.10}^{+0.16}$	\\
R G & log g & $4.05_{-0.18}^{+0.20}$ & $4.45_{-0.08}^{+0.09}$ & $3.72_{-0.11}^{+0.12}$ & $5.41_{-0.13}^{+0.21}$ & $4.37_{-0.22}^{+0.24}$ & $4.73_{-0.13}^{+0.13}$ & $4.53_{-0.46}^{+0.30}$ & $3.82_{-0.14}^{+0.16}$ & $3.67_{-0.10}^{+0.15}$	\\
R N & log g & $4.04_{-0.16}^{+0.18}$ & $4.48_{-0.08}^{+0.09}$ & $3.71_{-0.10}^{+0.11}$ & $5.35_{-0.10}^{+0.08}$ & $4.37_{-0.20}^{+0.23}$ & $4.73_{-0.12}^{+0.12}$ & $4.56_{-0.42}^{+0.27}$ & $3.91_{-0.16}^{+0.18}$ & $4.52_{-0.30}^{+0.32}$	\\ \hline
F C & R & $0.68_{-0.10}^{+0.10}$ & $0.48_{-0.03}^{+0.03}$ & $0.94_{-0.04}^{+0.05}$ & $0.66_{-0.06}^{+0.06}$ & $0.79_{-0.10}^{+0.10}$ & $0.71_{-0.04}^{+0.04}$ & $0.85_{-0.10}^{+0.11}$ & $0.69_{-0.04}^{+0.04}$ & $0.63_{-0.12}^{+0.13}$	\\
F G & R & $0.69_{-0.10}^{+0.09}$ & $0.48_{-0.03}^{+0.03}$ & $0.93_{-0.04}^{+0.04}$ & $0.66_{-0.06}^{+0.06}$ & $0.79_{-0.10}^{+0.09}$ & $0.71_{-0.04}^{+0.04}$ & $0.86_{-0.09}^{+0.10}$ & $0.69_{-0.04}^{+0.04}$ & $0.63_{-0.12}^{+0.12}$	\\
F N & R & $0.68_{-0.09}^{+0.09}$ & $0.48_{-0.03}^{+0.03}$ & $0.93_{-0.04}^{+0.04}$ & $0.66_{-0.05}^{+0.06}$ & $0.79_{-0.09}^{+0.09}$ & $0.71_{-0.04}^{+0.04}$ & $0.85_{-0.09}^{+0.09}$ & $0.68_{-0.03}^{+0.03}$ & $0.63_{-0.12}^{+0.12}$	\\
R C & R & $0.72_{-0.11}^{+0.10}$ & $0.49_{-0.04}^{+0.04}$ & $0.99_{-0.05}^{+0.06}$ & $0.61_{-0.06}^{+0.06}$ & $0.69_{-0.09}^{+0.09}$ & $0.69_{-0.05}^{+0.05}$ & $0.84_{-0.13}^{+0.15}$ & $0.81_{-0.06}^{+0.06}$ & $0.52_{-0.10}^{+0.11}$	\\
R G & R & $0.72_{-0.10}^{+0.10}$ & $0.50_{-0.04}^{+0.04}$ & $0.99_{-0.06}^{+0.06}$ & $0.60_{-0.06}^{+0.06}$ & $0.68_{-0.09}^{+0.09}$ & $0.69_{-0.05}^{+0.05}$ & $0.84_{-0.12}^{+0.14}$ & $0.86_{-0.06}^{+0.06}$ & $0.53_{-0.10}^{+0.11}$	\\
R N & R & $0.72_{-0.09}^{+0.09}$ & $0.50_{-0.03}^{+0.04}$ & $0.99_{-0.05}^{+0.05}$ & $0.61_{-0.05}^{+0.06}$ & $0.68_{-0.08}^{+0.08}$ & $0.69_{-0.05}^{+0.05}$ & $0.81_{-0.11}^{+0.12}$ & $0.81_{-0.05}^{+0.06}$ & $0.40_{-0.05}^{+0.07}$	\\ \hline
F C & $\mathrm{T_{eff}}$ & $1333.79_{-31.91}^{+32.71}$ & $1464.51_{-26.14}^{+26.97}$ & $1169.07_{-21.85}^{+22.71}$ & $1295.39_{-36.28}^{+37.30}$ & $1124.45_{-24.47}^{+25.38}$ & $1049.15_{-26.50}^{+30.15}$ & $853.72_{-47.68}^{+52.44}$ & $847.62_{-20.41}^{+21.15}$ & $717.17_{-30.67}^{+32.00}$	\\
F G & $\mathrm{T_{eff}}$ & $1325.20_{-28.88}^{+29.94}$ & $1461.66_{-24.90}^{+26.13}$ & $1176.50_{-22.57}^{+21.59}$ & $1297.46_{-37.22}^{+37.79}$ & $1124.28_{-23.83}^{+24.38}$ & $1047.87_{-25.79}^{+29.33}$ & $852.41_{-45.69}^{+50.57}$ & $845.48_{-19.99}^{+20.98}$ & $715.34_{-30.18}^{+31.38}$	\\
F N & $\mathrm{T_{eff}}$ & $1328.49_{-29.06}^{+29.28}$ & $1465.75_{-24.78}^{+24.85}$ & $1171.62_{-22.00}^{+21.58}$ & $1295.48_{-35.03}^{+36.89}$ & $1120.24_{-22.77}^{+24.30}$ & $1048.70_{-24.88}^{+27.85}$ & $856.36_{-43.37}^{+48.99}$ & $852.59_{-18.69}^{+20.49}$ & $716.90_{-29.83}^{+30.65}$	\\
R C & $\mathrm{T_{eff}}$ & $1220.51_{-34.74}^{+34.67}$ & $1349.59_{-36.57}^{+39.18}$ & $1061.97_{-25.82}^{+26.21}$ & $1227.07_{-47.49}^{+48.33}$ & $1116.11_{-37.32}^{+38.24}$ & $968.86_{-34.22}^{+34.15}$ & $787.84_{-61.32}^{+71.06}$ & $712.02_{-22.57}^{+26.18}$ & $722.01_{-40.77}^{+42.74}$	\\
R G & $\mathrm{T_{eff}}$ & $1218.15_{-33.68}^{+32.98}$ & $1334.25_{-33.94}^{+32.75}$ & $1062.43_{-26.39}^{+28.11}$ & $1237.30_{-48.99}^{+52.22}$ & $1115.92_{-35.27}^{+36.39}$ & $967.32_{-32.82}^{+33.55}$ & $787.84_{-59.38}^{+65.91}$ & $691.80_{-20.36}^{+26.09}$ & $719.76_{-39.73}^{+41.50}$	\\
R N & $\mathrm{T_{eff}}$ & $1217.48_{-32.82}^{+31.47}$ & $1340.13_{-32.94}^{+31.06}$ & $1059.99_{-24.68}^{+25.65}$ & $1221.94_{-41.75}^{+43.60}$ & $1116.21_{-32.88}^{+34.58}$ & $964.77_{-30.74}^{+33.75}$ & $798.66_{-54.31}^{+63.54}$ & $710.72_{-21.75}^{+23.84}$ & $844.14_{-53.89}^{+53.45}$	\\ \hline
F C & log \ch{H2O} & $-3.97_{-0.11}^{+0.35}$ & $-3.81_{-0.05}^{+0.05}$ & $-3.91_{-0.04}^{+0.04}$ & $-3.64_{-0.05}^{+0.05}$ & $-3.76_{-0.05}^{+0.05}$ & $-3.44_{-0.04}^{+0.04}$ & $-3.60_{-0.10}^{+0.10}$ & $-3.02_{-0.06}^{+0.07}$ & $-3.59_{-0.05}^{+0.06}$	\\
F G & log \ch{H2O} & $-4.00_{-0.08}^{+0.11}$ & $-3.81_{-0.05}^{+0.05}$ & $-3.85_{-0.07}^{+0.07}$ & $-3.63_{-0.05}^{+0.06}$ & $-3.75_{-0.05}^{+0.06}$ & $-3.44_{-0.04}^{+0.04}$ & $-3.60_{-0.09}^{+0.10}$ & $-3.02_{-0.06}^{+0.06}$ & $-3.59_{-0.05}^{+0.05}$	\\
F N & log \ch{H2O} & $-4.00_{-0.09}^{+0.15}$ & $-3.81_{-0.04}^{+0.05}$ & $-3.89_{-0.05}^{+0.06}$ & $-3.63_{-0.05}^{+0.05}$ & $-3.74_{-0.05}^{+0.05}$ & $-3.44_{-0.04}^{+0.04}$ & $-3.60_{-0.09}^{+0.10}$ & $-3.07_{-0.05}^{+0.05}$ & $-3.59_{-0.05}^{+0.05}$	\\
R C & log \ch{H2O} & $-4.08_{-0.08}^{+0.09}$ & $-3.82_{-0.06}^{+0.08}$ & $-3.93_{-0.04}^{+0.05}$ & $-3.61_{-0.05}^{+0.06}$ & $-3.50_{-0.12}^{+0.15}$ & $-3.45_{-0.05}^{+0.05}$ & $-3.63_{-0.15}^{+0.14}$ & $-3.24_{-0.05}^{+0.06}$ & $-3.58_{-0.06}^{+0.07}$	\\
R G & log \ch{H2O} & $-4.08_{-0.07}^{+0.07}$ & $-3.84_{-0.05}^{+0.06}$ & $-3.93_{-0.04}^{+0.05}$ & $-3.59_{-0.06}^{+0.10}$ & $-3.50_{-0.12}^{+0.14}$ & $-3.45_{-0.05}^{+0.05}$ & $-3.63_{-0.14}^{+0.13}$ & $-3.26_{-0.04}^{+0.05}$ & $-3.58_{-0.06}^{+0.06}$	\\
R N & log \ch{H2O} & $-4.09_{-0.07}^{+0.07}$ & $-3.83_{-0.05}^{+0.06}$ & $-3.93_{-0.04}^{+0.04}$ & $-3.61_{-0.05}^{+0.05}$ & $-3.49_{-0.11}^{+0.13}$ & $-3.45_{-0.05}^{+0.05}$ & $-3.61_{-0.13}^{+0.12}$ & $-3.24_{-0.05}^{+0.06}$ & $-3.29_{-0.13}^{+0.14}$	\\ \hline
F C & log \ch{CH4} & $-5.09_{-0.17}^{+0.61}$ & $-4.86_{-0.07}^{+0.06}$ & $-4.92_{-0.07}^{+0.06}$ & $-4.09_{-0.07}^{+0.07}$ & $-4.37_{-0.06}^{+0.06}$ & $-3.73_{-0.06}^{+0.06}$ & $-3.75_{-0.15}^{+0.14}$ & $-3.26_{-0.08}^{+0.09}$ & $-3.94_{-0.06}^{+0.07}$	\\
F G & log \ch{CH4} & $-5.13_{-0.13}^{+0.19}$ & $-4.84_{-0.07}^{+0.08}$ & $-4.81_{-0.12}^{+0.14}$ & $-4.08_{-0.07}^{+0.08}$ & $-4.35_{-0.07}^{+0.09}$ & $-3.73_{-0.06}^{+0.06}$ & $-3.76_{-0.14}^{+0.13}$ & $-3.26_{-0.07}^{+0.08}$ & $-3.94_{-0.05}^{+0.06}$	\\
F N & log \ch{CH4} & $-5.13_{-0.13}^{+0.25}$ & $-4.85_{-0.07}^{+0.07}$ & $-4.89_{-0.08}^{+0.12}$ & $-4.08_{-0.07}^{+0.07}$ & $-4.35_{-0.06}^{+0.06}$ & $-3.73_{-0.05}^{+0.06}$ & $-3.75_{-0.14}^{+0.13}$ & $-3.39_{-0.08}^{+0.08}$ & $-3.94_{-0.05}^{+0.06}$	\\
R C & log \ch{CH4} & $-5.23_{-0.12}^{+0.14}$ & $-4.84_{-0.10}^{+0.13}$ & $-4.94_{-0.07}^{+0.07}$ & $-4.06_{-0.07}^{+0.08}$ & $-3.95_{-0.16}^{+0.19}$ & $-3.76_{-0.07}^{+0.07}$ & $-3.80_{-0.27}^{+0.19}$ & $-3.73_{-0.08}^{+0.11}$ & $-3.94_{-0.06}^{+0.08}$	\\
R G & log \ch{CH4} & $-5.24_{-0.11}^{+0.11}$ & $-4.89_{-0.08}^{+0.09}$ & $-4.94_{-0.07}^{+0.07}$ & $-4.03_{-0.09}^{+0.14}$ & $-3.95_{-0.16}^{+0.18}$ & $-3.76_{-0.07}^{+0.07}$ & $-3.79_{-0.25}^{+0.18}$ & $-3.79_{-0.06}^{+0.08}$ & $-3.94_{-0.06}^{+0.08}$	\\
R N & log \ch{CH4} & $-5.24_{-0.11}^{+0.11}$ & $-4.86_{-0.08}^{+0.09}$ & $-4.94_{-0.07}^{+0.07}$ & $-4.07_{-0.06}^{+0.06}$ & $-3.94_{-0.15}^{+0.17}$ & $-3.76_{-0.07}^{+0.07}$ & $-3.77_{-0.23}^{+0.16}$ & $-3.73_{-0.08}^{+0.10}$ & $-3.47_{-0.18}^{+0.18}$	\\ \hline
F C & log \ch{NH3} & $-5.63_{-1.11}^{+0.70}$ & - & - & - & - & $-5.39_{-0.40}^{+0.18}$ & $-4.97_{-0.21}^{+0.18}$ & - & -	\\
F G & log \ch{NH3} & $-5.76_{-1.32}^{+0.35}$ & - & - & - & - & $-5.39_{-0.37}^{+0.18}$ & $-4.97_{-0.20}^{+0.18}$ & - & -	\\
F N & log \ch{NH3} & $-5.77_{-1.79}^{+0.47}$ & - & - & - & - & $-5.39_{-0.33}^{+0.17}$ & $-4.97_{-0.20}^{+0.18}$ & - & -	\\
R C & log \ch{NH3} & $-6.09_{-3.08}^{+0.46}$ & - & - & - & - & $-5.65_{-3.17}^{+0.31}$ & $-5.02_{-0.34}^{+0.24}$ & - & -	\\
R G & log \ch{NH3} & $-6.13_{-3.12}^{+0.43}$ & - & - & - & - & $-5.65_{-3.21}^{+0.31}$ & $-5.01_{-0.32}^{+0.23}$ & - & -	\\
R N & log \ch{NH3} & $-6.22_{-3.22}^{+0.47}$ & - & - & - & - & $-5.62_{-2.82}^{+0.28}$ & $-5.00_{-0.31}^{+0.22}$ & - & -	\\ \hline
F C & log \ch{CO2} & - & - & - & - & - & - & - & - & -	\\
F G & log \ch{CO2} & - & - & - & - & - & - & - & - & -	\\
F N & log \ch{CO2} & - & - & - & - & - & - & - & - & -	\\
R C & log \ch{CO2} & - & - & - & - & - & - & - & - & -	\\
R G & log \ch{CO2} & - & - & - & - & - & - & - & - & -	\\
R N & log \ch{CO2} & - & - & - & - & - & - & - & - & -	\\ \hline
F C & log \ch{CO} & - & - & $-3.69_{-0.29}^{+0.25}$ & $-2.85_{-0.21}^{+0.19}$ & - & - & - & - & -	\\
F G & log \ch{CO} & - & - & $-3.66_{-0.23}^{+0.19}$ & $-2.83_{-0.21}^{+0.18}$ & - & - & - & - & -	\\
F N & log \ch{CO} & - & - & $-3.69_{-0.24}^{+0.21}$ & $-2.84_{-0.20}^{+0.18}$ & - & - & - & - & -	\\
R C & log \ch{CO} & - & - & $-3.80_{-0.26}^{+0.24}$ & $-2.87_{-0.24}^{+0.20}$ & - & - & - & - & -	\\
R G & log \ch{CO} & - & - & $-3.80_{-0.26}^{+0.23}$ & $-2.82_{-0.23}^{+0.21}$ & - & - & - & - & -	\\
R N & log \ch{CO} & - & - & $-3.81_{-0.25}^{+0.22}$ & $-2.84_{-0.21}^{+0.19}$ & - & - & - & - & -	\\ \hline
F C & log \ch{H2S} & - & - & - & - & - & - & - & - & -	\\
F G & log \ch{H2S} & - & - & - & - & - & - & - & - & -	\\
F N & log \ch{H2S} & - & - & - & - & - & - & - & - & -	\\
R C & log \ch{H2S} & - & - & - & - & - & - & - & - & -	\\
R G & log \ch{H2S} & - & - & - & - & - & - & - & - & -	\\
R N & log \ch{H2S} & - & - & - & - & - & - & - & - & -	\\ \hline
F C & log \ch{K} & $-7.29_{-0.16}^{+0.17}$ & $-6.46_{-0.10}^{+0.10}$ & $-6.78_{-0.08}^{+0.08}$ & $-7.04_{-0.11}^{+0.11}$ & $-6.51_{-0.08}^{+0.08}$ & $-6.87_{-0.05}^{+0.06}$ & $-7.17_{-0.12}^{+0.13}$ & $-6.68_{-0.08}^{+0.08}$ & $-7.07_{-0.11}^{+0.11}$	\\
F G & log \ch{K} & $-7.32_{-0.14}^{+0.15}$ & $-6.46_{-0.10}^{+0.10}$ & $-6.79_{-0.07}^{+0.07}$ & $-7.02_{-0.11}^{+0.12}$ & $-6.52_{-0.08}^{+0.08}$ & $-6.87_{-0.05}^{+0.06}$ & $-7.16_{-0.11}^{+0.12}$ & $-6.68_{-0.08}^{+0.08}$ & $-7.07_{-0.11}^{+0.11}$	\\
F N & log \ch{K} & $-7.31_{-0.15}^{+0.15}$ & $-6.44_{-0.10}^{+0.09}$ & $-6.79_{-0.07}^{+0.07}$ & $-7.04_{-0.11}^{+0.11}$ & $-6.49_{-0.08}^{+0.07}$ & $-6.87_{-0.05}^{+0.05}$ & $-7.15_{-0.11}^{+0.12}$ & $-6.86_{-0.12}^{+0.10}$ & $-7.05_{-0.11}^{+0.11}$	\\
R C & log \ch{K} & - & $-6.47_{-0.92}^{+0.43}$ & $-6.83_{-1.93}^{+0.49}$ & - & - & - & - & - & -	\\
R G & log \ch{K} & - & $-7.27_{-2.17}^{+0.60}$ & $-6.87_{-1.92}^{+0.51}$ & - & - & - & - & $-5.99_{-3.27}^{+0.42}$ & -	\\
R N & log \ch{K} & - & $-6.69_{-1.18}^{+0.46}$ & $-6.90_{-1.89}^{+0.51}$ & - & - & - & - & - & -	\\ \hline
F C & log \ch{CrH} & $-9.01_{-0.18}^{+0.32}$ & - & $-9.47_{-0.21}^{+0.17}$ & - & - & - & - & - & -	\\
F G & log \ch{CrH} & $-9.06_{-0.15}^{+0.14}$ & - & $-9.40_{-0.20}^{+0.16}$ & - & - & - & - & - & -	\\
F N & log \ch{CrH} & $-9.06_{-0.16}^{+0.18}$ & - & - & - & - & - & - & - & -	\\
R C & log \ch{CrH} & - & - & - & - & - & - & - & - & -	\\
R G & log \ch{CrH} & - & - & - & - & - & - & - & - & -	\\
R N & log \ch{CrH} & - & - & - & - & - & - & - & - & $-7.16_{-0.50}^{+0.30}$	\\ \hline
F C & log \ch{FeH} & - & - & - & - & $-9.44_{-0.20}^{+0.17}$ & $-9.60_{-0.12}^{+0.11}$ & $-10.25_{-0.76}^{+0.33}$ & $-10.12_{-0.96}^{+0.38}$ & -	\\
F G & log \ch{FeH} & - & - & - & - & $-9.42_{-0.19}^{+0.16}$ & $-9.60_{-0.12}^{+0.11}$ & $-10.24_{-0.72}^{+0.32}$ & $-10.11_{-0.92}^{+0.38}$ & -	\\
F N & log \ch{FeH} & - & - & - & - & $-9.42_{-0.18}^{+0.16}$ & $-9.62_{-0.14}^{+0.11}$ & $-10.25_{-0.69}^{+0.31}$ & - & -	\\
R C & log \ch{FeH} & - & - & - & - & $-7.27_{-0.20}^{+0.20}$ & $-7.96_{-0.18}^{+0.15}$ & - & $-7.79_{-0.25}^{+0.19}$ & -	\\
R G & log \ch{FeH} & - & - & - & - & $-7.27_{-0.19}^{+0.18}$ & $-7.96_{-0.17}^{+0.15}$ & - & $-8.69_{-2.13}^{+0.90}$ & -	\\
R N & log \ch{FeH} & - & - & - & - & $-7.26_{-0.17}^{+0.17}$ & $-7.96_{-0.15}^{+0.13}$ & - & $-7.79_{-0.24}^{+0.17}$ & -	\\ \hline
F C & log \ch{CaH} & - & - & - & - & $3.52_{-0.12}^{+0.12}$ & - & - & - & -	\\
F G & log \ch{CaH} & - & - & - & - & - & - & - & - & -	\\
F N & log \ch{CaH} & - & - & - & - & $-3.59_{-4.94}^{+0.91}$ & - & - & - & -	\\
R C & log \ch{CaH} & - & - & - & - & - & - & - & - & -	\\
R G & log \ch{CaH} & - & - & - & - & - & - & - & - & -	\\
R N & log \ch{CaH} & - & - & - & - & - & - & - & - & -	\\ \hline
F C & log \ch{TiH} & - & - & - & - & - & $-10.50_{-0.55}^{+0.28}$ & $-10.24_{-0.52}^{+0.32}$ & $-10.09_{-0.74}^{+0.37}$ & -	\\
F G & log \ch{TiH} & - & - & - & - & - & $-10.50_{-0.52}^{+0.28}$ & $-10.24_{-0.51}^{+0.31}$ & $-10.08_{-0.69}^{+0.35}$ & -	\\
F N & log \ch{TiH} & - & - & - & - & - & $-10.54_{-0.57}^{+0.30}$ & $-10.24_{-0.48}^{+0.31}$ & - & -	\\
R C & log \ch{TiH} & - & $-8.60_{-1.83}^{+0.45}$ & - & - & - & - & - & - & -	\\
R G & log \ch{TiH} & - & - & - & - & - & - & - & - & -	\\
R N & log \ch{TiH} & - & $-8.93_{-1.78}^{+0.58}$ & - & - & - & - & - & - & -	\\ \hline
F C & $p_{\mathrm{t}}$ & - & - & - & - & - & - & - & - & -	\\
F G & $p_{\mathrm{t}}$ & $0.54_{-1.60}^{+0.57}$ & $0.58_{-1.60}^{+0.54}$ & $0.55_{-0.67}^{+0.17}$ & $0.50_{-1.76}^{+0.96}$ & $0.47_{-1.51}^{+0.66}$ & $0.12_{-1.44}^{+1.17}$ & $0.08_{-1.39}^{+1.16}$ & $0.14_{-1.47}^{+1.13}$ & $0.20_{-1.49}^{+1.03}$	\\
F N & $p_{\mathrm{t}}$ & $0.33_{-1.46}^{+0.80}$ & $0.42_{-1.52}^{+0.77}$ & $0.53_{-1.33}^{+0.44}$ & $0.29_{-1.54}^{+1.16}$ & $0.28_{-1.44}^{+0.90}$ & $0.36_{-1.58}^{+0.89}$ & $0.09_{-1.39}^{+1.14}$ & $0.40_{-0.16}^{+0.12}$ & $0.19_{-1.47}^{+1.03}$	\\
R C & $p_{\mathrm{t}}$ & - & - & - & - & - & - & - & - & -	\\
R G & $p_{\mathrm{t}}$ & $0.22_{-1.38}^{+0.92}$ & $0.41_{-1.52}^{+0.73}$ & $0.32_{-1.46}^{+0.82}$ & $0.76_{-1.86}^{+0.68}$ & $0.22_{-1.43}^{+1.00}$ & $0.09_{-1.37}^{+1.15}$ & $0.07_{-1.36}^{+1.11}$ & $0.25_{-1.45}^{+0.94}$ & $0.27_{-1.51}^{+0.93}$	\\
R N & $p_{\mathrm{t}}$ & $0.22_{-1.33}^{+0.92}$ & $0.25_{-1.40}^{+0.92}$ & $0.29_{-1.40}^{+0.86}$ & $0.09_{-1.37}^{+1.16}$ & $0.20_{-1.39}^{+0.99}$ & $0.11_{-1.36}^{+1.12}$ & $0.09_{-1.35}^{+1.10}$ & $0.23_{-1.42}^{+0.96}$ & $0.23_{-1.43}^{+0.90}$	\\ \hline
F C & $p_{\mathrm{b}}$ & - & - & - & - & - & - & - & - & -	\\
F G & $p_{\mathrm{b}}$ & $0.50_{-0.32}^{+0.32}$ & $0.50_{-0.32}^{+0.32}$ & $0.49_{-0.31}^{+0.32}$ & $0.50_{-0.33}^{+0.33}$ & $0.49_{-0.32}^{+0.33}$ & $0.51_{-0.34}^{+0.33}$ & $0.50_{-0.33}^{+0.33}$ & $0.50_{-0.34}^{+0.33}$ & $0.50_{-0.33}^{+0.33}$	\\
F N & $p_{\mathrm{b}}$ & $0.50_{-0.32}^{+0.31}$ & $0.51_{-0.32}^{+0.31}$ & $0.49_{-0.31}^{+0.31}$ & $0.50_{-0.33}^{+0.32}$ & $0.50_{-0.32}^{+0.32}$ & $0.52_{-0.34}^{+0.32}$ & $0.50_{-0.33}^{+0.33}$ & $0.54_{-0.31}^{+0.29}$ & $0.50_{-0.32}^{+0.32}$	\\
R C & $p_{\mathrm{b}}$ & - & - & - & - & - & - & - & - & -	\\
R G & $p_{\mathrm{b}}$ & $0.50_{-0.31}^{+0.32}$ & $0.51_{-0.32}^{+0.31}$ & $0.50_{-0.32}^{+0.31}$ & $0.50_{-0.33}^{+0.33}$ & $0.50_{-0.32}^{+0.32}$ & $0.50_{-0.32}^{+0.32}$ & $0.50_{-0.32}^{+0.33}$ & $0.50_{-0.32}^{+0.32}$ & $0.51_{-0.33}^{+0.32}$	\\
R N & $p_{\mathrm{b}}$ & $0.51_{-0.31}^{+0.30}$ & $0.50_{-0.31}^{+0.31}$ & $0.51_{-0.31}^{+0.31}$ & $0.51_{-0.33}^{+0.32}$ & $0.51_{-0.31}^{+0.31}$ & $0.51_{-0.32}^{+0.31}$ & $0.50_{-0.31}^{+0.32}$ & $0.50_{-0.32}^{+0.32}$ & $0.50_{-0.31}^{+0.32}$	\\ \hline
F C & $\tau$ & - & - & - & - & - & - & - & - & -	\\
F G & $\tau$ & $-2.19_{-1.81}^{+2.29}$ & $-2.19_{-1.86}^{+2.50}$ & $0.4_{-3.43}^{+0.55}$ & $-2.55_{-1.63}^{+2.78}$ & $-2.17_{-1.86}^{+2.69}$ & $-2.99_{-1.34}^{+1.37}$ & $-2.76_{-1.49}^{+1.52}$ & $-2.86_{-1.43}^{+1.49}$ & $-2.68_{-1.52}^{+1.68}$	\\
F N & $\tau$ & $-2.63_{-1.53}^{+2.03}$ & $-2.66_{-1.50}^{+2.18}$ & $-1.62_{-2.19}^{+2.41}$ & $-2.77_{-1.46}^{+1.97}$ & $-2.68_{-1.48}^{+1.85}$ & $-2.77_{-1.49}^{+1.88}$ & $-2.76_{-1.48}^{+1.52}$ & $0.04_{-0.15}^{+0.16}$ & $-2.66_{-1.53}^{+1.68}$	\\
R C & $\tau$ & - & - & - & - & - & - & - & - & -	\\
R G & $\tau$ & $-2.64_{-1.50}^{+1.66}$ & $-2.56_{-1.56}^{+2.31}$ & $-2.51_{-1.58}^{+2.06}$ & $-2.21_{-1.85}^{+2.90}$ & $-2.78_{-1.41}^{+1.60}$ & $-2.92_{-1.35}^{+1.41}$ & $-2.66_{-1.52}^{+1.57}$ & $-2.67_{-1.50}^{+1.71}$ & $-2.52_{-1.60}^{+1.83}$	\\
R N & $\tau$ & $-2.70_{-1.43}^{+1.57}$ & $-2.79_{-1.38}^{+1.66}$ & $-2.60_{-1.50}^{+1.76}$ & $-2.92_{-1.36}^{+1.41}$ & $-2.76_{-1.43}^{+1.58}$ & $-2.91_{-1.34}^{+1.42}$ & $-2.64_{-1.51}^{+1.52}$ & $-2.66_{-1.50}^{+1.67}$ & $-2.48_{-1.59}^{+1.82}$	\\ \hline
F C & $Q_0$ & - & - & - & - & - & - & - & - & -	\\
F G & $Q_0$ & - & - & - & - & - & - & - & - & -	\\
F N & $Q_0$ & $9.83_{-7.48}^{+32.86}$ & $10.06_{-7.70}^{+33.82}$ & $9.56_{-7.25}^{+31.66}$ & $9.91_{-7.67}^{+35.19}$ & $9.89_{-7.57}^{+32.32}$ & $10.78_{-8.51}^{+36.02}$ & $10.24_{-8.02}^{+34.98}$ & $18.68_{-14.39}^{+35.32}$ & $10.18_{-7.87}^{+34.11}$	\\
R C & $Q_0$ & - & - & - & - & - & - & - & - & -	\\
R G & $Q_0$ & - & - & - & - & - & - & - & - & -	\\
R N & $Q_0$ & $9.85_{-7.43}^{+30.42}$ & $10.32_{-7.86}^{+31.79}$ & $9.96_{-7.57}^{+30.88}$ & $10.25_{-7.99}^{+35.31}$ & $10.08_{-7.68}^{+32.27}$ & $10.14_{-7.78}^{+32.93}$ & $10.13_{-7.81}^{+33.77}$ & $10.22_{-7.88}^{+33.94}$ & $9.78_{-7.45}^{+31.84}$	\\ \hline
F C & $a_0$ & - & - & - & - & - & - & - & - & -	\\
F G & $a_0$ & - & - & - & - & - & - & - & - & -	\\
F N & $a_0$ & $0.70_{-0.13}^{+0.10}$ & $0.70_{-0.13}^{+0.10}$ & $0.70_{-0.12}^{+0.10}$ & $0.70_{-0.13}^{+0.10}$ & $0.70_{-0.13}^{+0.10}$ & $0.70_{-0.13}^{+0.10}$ & $0.70_{-0.13}^{+0.10}$ & $0.69_{-0.09}^{+0.08}$ & $0.70_{-0.13}^{+0.10}$	\\
R C & $a_0$ & - & - & - & - & - & - & - & - & -	\\
R G & $a_0$ & - & - & - & - & - & - & - & - & -	\\
R N & $a_0$ & $0.70_{-0.12}^{+0.10}$ & $0.70_{-0.12}^{+0.10}$ & $0.70_{-0.12}^{+0.10}$ & $0.70_{-0.13}^{+0.10}$ & $0.70_{-0.12}^{+0.10}$ & $0.70_{-0.13}^{+0.10}$ & $0.70_{-0.13}^{+0.10}$ & $0.70_{-0.13}^{+0.10}$ & $0.70_{-0.13}^{+0.10}$ \\ \hline
F C & $a$ & - & - & - & - & - & - & - & - & -	\\
F G & $a$ & - & - & - & - & - & - & - & - & -	\\
F N & $a$ & $0.35_{-0.85}^{+0.86}$ & $0.35_{-0.84}^{+0.84}$ & $0.41_{-0.81}^{+0.80}$ & $0.33_{-0.88}^{+0.88}$ & $0.36_{-0.86}^{+0.86}$ & $0.15_{-0.89}^{+1.03}$ & $0.30_{-0.88}^{+0.91}$ & $-0.81_{-0.11}^{+0.14}$ & $0.34_{-0.89}^{+0.88}$	\\
R C & $a$ & - & - & - & - & - & - & - & - & -	\\
R G & $a$ & - & - & - & - & - & - & - & - & -	\\
R N & $a$ & $0.34_{-0.81}^{+0.83}$ & $0.35_{-0.84}^{+0.84}$ & $0.37_{-0.83}^{+0.83}$ & $0.30_{-0.88}^{+0.89}$ & $0.33_{-0.86}^{+0.86}$ & $0.30_{-0.86}^{+0.88}$ & $0.32_{-0.87}^{+0.88}$ & $0.32_{-0.87}^{+0.88}$ & $0.32_{-0.88}^{+0.87}$ \\
\enddata
\end{deluxetable*}
\end{longrotatetable}

\clearpage

\section{Impact of prior choice}
\label{sect: Impact of prior choice}

As mentioned in Section~\ref{sec:priors}, we retrieve for mixing ratios of the following species: \ch{H2O}, \ch{CH4}, \ch{NH3}, \ch{CO2}, \ch{CO}, \ch{H2S}, \ch{CrH}, \ch{FeH}, \ch{CaH}, \ch{TiH}, and \ch{K}. In the case of the L5 brown dwarf, we first constrained both \ch{CO} and \ch{CH4} for the restricted-wavelength, reduced, non-gray retrieval, which goes against general expectation (e.g, \citet{Fegley1996ApJ} and \citet{Hubeny2007ApJ}). To check this issue, we repeat the retrieval but exclude \ch{CH4} as chemical species to retrieve for. Discarding \ch{CH4} resulted in an decreased value of $\log{g}$ from $6.19_{-0.25}^{+0.20}$ to $5.83_{-0.22}^{+0.30}$ in our retrieval, without changing other quantities remarkably. Thus, high $\log{g}$ values for non-gray cloud retrievals may therefore also be reasoned on the basis of the selected prior values. Still, no clouds parameters can be constrained. \\ 

\begin{figure}[!h]
\begin{center}
\includegraphics[width=\columnwidth]{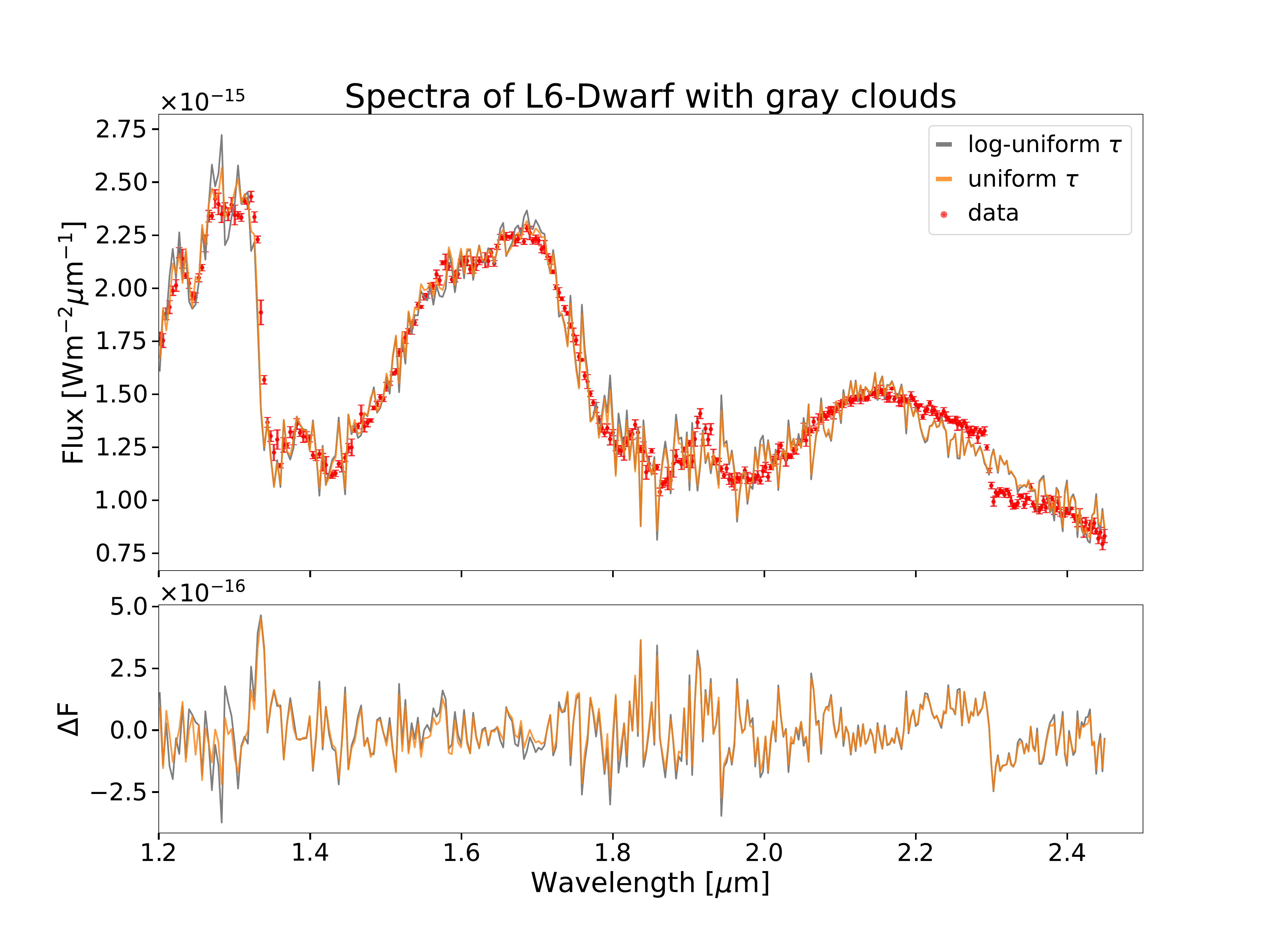}
\end{center}
\vspace{-0.15in}
\caption{Median restricted spectra (F) and residuals ($\Delta$F) associated with the L6 dwarf of our curated sample, comparing the gray-cloud model with the original log-uniform prior for $\tau$ (gray line) and the adjusted uniform prior (orange line). Data are shown as dots with associated uncertainties.}
\label{fig:L6 spectra tau adjustment}
\end{figure}

\begin{figure*}
\begin{center}
\includegraphics[width=\columnwidth]{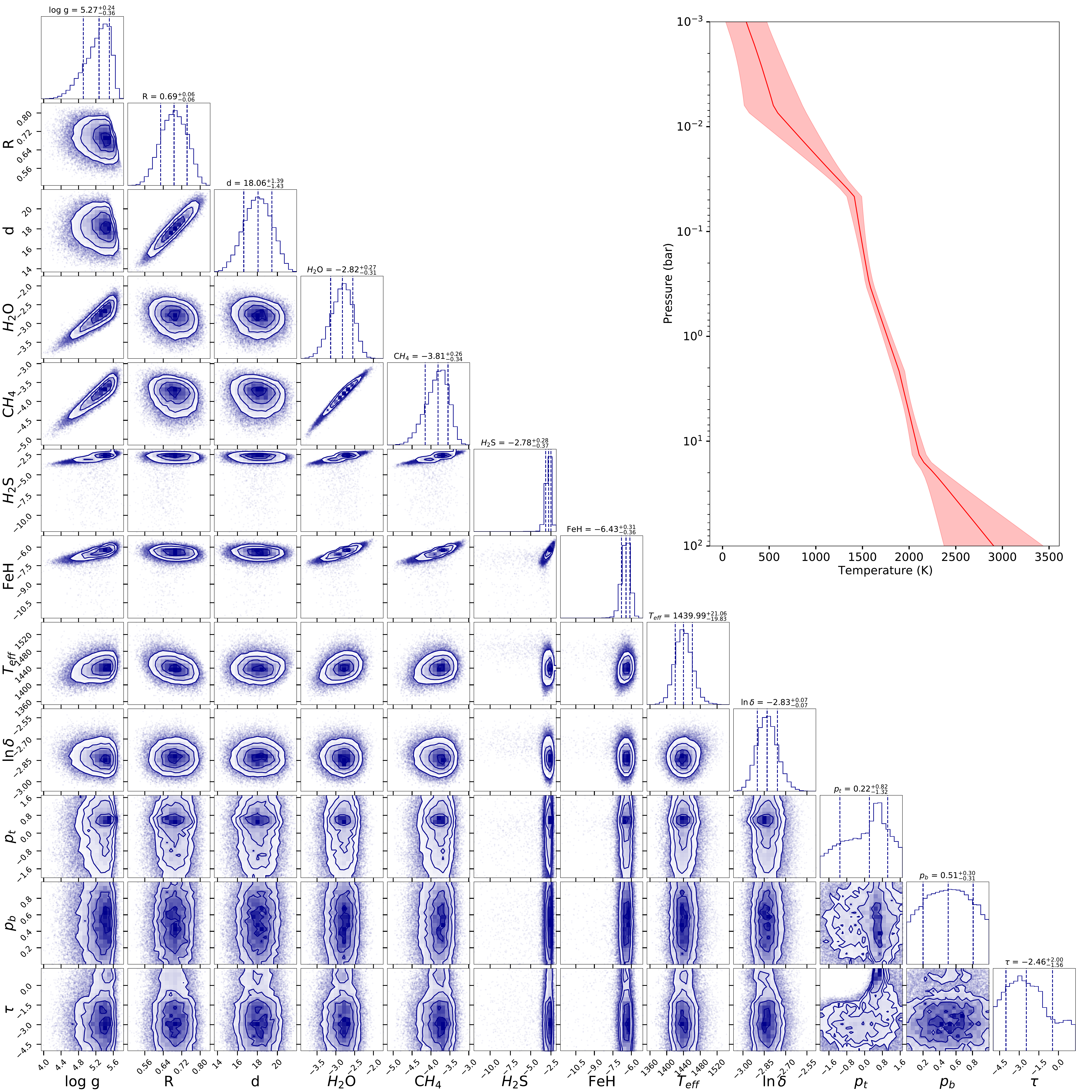}
\includegraphics[width=\columnwidth]{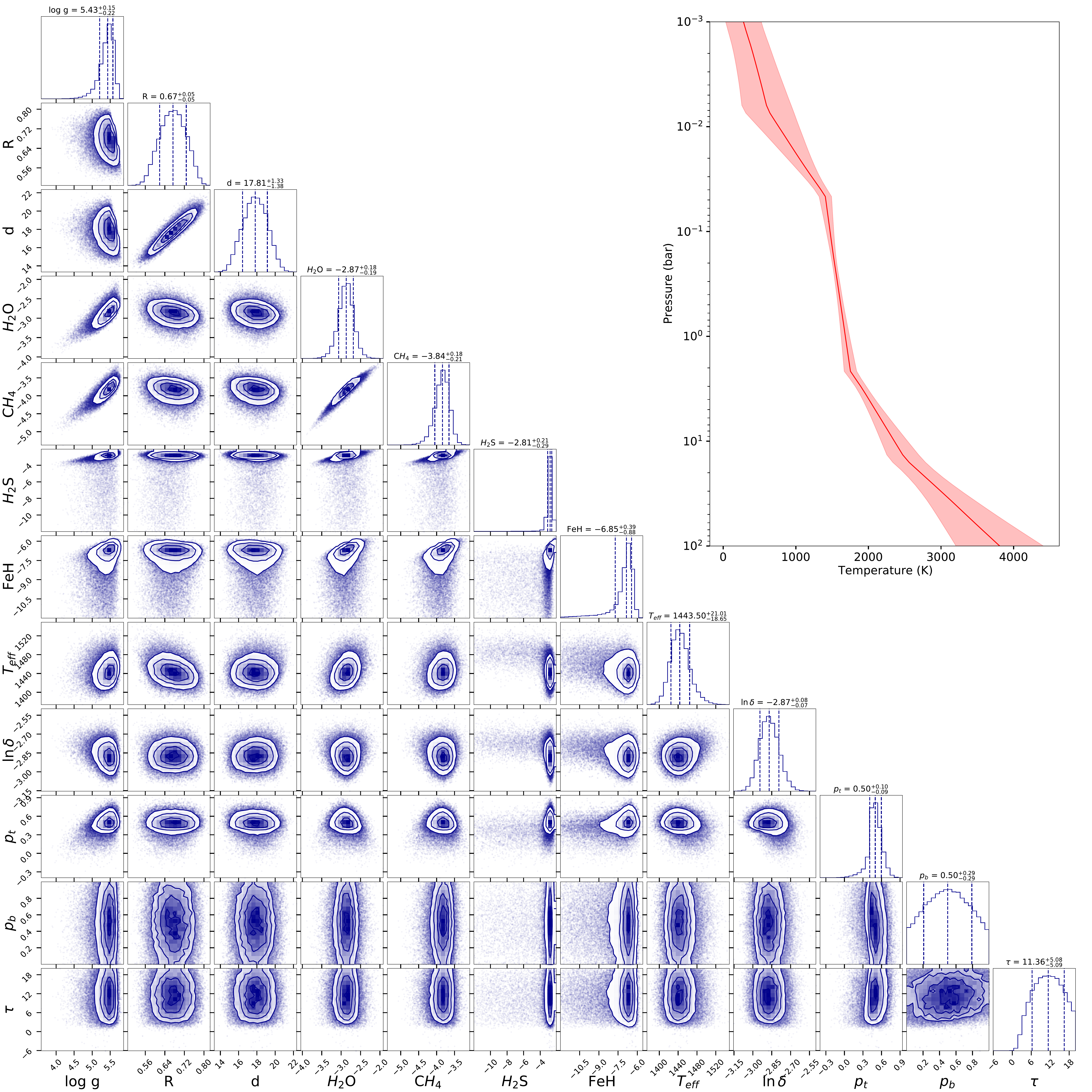}
\end{center}
\vspace{-0.1in}
\caption{Joint posterior distributions from the free-chemistry retrieval analysis of the spectrum with a restricted wavelength range and gray clouds for the L6 standard brown dwarf of our curated sample. Left panel: log-uniform prior for $\tau$.  Right panel: uniform prior for $\tau$. See Figures~\ref{fig:posteriors_appendix} \& \ref{fig:posteriors_appendix2} for details on the posterior plots. The graphs in both upper right corners show the retrieved temperature-pressure profiles. The solid line corresponds to the median profile, while the shaded area correspond to the 1-$\sigma$ confidence intervals.}
\label{fig:L6 tau adjustment}
\end{figure*}

In Section~\ref{sec:lt_clouds} we describe the inability to retrieve cloud properties, especially that in most instances only upper limits for the cloud properties are obtained. To investigate the potential impact of our prior choice of the optical depth on the other retrieval parameters, we show the spectra and joint posterior distributions from the free-chemistry retrieval analysis of the spectrum with a restricted wavelength range and gray clouds for the L6 standard brown dwarf of our curated sample (see Figures~\ref{fig:L6 spectra tau adjustment} \& \ref{fig:L6 tau adjustment}).

\begin{figure}
\centering
\includegraphics[width=\columnwidth]{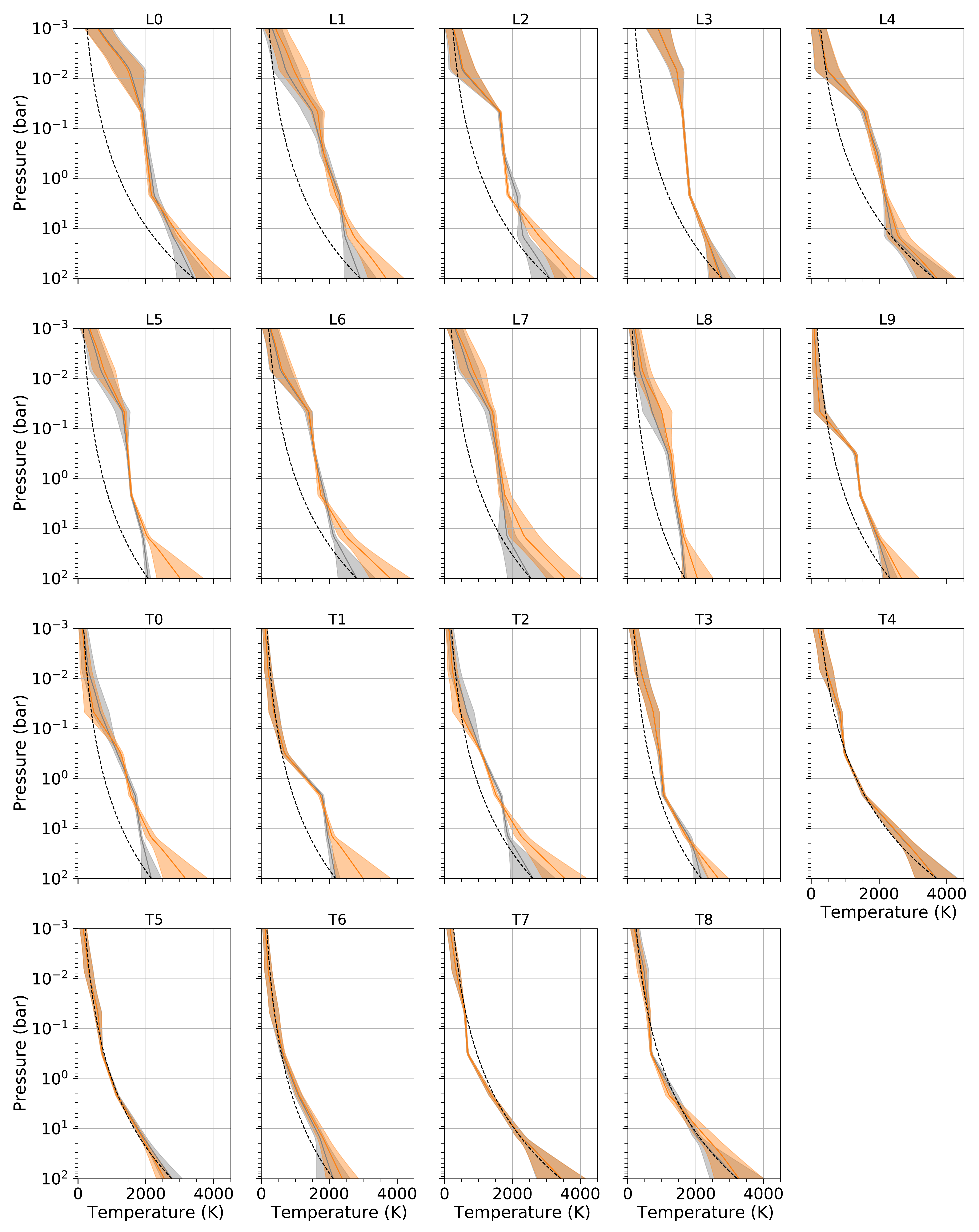}
\caption{Retrieved median temperature-pressure profiles with their associated $1\sigma$ uncertainties, analogous to Figure~\ref{fig:T-P sequence}. In each panel, we compare the gray-cloud models with the original prior of optical depth (gray line) and the adjusted one (orange line) models. Adiabatic profiles (black dashed line) are indicated for comparison.}
\label{fig:T-P sequence newtau}
\end{figure}

\begin{figure}
\begin{center}
\includegraphics[width=\columnwidth]{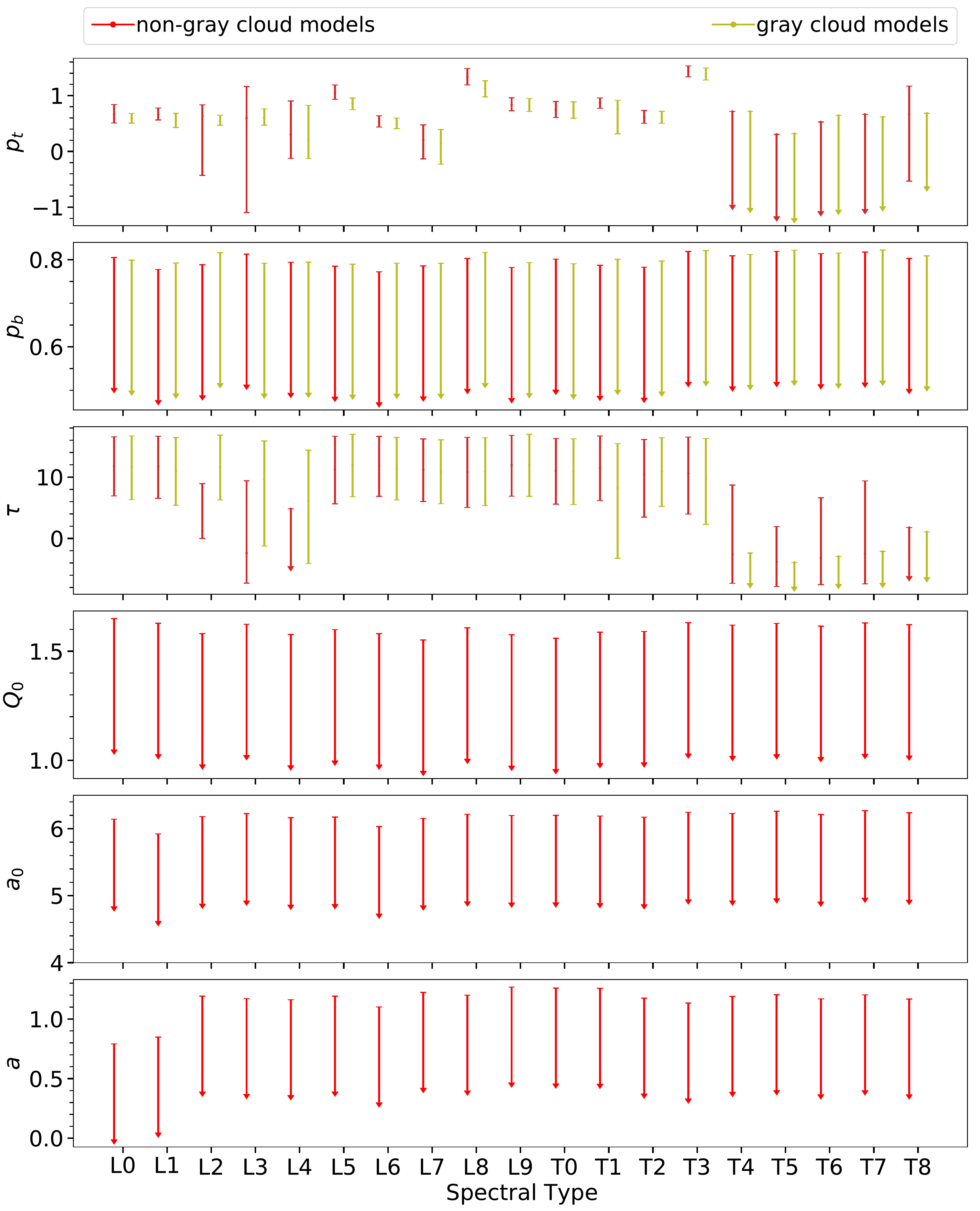}
\end{center}
\vspace{-0.1in}
\caption{Retrieved cloud parameters from our suite of brown dwarf retrievals across the L-T sequence for the cloud models and a restricted wavelength range. In each panel, gray-cloud (olive-green line) and non-gray-cloud (red line) models are compared. Parameters $p_{\mathrm{t}}$ and $p_{\mathrm{b}}$ represent the cloud top and bottom pressures and $\tau$ is optical depth. $Q_0$ is the proxy for the cloud particle composition, $a$ is the monodisperse particle radius and $a_0$ is the power-law index that describes wavelength variation.}
\label{fig:cloud sequence newtau}
\end{figure}

Interestingly, changing the prior of $\tau$ to a uniform distribution (between -10 and 20) results in now constraining an optical depth. The same behavior is also found for most of the other gray and non-gray cloud retrievals. Figure~\ref{fig:cloud sequence newtau} shows the retrieved cloud parameters from our suite of brown dwarfs across the L-T sequence with the gray and non-gray cloud model and a restricted wavelength range. Values of $\tau$ and $p_t$ are mostly constrained up to T3 dwarf whereas for later T dwarfs only upper and lower limits can be found.

Not only do we now retrieve optical depths that indicate a cloud existence, but also the temperature-pressure profile changes at the lower atmosphere without significantly changing values of the other retrieved quantities (see Figure~\ref{fig:T-P sequence newtau}). This indicates that we are still having a prior dependency when considering retrieving for clouds. Thus, further investigations are needed.

\end{document}